\PassOptionsToPackage{pdfpagelabels=false}{hyperref}
\documentclass{aa}
\usepackage[varg]{txfonts}
\pdfoutput=1
\usepackage{graphicx}
\usepackage{newtxtext,newtxmath}
\usepackage{xspace}
\usepackage{bm}
\usepackage{longtable}
\RequirePackage[colorlinks=true,linkcolor=blue,citecolor=blue,urlcolor=blue]{hyperref}

\newcommand{\msun}{\,M$_{\odot}$\xspace}

\begin{document}

\title{Introducing the TNG-Cluster Simulation: overview and physical properties of the gaseous intracluster medium}
\titlerunning{Introducing the TNG-Cluster Simulation}
\author{Dylan Nelson\inst{1}\thanks{E-mail: dnelson@uni-heidelberg.de}
\and Annalisa Pillepich\inst{2}
\and Mohammadreza Ayromlou\inst{1}
\and Wonki Lee$^{4,5}$
\and Katrin Lehle\inst{1}
\and \\Eric Rohr\inst{2}
\and Nhut Truong\inst{3,2}
}

\institute{Universit\"{a}t Heidelberg, Zentrum f\"{u}r Astronomie, ITA, Albert-Ueberle-Str. 2, 69120 Heidelberg, Germany \label{1}
\and Max-Planck-Institut f\"{u}r Astronomie, K\"{o}nigstuhl 17, 69117 Heidelberg, Germany \label{2}
\and NASA/Goddard Space Flight Center, Greenbelt, MD 20771, USA \label{3}
\and Yonsei University, Department of Astronomy, Seoul, Republic of Korea \label{4}
\and Harvard-Smithsonian Center for Astrophysics, 60 Garden Street, Cambridge, MA, 02138, USA \label{5}
}

\date{}

\abstract{
We introduce the new TNG-Cluster project, an addition to the IllustrisTNG suite of cosmological magnetohydrodynamical simulations of galaxy formation. Our objective is to significantly increase the statistical sampling of the most massive and rare objects in the Universe: galaxy clusters with $\log{(M_{\rm 200c} / \rm{M}_\odot)} \gtrsim 14.3 - 15.4$ at $z=0$. To do so, we re-simulate 352 cluster regions drawn from a 1 Gpc volume, thirty-six times larger than TNG300, keeping entirely fixed the IllustrisTNG physical model as well as the numerical resolution. This new sample of hundreds of massive galaxy clusters enables studies of the assembly of high-mass ellipticals and their supermassive black holes (SMBHs), brightest cluster galaxies (BCGs), satellite galaxy evolution and environmental processes, jellyfish galaxies, intracluster medium (ICM) properties, cooling and active galactic nuclei (AGN) feedback, mergers and relaxedness, magnetic field amplification, chemical enrichment, and the galaxy-halo connection at the high-mass end, with observables from the optical to radio synchrotron and the Sunyaev-Zeldovich (SZ) effect, to X-ray emission, as well as their cosmological applications. We present an overview of the simulation, the cluster sample, selected comparisons to data, and a first look at the diversity and physical properties of our simulated clusters and their hot ICM.
}


\maketitle


\section{Introduction}

Galaxy clusters are the ultimate outcome of the hierarchical assembly of structure in the standard cosmological framework, $\Lambda$CDM. Their stellar bodies, supermassive black holes (SMBHs), dark matter halos, and gaseous atmospheres are all constructed from the collective merging of thousands, if not tens of thousands, of smaller constituents. Clusters are the most massive virialized structures in the Universe. They host the most massive galaxies, the most massive SMBHs, and the largest populations of satellite galaxies. The dynamics and physical processes at play in the hot plasma of the intracluster medium (ICM) are complex, diverse, and observable. Clusters are cosmologically rare, yet provide a key observational probe of our cosmological model. 

The hot, ionized gas of the ICM can be studied across a number of multi-wavelength tracers, from cluster cores to their outskirts \citep{reiprich13}. These include X-ray emission \citep{pratt09,vikhlinin09} with instruments such as Chandra, XMM-Newton, and Suzaku \citep{bautz09,sato12}, plus all-sky surveys from ROSAT and now eROSITA \citep{merloni12}. Future X-ray imaging spectrometers including the recently launched XRISM \citep{xrism20}, the Line Emission Mapper \citep[LEM;][]{kraft22}, and ATHENA X-IFU \citep{nandra13} will reveal the kinematics and physical properties of this hot gas \citep{simionescu19}.

The ionized ICM also imprints the SZ effect \citep{sunyaev70}, quantified to great success over the past decade with survey machines including Planck, ACT, and SPT \citep{planck2013_xx}. Future high spatial resolution observations of the SZ effect in the ICM will reveal new intracluster physics \citep{romero17,mroczkowski19}, e.g. with ALMA \citep{kitayama23}, NIKA2 \citep{perotto22}, and TolTEC \citep{montana19}, plus survey-scale programs including CMB-S4 \citep{raghunathan22} and Simons Observatory \citep[SO;][]{simons19}. The ICM is also visible at radio wavelengths due to synchrotron emission from relativistic particle populations. This tracer has led to recent discoveries with LOFAR \citep[radio megahalos;][]{cuciti22} and MeerKAT \citep{knowles22}.

The ICM is not exclusively hot: it is a multi-phase gaseous halo also observed to contain cooler phases such as H$\alpha$ \citep{fabian03b,crawford05}, MgII \citep{anand22}, and even molecular CO \citep{salome06}. Cool gas in cluster cores is, perhaps surprisingly, common, at least in some systems \citep{olivares19}. Indeed, the central thermodynamical properties of clusters suggest a possible dichotomy, and/or diversity, between `cool-core' and `non-cool-core' clusters \citep{mccarthy04,cavagnolo08,rafferty08}.

At all halo mass scales, astrophysical feedback processes within galaxies imprint signatures in the physical properties, and observables, of hot halos \citep{nelson18b,truong20,truong21,oppenheimer20,ayromlou23}. In clusters, accreting SMBHs i.e. active galactic nuclei (AGN) inject large amounts of feedback energy into the inner ICM \citep{hlavaceklarrondo22}. As a result, cluster thermodynamics are set by physics beyond gravitational dynamics alone \citep{valageas99,wu00,lewis00,babul02}. The observational manifestations of AGN feedback in clusters include relativistic and radio-emitting jets, together with associated cavities/bubbles visible in X-ray emission \citep{fabian12}. These heating sources offset radiative cooling in the ICM \citep{voit05} and lead to an approximate global thermal balance, preventing runaway cooling flows \citep{mcnamara12}.

As a result, the brightest central/cluster galaxies (BCGs) that form at the centers of clusters are predominantly massive, quenched, ellipticals \citep{delucia07}. These galaxies have recent formation (i.e. assembly) redshifts, yet are primarily built via accreted, ex-situ stars which have formed already at $z \gtrsim 2$ \citep{rodriguezgomez15,thomas10}. Their stellar light extends for hundreds of kilo parsecs, producing the intra-cluster light \citep[ICL;][]{lin04}. Clusters host the heaviest SMBHs in the Universe, with $M \gtrsim 10^{10}$\msun \citep{vandenbosch12}. These are radiatively inefficient, low accretion rate SMBHs preferentially in the centers of cool-core clusters \citep{hlavaceklarrondo11}.

Clusters are also observed to contain hundreds or even thousands of luminous satellite galaxies \citep{hansen09}. Galaxy evolution within clusters produces strong environmental effects including ram-pressure and tidal stripping \citep{moore96,yun19,rohr23}, quenching \citep{wetzel13}, and morphological transformation \citep{dressler80}. Satellite properties are biased with respect to their field counterparts, in star formation activity and colors \citep{kauffmann04}, metallicity \citep{pasquali10}, and stellar to (sub)halo mass ratio \citep{niemiec17}, indicative of environmental effects. Satellite galaxy populations in high-mass host halos therefore probe a broad range of physical processes \citep{bahe15,ayromlou19,martinnavarro21}.

Clusters have typically been observed at low redshift, where they can be detected in abundance and characterized in detail. However, their progenitors at $z \gtrsim 1-2$ reveal the assembly of these rare overdensities \citep{overzier16}. Massive protoclusters have already formed their hot ICM by $z \sim 2$ \citep{dimascolo23}, and the cooler gas phases in and around protoclusters may produce observed large-scale Lyman alpha emission \citep{steidel00,byrohl23}. At earlier evolutionary stages, prior to virialization, protoclusters are characterized by numerous individual galaxies evolving within a large-scale overdensity of the cosmic web. Galaxies in such dense regions have preferentially early formation redshifts and experience strong environmental effects \citep{thomas05,postman05}. At even earlier times, during the epoch of the first galaxies, protoclusters likely host the most massive galaxies that exist, as now being detected at $z \gtrsim 6$ in the JWST era \citep{labbe23,looser23,lovell23}.

The multi-wavelength characterization and modeling of galaxy clusters have become a central theme of recent large survey programs \citep{chexmate21}. Cluster samples have been assembled using a variety of observational tracers, based on their constituent gas (mainly X-ray and SZ), stars (optical), and dark matter (weak lensing). It is now appreciated that strong selection functions, particularly with X-ray selected samples, may have biased our view of galaxy cluster physics \citep{donahue01,rossetti16,rossetti17,lovisari17,popesso23}. Potential biases are also critical if clusters are to be used as cosmological probes.

A cornerstone of precision cosmology is the detection and characterization of galaxy clusters, across the electromagnetic spectrum \citep{allen11}. In the microwave, for example, Simons Observatory (SO) will detect tens of thousands of clusters via the thermal and kinetic SZ effects, enabling cosmological parameter constraints from tSZ cluster counts \citep{ade19}. All uses of galaxy clusters for cosmology are limited by theoretical uncertainties in mapping from observable properties of clusters to total halo mass \citep[mass-observable relations; e.g.][]{zhang08,zhang10,vikhlinin09,hoekstra15,mantz16}. In the optical, the recently launched Euclid mission will measure the abundance and properties of clusters as cosmological probes \citep{amendola18}. In the X-ray, the eROSITA all-sky survey telescope will use cluster number counts for Stage IV cosmological parameter constraints \citep{pillepich12}. Achieving these goals requires cosmological hydrodynamical simulations, which can produce high realism synthetic observations through forward modeling, while self-consistently accounting for the complexities of cluster baryonic physics such as feedback. 

This is a challenging task. Despite their broad observational importance and extensive characterization, our theoretical understanding of clusters is far from complete \citep{kravtsov12}. In part, this is because the evolution of the baryonic content of these halos is subject to a complex interplay of gravitational, (magneto)hydrodynamical, radiative, and chemical processes \citep{donahue22}. Clusters are baryonically closed systems \citep{ayromlou23}, but this does not imply they are static. Their virial shocks are located at $\sim 2-3 \times r_{\rm vir}$ \citep{voit03}, the extent of the hot ICM \citep{molnar09,vazza10b,zinger18}. Gas inflows from the intergalactic medium into clusters -- cosmic accretion -- often in the form of large-scale filaments \citep{malavasi23}. 

In the opposite direction, and fighting against the gravitational potential well, feedback-driven outflows can reshape gaseous halos. In clusters, AGN can launch powerful outflows, in the form of relativistic jets \citep{blandford19} and high-velocity winds \citep{yuan14}. The physical drivers behind AGN feedback remain an ongoing topic of study \citep{liska22}. The relationship of AGN feedback to the structure, morphology, and evolution of the gaseous ICM is likewise an active area of research \citep{werner19,altamura23}.

In addition to feedback, cosmic structure formation also produces complex gas kinematics in the ICM. Mergers of substructure (i.e. other halos) drive halo-scale motions and turbulence \citep{iapichino12}. The resulting cascade of energy to small scales produces observable velocity structure \citep{hitomi18} and amplifies magnetic fields via the turbulent dynamo \citep{kazantsev68}. All contribute non-thermal support to cluster atmospheres, producing a so-called hydrostatic bias \citep{miraldaescude95}. Capturing the interplay of such a broad range of physical ingredients is non-trivial.

To model the complex and non-linear interactions between the relevant physical processes in clusters requires numerical calculations \citep{katz93,evrard94,eke98}. Simulations of galaxy clusters are, however, challenging. Their large masses and volumes make high mass or spatial resolution computationally expensive, but as end-products of hierarchical assembly, they are built up from smaller galaxies and halos, making high resolution mandatory. Beyond resolution, accurate solvers and numerical methods for galaxy cluster physics also remain a computational challenge \citep{zuhone15,kannan17,ehlert18,steinwandel23}.

There are many cosmological hydrodynamical simulation projects focused on galaxy clusters with $M_{\rm 200c} > 10^{14}$\msun. Simulations with high $\sim 10^6$\msun baryon resolution, similar to typical large-volume cosmological boxes, include Hydrangea \citep{bahe17b} and C-EAGLE \citep{barnes17b}. Current $\sim (100\,\rm{Mpc})^3$ large-volume simulations at this resolution such as Illustris, EAGLE, TNG100, and SIMBA all contain a handful of such high-mass halos. Simulations of clusters with even better $\lesssim 10^5$\msun baryon resolution, required to model smaller satellite systems, are prohibitively expensive, and only the $\sim 10^{14}$\msun clusters from TNG50 \citep{pillepich19,nelson19b} and Romulus-C \citep{tremmel18} exist.

At the $\sim 10^7 - 10^8$\msun baryon resolution level there are large-volume simulations such as TNG300, Magneticum 2hr \citep{dolag16}, and the recent MillenniumTNG \citep{pakmor23} and FLAMINGO \citep{schaye23} simulations, and numerous zoom suites including Rhapsody-G \citep{hahn17}, DIANOGA \citep{rasia15,bassini20}, YZiCS \citep{choi17}, and The Three Hundred \citep[][including realizations with multiple galaxy formation models]{cui18,cui22}. At lower $\gtrsim 10^9$\msun baryon resolution, simulations no longer capture galaxy formation physics, and are often aimed at large-scale and cosmological applications, e.g. MACSIS \citep{barnes17a}, cosmo-OWLS \citep{lebrun14}, and BAHAMAS \citep{mccarthy17}. Of all these projects, we note that The Three Hundred is most similar to TNG-Cluster in its approach and objectives. 

However, there are significant differences in the cluster simulation projects above, beyond numerical resolution alone. In particular, the statistics i.e. sample sizes, numerical methodologies, and range of included physics all vary considerably. In order to simulate galaxies, and galaxy clusters, the adopted physical model is crucial. For example, some previous cosmological cluster simulations, including `non-radiative' models, neglect black hole feedback from a central engine. Without SMBH feedback galaxy formation at the high-mass end is widely unrealistic, and the properties of the ICM may become unreliable. Overall, the range of scientific applications and science questions that can be addressed with a given galaxy cluster simulation depends on its scope, realism, and physical richness.

Here we introduce a new large simulation project, TNG-Cluster.\footnote{\url{www.tng-project.org/cluster}} This is a suite of fully cosmological galaxy cluster simulations, including magnetohydrodynamics and the complete baryonic physics model of the IllustrisTNG simulation project. We simulate 352 high-mass halos with $M_{\rm 200c} \sim 10^{14.5 - 15.4}$\msun down to $z=0$, with three principal scientific goals: (i) to understand the gas physics of the ICM, (ii) to provide the necessary observables for precision cosmology in the next decade, and (iii) to explore the assembly of the most massive galaxies, their SMBHs, and their surrounding satellite galaxy populations.

In this introductory paper we give an overview of TNG-Cluster and its outcomes. In a series of companion papers we further showcase first science results from TNG-Cluster: a census of ICM kinematics \citep{ayromlou24}, their inference mock X-ray spectroscopic observations \citep[e.g. XRISM/LEM;][]{truong24}, the realization of cool-core and non cool-core systems \citep{lehle24}, the circumgalactic medium by cluster massive satellites \citep{rohr24} and the formation of radio relics in massive mergers \citep{lee24}. The TNG-Cluster simulation will be made publicly available in the future (\textcolor{blue}{Pillepich et al. in preparation}).

This paper is structured as follows. Section \ref{sec_methods} discusses our methodology -- the model and simulations (Section \ref{sec_sims}), the zoom and full-box reconstruction (Section \ref{sec_resims}), the simulation output (Section \ref{sec_methods2}), and analysis details (Section \ref{sec_methods3}). In Section \ref{sec_sample} we present the TNG-Cluster sample. Section \ref{sec_gas} explores the properties of the gaseous ICM, while Section \ref{sec_stars} considers the stellar and SMBH components. We summarize our findings in Section \ref{sec_conclusions}.


\section{Simulations} \label{sec_methods}

\subsection{The TNG model and simulations} \label{sec_sims}

The TNG100 and TNG300 volumes of the IllustrisTNG project \citep[hereafter, TNG;][]{pillepich18b, nelson18a, naiman18, marinacci18, springel18}, together with the high-resolution TNG50 simulation \citep{nelson19b,pillepich19}, are a series of three large cosmological volumes, simulated with gravo-magnetohydrodynamics (MHD) and incorporating a comprehensive model for galaxy formation physics \citep{weinberger17,pillepich18a}.

The largest volume, TNG300, includes 2$\times$2500$^3$ resolution elements in a $\sim$\,300 Mpc box, giving it a baryon mass resolution of $1.1 \times 10^7$\msun. The gravitational softening lengths are 1.5 kpc at $z$\,=\,0 for the stars and dark matter, while gas cells have an adaptive gravitational softening with a minimum of 370 comoving pc (see Table \ref{simTable}). Adopting the fiducial TNG model, we extend this effort with our new project \textbf{TNG-Cluster}, which improves the statistics and sampling of the most massive dark matter halos, with total mass exceeding $\log{(M_{\rm 200c} / \rm{M}_\odot)} > 14.5$ at redshift zero.

All aspects of the physical model, including parameter values and the simulation code, are described in \citet{weinberger17} and \citet{pillepich18a} and are kept unchanged for the TNG-Cluster simulation. In particular, TNG uses the \textsc{Arepo} code \citep{springel10}, which solves for the coupled evolution under self-gravity and ideal, continuum MHD \citep{pakmor11,pakmor13}. Self-gravity is solved with a Tree-PM approach, whereas the fluid dynamics use a Godunov type finite-volume scheme on an unstructured, moving, Voronoi tessellation. The numerical scheme is second order accurate in both time and space \citep{pakmor16}.

The simulations include a physical model for the most important processes relevant for the formation and evolution of galaxies. Specifically: (i) gas radiative processes, including primordial/metal-line cooling and heating from the background radiation field, (ii) star formation (SF) in the dense interstellar medium (ISM), (iii) stellar population evolution and chemical enrichment following supernovae Ia, II, as well as AGB stars, with individual accounting for the nine elements H, He, C, N, O, Ne, Mg, Si, and Fe, (iv) supernova driven galactic-scale outflows or winds \citep[see][for details]{pillepich18a}, (v) the formation, coalescence, and growth of SMBHs, (vi) and dual-mode SMBH feedback operating in a thermal `quasar' state at high accretion rates and a kinetic `wind' state at low accretion rates \citep[see][for details]{weinberger17}.\footnote{TNG-Cluster adopts the same modification to the SF time-scale law as in TNG50, whereby gas cells with extremely short SF time-scales, in the densest environments, are converted into stars in a reasonable number of numerical time-steps \citep[see Section 2.2 of][]{nelson19b}.}

\begin{figure*}
\centering
\includegraphics[angle=0,width=5.7in]{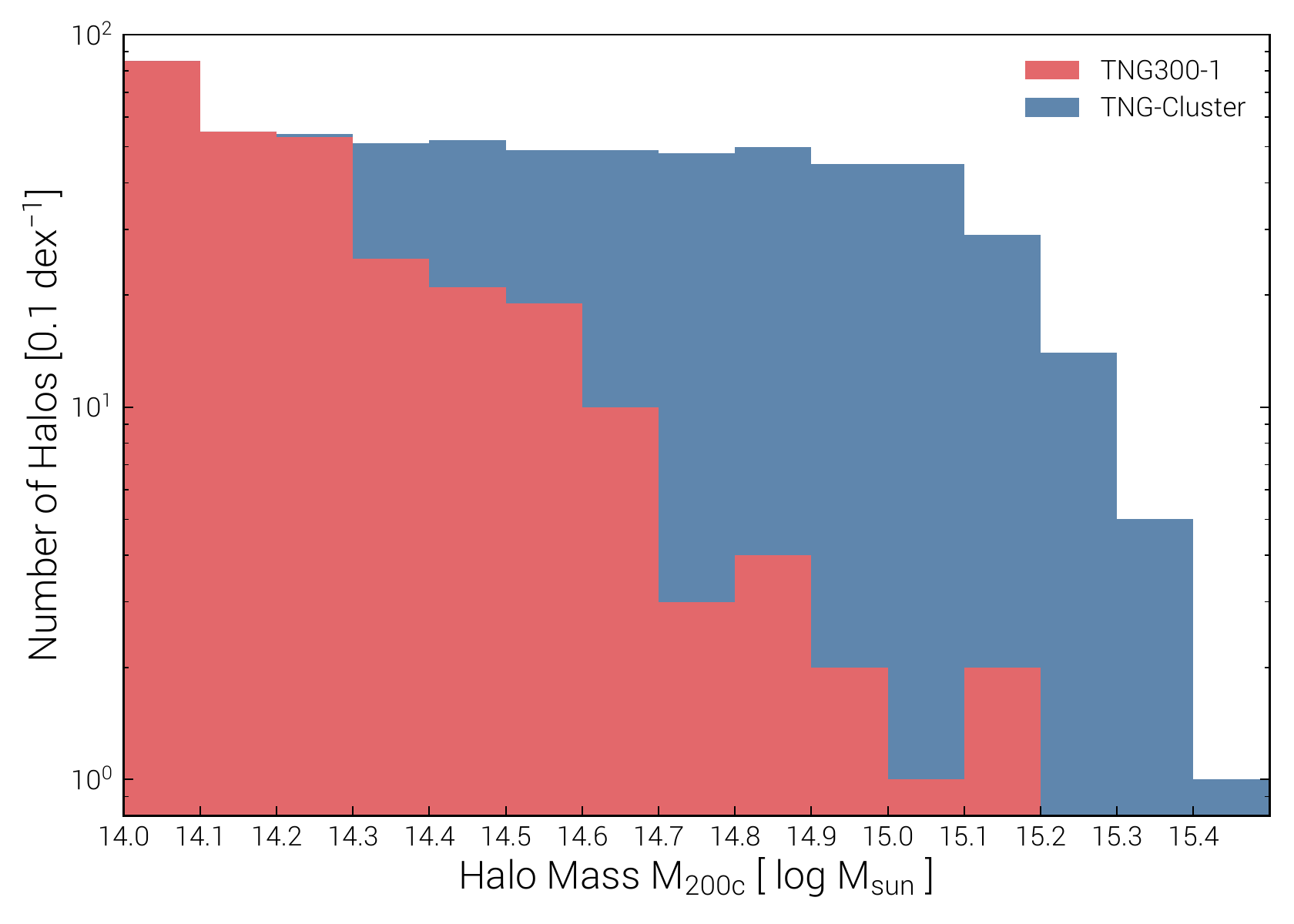}
\caption{ The $z=0$ cluster mass function of the new TNG-Cluster simulation project (blue) stacked on top of our existing cluster sample from TNG300 (orange), showing the number of halos per 0.1 dex bin in $M_{\rm 200c}$. We increase the statistics for halos $\gtrsim 10^{14.5}$\msun by an order of magnitude, from $\sim$ 30 objects to $\sim$ 350. Likewise, we increase the sampling of the most massive and rare $\gtrsim 10^{15}$\msun halos by a factor of \textit{thirty}, from just three halos in TNG300 to ninety in TNG-Cluster. The combined $M_{\rm 200c} > 10^{14}$\msun cluster sample increases from 280 to 636 halos.
 \label{fig_mass_function}}
\end{figure*}

We make one addition: coupling the Monte Carlo tracer particle population \citep{genel13} to an active, on-the-fly shock finder \citep{schaal15} such that accretion properties including temperature, density, and shock mach number history are stored for every tracer \citep[following][]{nelson13}. This change is entirely passive, and only provides new and previously unavailable information that can be used in future analyses.

One of the most important differences between the TNG-Cluster sample and other hydrodynamical simulations of galaxy clusters is its resolution and statistics. The TNG model for baryonic and feedback physics is also physically comprehensive, robust, and well-validated. In particular, the TNG model has been shown to produce realistic galaxy populations, consistent with a broad range of observational constraints, across an enormous dynamic range, from small dwarf galaxies ($M_\star \simeq 10^8$\msun) to massive galaxy clusters ($M_{\rm halo} \simeq 10^{15}$\msun). For example, in the cluster regime, the TNG model has been validated in the cool-core/non-cool-core dichotomy \citep{barnes18}, ICM metallicity profiles \citep{vog18a}, X-ray properties \citep{truong20}, satellite populations \citep{stevens19,stevens21,donnari20a,donnari20b,ayromlou21a}, the colors \citep{nelson18a}, quenching \citep{donnari19}, and density structure \citep{wang18} of central early-types, their stellar halos \citep{pulsoni20}, lensing signals \citep{renneby20} and Sunyaev-Zeldovich effect \citep{lim21}. As a result, the TNG-Cluster halos are simulated with a model that has a high degree of physical fidelity, for the thermodynamical properties of the ICM as well as the galaxy populations hosted therein. It is also the only cosmological simulation suite of massive clusters and their galaxies that self-consistently includes magnetic fields.

The TNG model is not perfect. Informative tensions and disagreements between previous IllustrisTNG simulations and observational data are discussed in the TNG public data release \citep{nelson19a}. As with all cosmological hydrodynamical simulations, the multi-scale, multi-physics nature of the physical phenomena involved require significant modeling approximations and uncertainties. At the cluster mass-scale, as an illustrative example, the TNG model does not include the relativistic jets launched from SMBHs, nor the non-thermal impact of cosmic rays. The use and interpretation of TNG-Cluster will benefit from careful comparisons with observations, a process we start here.

As in all TNG simulations, we adopt a \cite{planck2015_xiii} cosmology with $\Omega_{\Lambda,0}=0.6911$, $\Omega_{m,0}=0.3089$, $\Omega_{b,0}=0.0486$, $\sigma_8=0.8159$, $n_s=0.9667$ and $h=0.6774$. 

\subsection{Re-simulation and full volume reconstruction} \label{sec_resims}

{\renewcommand{\arraystretch}{1.3}
\begin{table*}
  \caption{Details of the TNG-Cluster simulation in comparison to its smaller volume siblings. The values are: volume, box side-length, number of initial gas cells, dark matter particles, and tracers; the mean baryonic cell mass and dark matter particle mass; the gravitational softening length minimum for the gas, and the softening for the collisionless components at $z=0$; the number of groups ($\log M_{\rm 200c} / \rm{M}_\odot$ from 13.0 to 14.0), lower mass clusters (14.0-14.5), intermediate-mass (14.5-15.0), and high mass ($>$15.0) clusters. $\dagger$ = effective full-volume equivalent. The TNG-Cluster numbers refer exclusively to the primary halos targeted for re-simulation.}
  \label{simTable}
  \begin{center}
    \begin{tabular}{lcllll}
     \hline\hline
     
 Run                        &                 & TNG50         & TNG100                & TNG300     & \textbf{TNG-Cluster} \\ \hline
 Volume                     & [\,Mpc$^3$\,]   & $51.7^3$          & $110.7^3$         & $302.6^3$ & $\mathbf{1003.8^3}$ \\
 $L_{\rm box}$              & [\,Mpc/$h$\,]   & $35$              & 75                & 205       & \textbf{680} \\
 $N_{\rm GAS}$              & -               & $2160^3$          & $1820^3$          & $2500^3$  & $\mathbf{8192^{3\dagger}}$ \\
 $N_{\rm DM}$               & -               & $2160^3$          & $1820^3$          & $2500^3$  & $\mathbf{8192^{3\dagger}}$ \\
 $N_{\rm TR}$               & -               & $2160^3$          & $2 \times 1820^3$ & $2500^3$  & $\mathbf{8192^{3\dagger}}$ \\
 $m_{\rm baryon}$           & [\,M$_\odot$\,] & $8.5 \times 10^4$ & $1.4 \times 10^6$ & $1.1 \times 10^7$ & $\mathbf{1.2 \times 10^7}$ \\
 $m_{\rm DM}$               & [\,M$_\odot$\,] & $4.5 \times 10^5$ & $7.5 \times 10^6$ & $5.9 \times 10^7$ & $\mathbf{6.1 \times 10^7}$ \\
 $\epsilon_{\rm gas,min}$   & [\,pc\,]        & $74$              & 185               & 370        & \textbf{370}  \\
 $\epsilon_{\rm DM,stars}$  & [\,pc\,]        & $288$             & 740               & 1480       & \textbf{1480} \\ \hline
 $N_{\rm groups}$           & $10^{13.0-14.0}$ & 23               & 168               & 3,545      & \textbf{0}    \\
 $N_{\rm clusters}$         & $10^{14.0-14.5}$ & 1                & 11                 & 239        & \textbf{57}    \\
 $N_{\rm clusters}$         & $10^{14.5-15.0}$ & 0                & 3                 & 38         & \textbf{207}  \\
 $N_{\rm clusters}$         & $> 10^{15.0}$    & 0                & 0                 & 3          & \textbf{92}  \\
 \hline
 
    \end{tabular}
  \end{center}
\end{table*}}

TNG-Cluster is a suite of several hundred multi-mass `zoom' re-simulations of cluster halos. To construct the initial conditions we first generate a random parent volume with $L_{\rm box}$ = 1 Gpc and $N_{\rm DM} = 2048^3$, giving a particle mass of $m_{\rm DM} \simeq 4 \times 10^9$\msun. We run a gravity only version of this parent simulation, TNG-Cluster-Dark, which has more than sufficient resolution to accurately identify massive clusters. We choose target halos by selecting on $z=0$ halo mass alone, making the sample unbiased in all other properties. The 352 halos are chosen randomly in 0.1 dex mass bins such that we (i) include all halos in the TNG-Cluster-Dark volume above $M_{\rm 200c} \ge 10^{15}$\msun, and (ii) compensate the drop-off of statistics in TNG300 from $10^{14.3}$ to $10^{15.0}$ to produce a flat distribution in halo mass over this range: see Figure \ref{fig_mass_function}.

\begin{figure*}
\centering
\includegraphics[angle=0,width=3.35in]{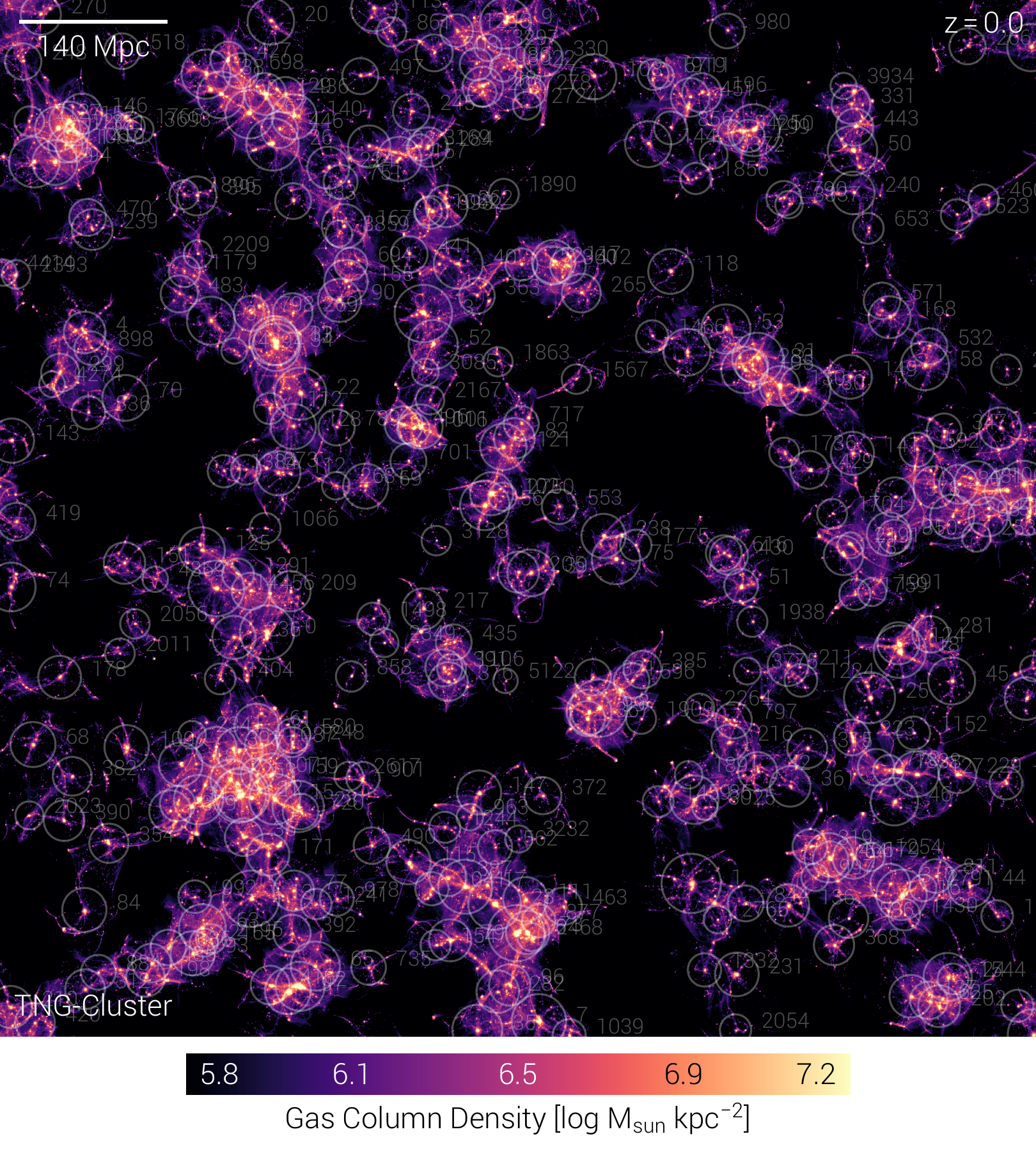}
\includegraphics[angle=0,width=3.35in]{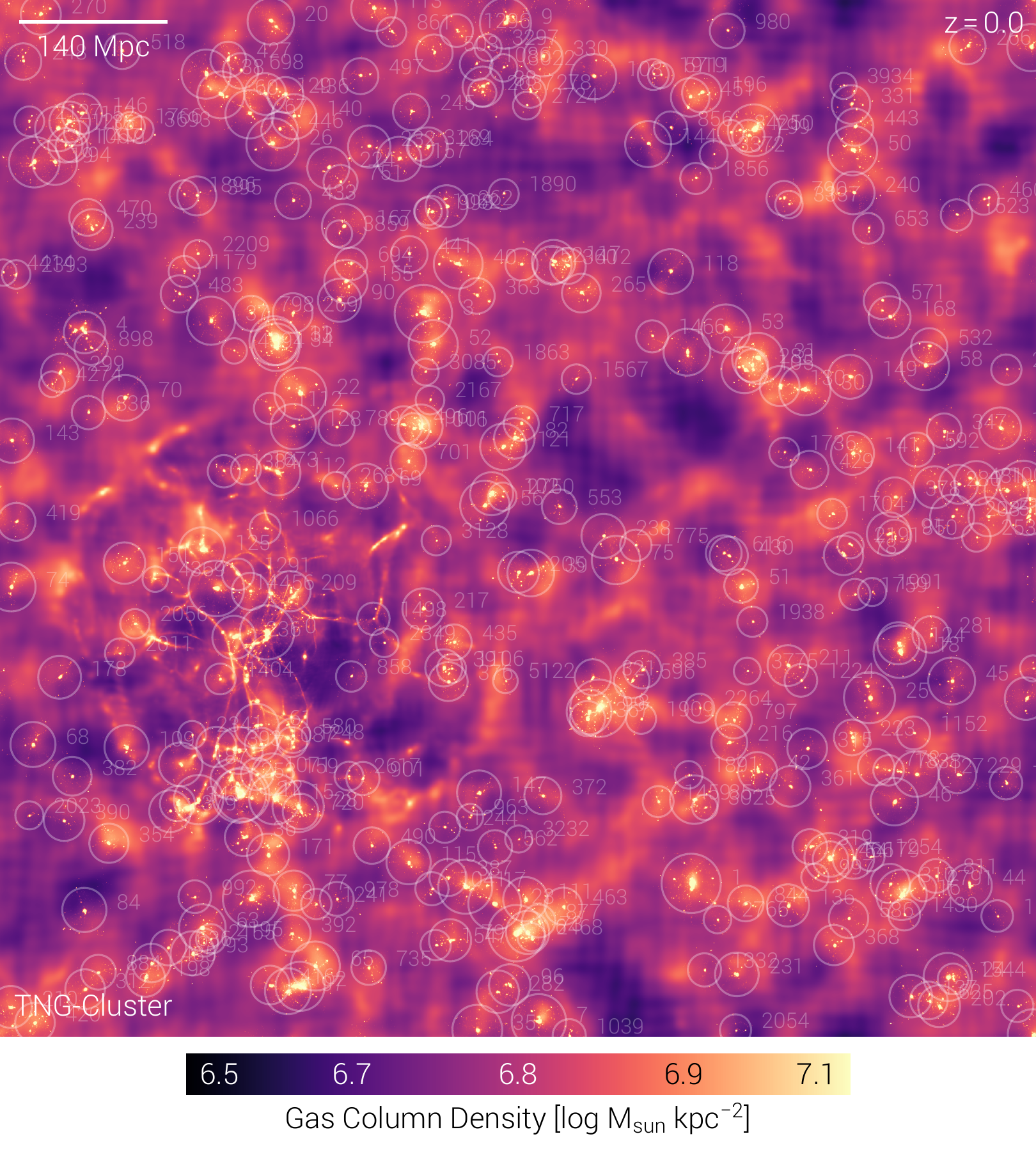}\\
\includegraphics[angle=0,width=3.35in]{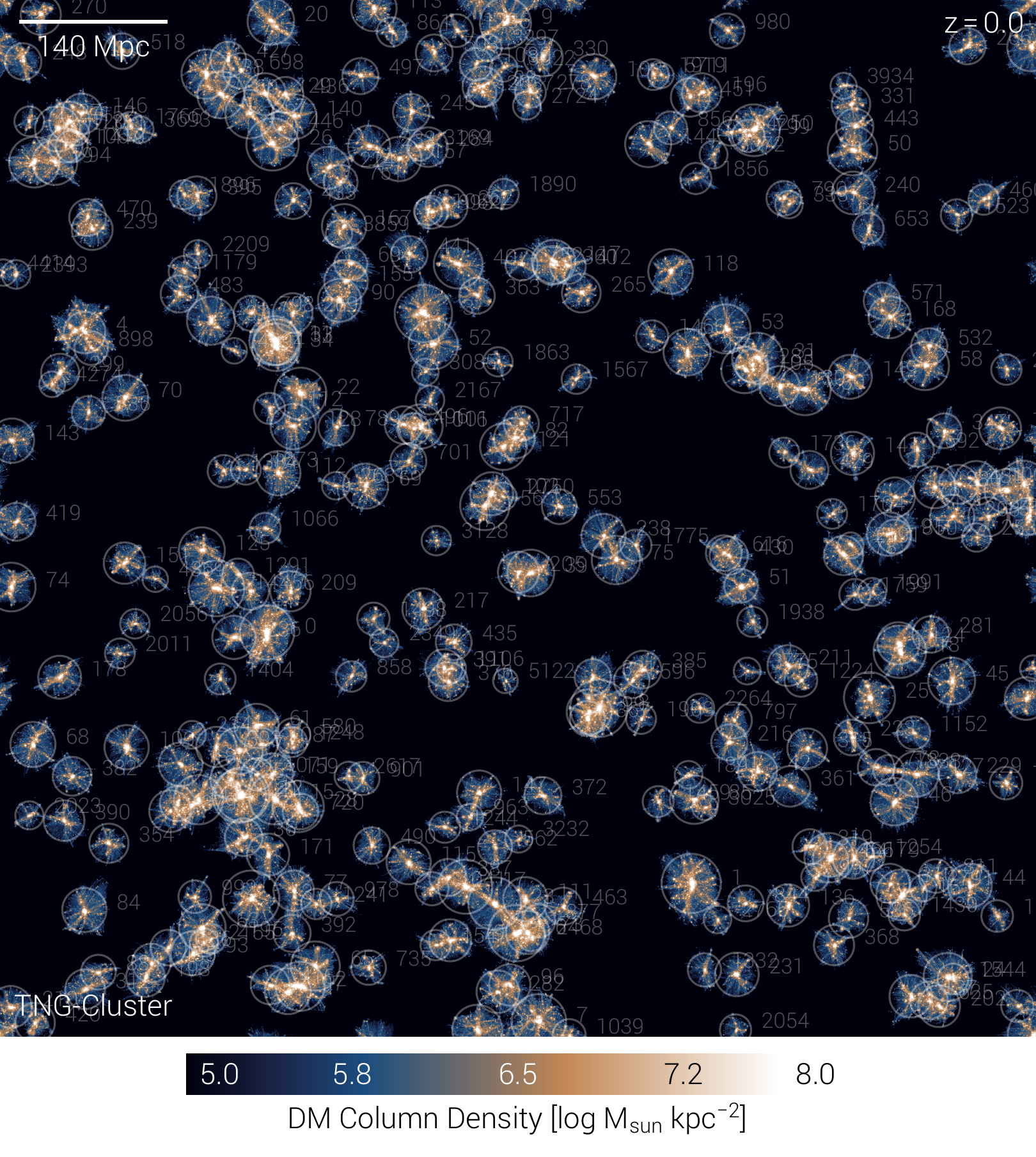}
\includegraphics[angle=0,width=3.35in]{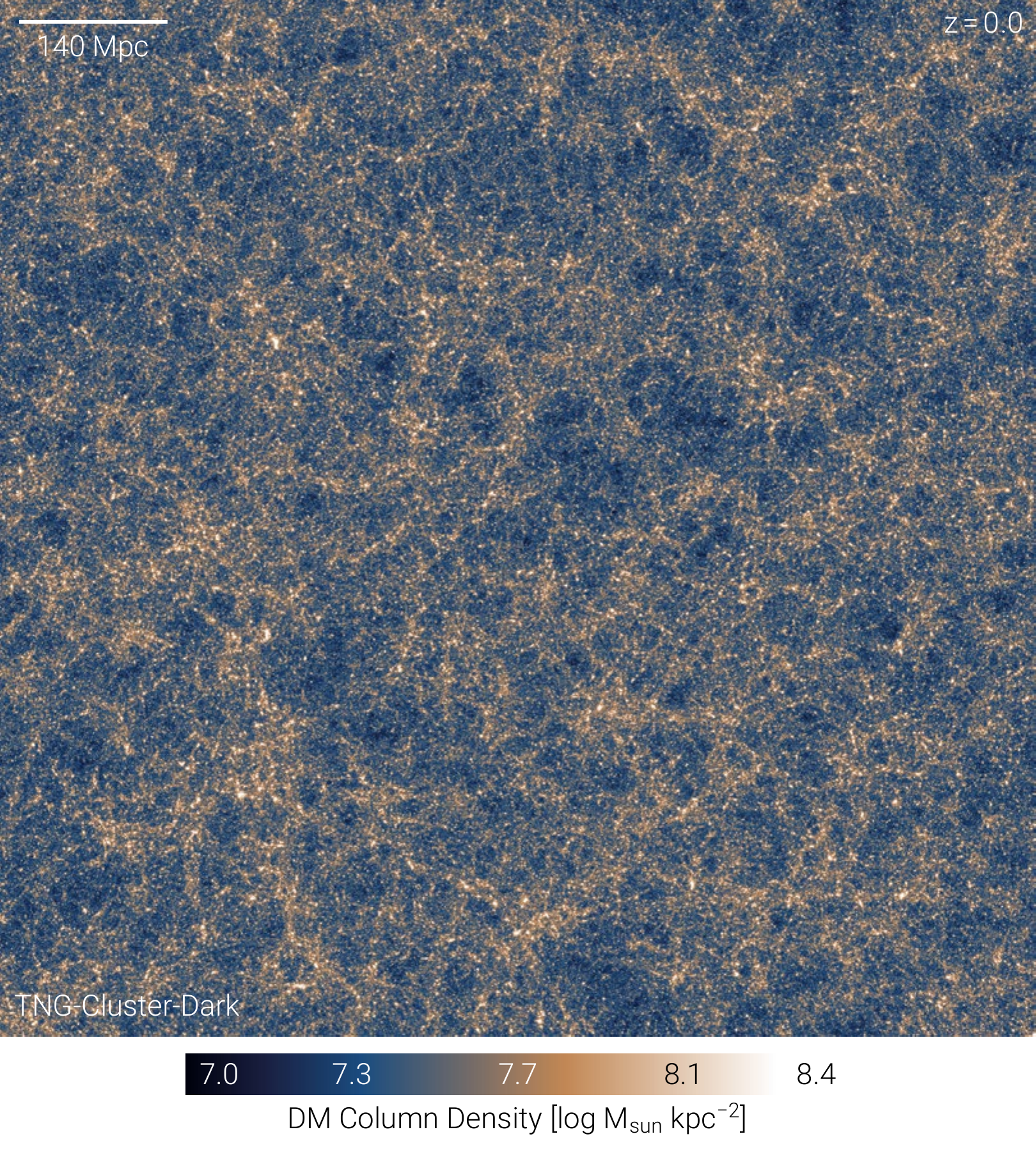}
\caption{Four visualizations of the TNG-Cluster volume at $z=0$. TNG-Cluster is a `virtual' simulation box constructed by stitching together the results of 352 individual zoom simulations. Each halo targeted for high-resolution simulation is marked by its Halo ID and a white circle demarcating $10r_{\rm vir}$. \textbf{Top left:} projected gas column density, showing high-resolution gas together with the first three intermediate/buffer regions. \textbf{Top right:} FoF-scope gas of all halos (which exist only in high resolution regions), together with all outer fuzz of the first (most massive) zoom target, which evidences the gradual coarsening of resolution as we move away from the central re-simulated halos. \textbf{Lower left:} high-resolution dark matter particles. The visible areas of nonzero mass density therefore show exactly the high-resolution regions of the zooms. \textbf{Lower right:} the parent, gravity-only volume of TNG-Cluster-Dark, from which massive halos were chosen for re-simulation. Shown in projected dark matter surface density, the large-scale structure is consistent with the three other views from the reconstructed, hydrodynamical TNG-Cluster volume.
 \label{fig_boxvis}}
\end{figure*}

For each target halo, we then select all member dark matter particles in its friends-of-friends group at $z=0$ and trace them back to their coordinates in the initial conditions (at $z=127$). We define this Lagrangian volume by constructing an adaptive oct-tree from $2^6 = 64$ up to $2^{11} = 2048$ grid cells per linear dimension, marking those cells that contain member particles. We then expand this marked region at the highest refinement level by progressively including neighboring cells until the total volume reaches three times its original value. This protects against contamination of the high-resolution region by low-resolution (i.e. high-mass) particles/cells at late times. The highest resolution grid is then refined by a further factor of four in linear extent, producing cells with 4$^3$ times smaller volume (lower mass). A particle set is created, with one total mass particle in each occupied cell -- eight discrete mass levels, progressively coarsening away from the region of interest. 

This particle set is then perturbed following the \cite{zeldovich70} approximation to imprint our chosen transfer function, using the same procedure as for the original TNG uniform mass simulation initial conditions \citep{pillepich18a} and the \textsc{N-GenIC} code \citep{springel05}, keeping the noise field consistent in the low-resolution grids and only adding additional small-scale power in the `zoom' region (we assess the size of these zoom regions, and the level of low-resolution `contamination', in Appendix~\ref{sec_app1}). The high-resolution volume has a spatial (mass) resolution that is 4 (64) times better than the original parent simulation, resulting in an effective resolution of $\sim 8192^3$, had the original volume been entirely covered by high resolution. This is in practice equivalent to TNG300-1, thereby extending the TNG300-1 cluster sample at roughly the same resolution.

With several hundred such multi-mass zoom simulations run to completion, we then reconstruct the parent volume at full resolution and with the hydrodynamical results, by stitching together the results of the individual zoom simulations. To do so we create the 100 saved snapshots and group catalogs (same configuration as TNG) by shifting the spatial coordinates of the box centered zooms to the correct position within the parent volume, offsetting particle IDs to maintain global uniqueness, converting the simulations back to the fiducial TNG unit system, and writing a file-level structure that enables the recovery of the exact particle set and/or group catalog of any of the original zooms. This reconstruction is ultimately for convenience: we label the resulting simulation \textbf{TNG-Cluster}. This `virtual box' has an identical data structure to all other full volume TNG simulations to date, enabling seamless analysis and the application of all existing tools, codes, and post-processing pipelines.

Figure \ref{fig_boxvis} visualizes our reconstruction of the TNG-Cluster volume from the series of independent zoom simulations. The total number of high resolution dark matter particles in TNG-Cluster is $\sim 2480^3$, making data storage and analysis expense roughly the same as for TNG300-1 ($N_{\rm DM} = 2500^3$).

\subsection{Simulation output and analysis} \label{sec_methods2}

\begin{figure*}
\centering
\includegraphics[angle=0,width=6.4in]{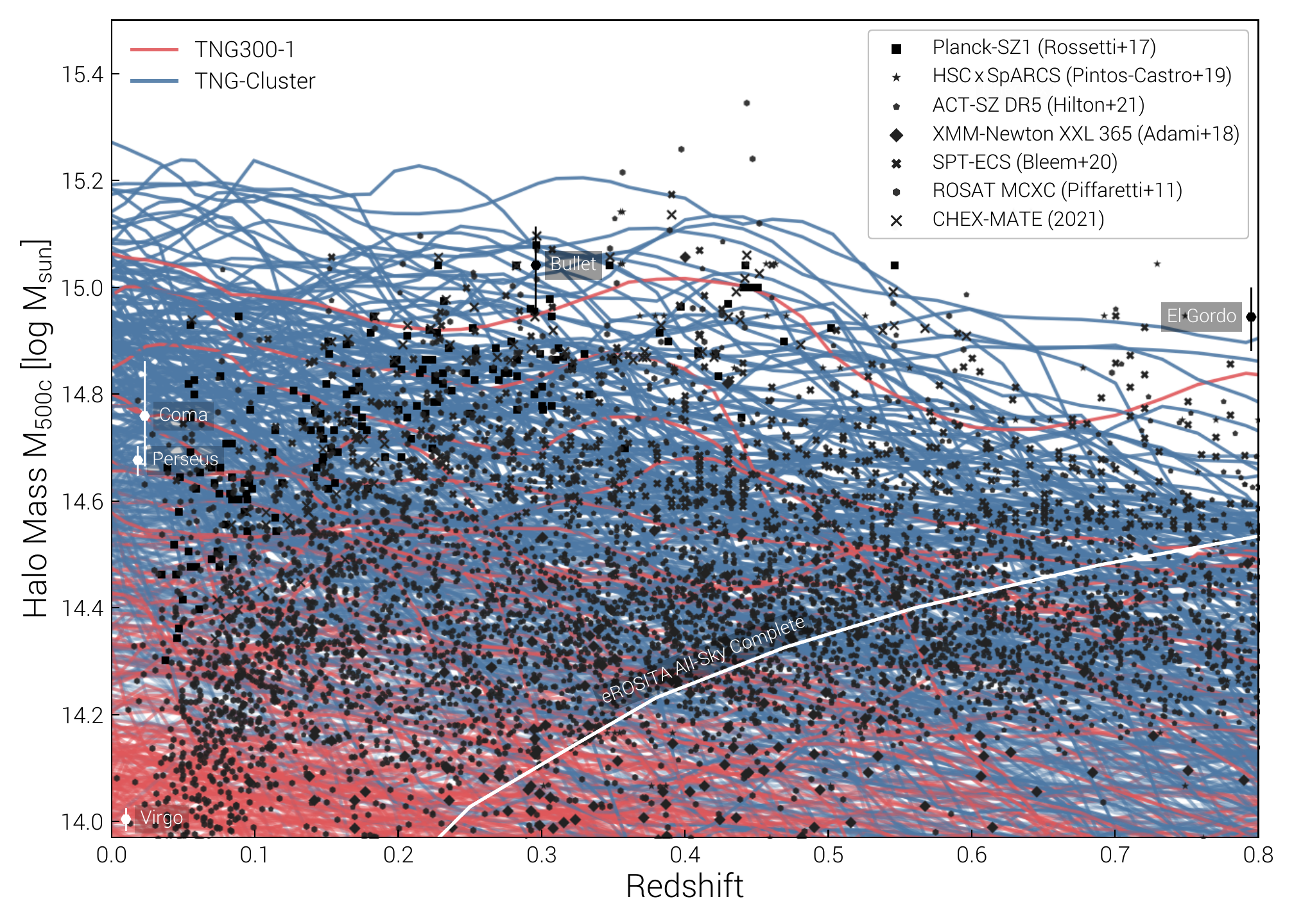}
\caption{ The combined sample of TNG-Cluster (blue) and TNG300 (red), showing $M_{\rm 500c}$ halo mass as a function of redshift. The mass assembly history of each $z=0$ cluster is shown as a single line. For reference, we compare to observational samples of galaxy clusters from large surveys/survey instruments including Planck, ACT, ROSAT, SPT (black markers; see text for details). Several notable low-redshift named clusters are also included for comparison in this mass-redshift plane -- Virgo, Coma, Perseus, the Bullet Cluster, and El Gordo (points with errorbars; see text for details). The TNG-Cluster sample, in blue, is essential to compare to the most massive detected galaxy clusters, from $z=0$ to cosmic noon.
 \label{fig_sim_sample}}
\end{figure*}

The output data of TNG-Cluster is identical to that of the previous TNG simulations, for both halo and galaxy catalogs and particle-level snapshots, with a few additions.

In particular, dark matter halos and galaxies in TNG-Cluster are identified as in all the TNG simulations, with a two-step process. First, the friends-of-friends algorithm \citep{davis85} identifies associations of dark matter particles. Then, the \textsc{SUBFIND} algorithm \citep{springel01} is applied to each halo in order to locate and characterize gravitationally-bound subhalos. Galaxies are subhalos containing non-zero stellar mass. Subhalos are tracked through time using the \textsc{SubLink} merger trees \citep{rodriguezgomez15}. 

Data is stored at 100 snapshots between $z=15$ and $z=0$, with 20 full and 80 mini snapshots \citep{nelson19a}. In addition to all quantities from the previous TNG simulations, we also store additional fields for tracer particles (see Section~\ref{sec_methods}). We also record details of the feedback energetics of each SMBH along their main evolutionary branches, as well as the timings and properties of the kinetic wind ejections from each SMBH.

When presenting results as a function of galaxy stellar mass, we define this quantity as the total stellar mass within a 3D spherical aperture of 30 pkpc. We also present results as a function of both $M_{\rm 200c}$ and $M_{\rm 500c}$, which are the total halo masses within $R_{\rm 200c}$ and $R_{\rm 500c}$, respectively. Both are defined as the radius within which the average mass density is 200 (500) times the critical density.

\subsection{Observables} \label{sec_methods3}

From the simulation outputs, we calculate several observables. For X-ray emission we compute broadband luminosities, typically $0.5-2.0$\,keV, using the collisional ionization tables of AtomDB from the Astrophysical Plasma Emission Code \citep[APEC v3.0.9-201t;][]{smith01}. We derive the X-ray emissivity of each gas cell based on its density, temperature, and metallicity, adopting the default \citet{anders89} abundances. We label C, N, O, Ne, Mg, Al, Si, S, Ar, Ca, Fe, and Ni as metals and do not include other trace elements beyond H and He into this category \citep[e.g. versus SOXS;][]{zuhone23}. For the high-temperature ICM plasma of our simulated clusters, there is negligible difference versus a simpler calculation of free-free (bremsstrahlung) emission alone \citep[e.g.][]{navarro95}.

As a proxy for the thermal Sunyaev-Zeldovich signal \citep{sunyaev70}, we compute the Compton y-parameter

\begin{equation}
y = \int \sigma_{\rm T} \frac{k_{\rm B} T}{m_{\rm e} c^2} n_{\rm e} \rm{d}l \propto \int P_{\rm e} \rm{d}l
\end{equation}

\noindent which is dimensionless, and proportional to the line-of-sight integral ($\rm{d}l$) of electron pressure. Here $\sigma_{\rm T}$ is the Thomson cross-section, $k_{\rm B}$ is Boltzmann's constant, $T$ is gas temperature, $m_{\rm e}$ is the mass of the electron, $c$ is the speed of light, and $n_{\rm e}$ is electron number density. In practice, to compute $y$ we calculate and sum $\Upsilon = \sigma_{\rm T} (k_{\rm B} T) / (m_{\rm e} c^2) \times m / (\mu_{\rm e} m_{\rm H})$ for gas cells, where $m$ is the total gas cell mass, $\mu_{\rm e}$ is the mean molecular weight per free electron, and $m_{\rm H}$ is the atomic mass of the hydrogen atom \citep[following][]{roncarelli07,kay12,mccarthy14}.

We typically consider X-ray luminosities and integrated $Y_{\rm SZ}$ parameters integrated within spherical 3D apertures of $R_{\rm 500c}$, in which case contributions from all gas cells are summed directly. When measuring instead in 2D projection, or creating maps of X-ray emission or SZ signal, we adopt the usual cubic-spline kernel projection approach \citep[following][]{nelson16} with a line-of-sight projection depth equal to the (cosmologically large) full box depth, thereby including reasonable 2-halo contributions as well as projection effects.

In both cases star-forming gas, which is pressurized by our two-phase sub-grid model, is assigned its cold phase temperature of 1000\,K, and so does not contribute to X-ray or SZ signals.

We also make a simple computation of the radio synchrotron emission from our simulated clusters \citep[following][]{xu12,marinacci18}. We track the evolving magnetic energy in each gas cell, but do not directly follow relativistic particles. We therefore assume a relativistic electron population following a power-law in number density per energy bin ${\rm d}n(\gamma) / {\rm d}\gamma \propto \gamma^{-p}$ where $\gamma$ is the Lorentz factor and $p=1.7$ is the spectral index. We further assume that only electrons between $\gamma_{\rm min} = 300$ to $\gamma_{\rm max} = 1.5 \times 10^4$ contribute, that the energy density ratio between relativistic protons and electrons is $k=10$ \citep{vazza14}, and that the total energy density in relativistic particles (electrons plus protons) is a ratio $\eta = 1$ of the magnetic energy density, i.e. equipartition, as inferred to be the case for galaxy clusters \citep{pinzke10}. The resulting synchrotron power is derived for a VLA-like configuration with a central frequency of 1400 MHz and a bandpass of 20 MHz \citep[see appendices A1 and A2 of][]{marinacci18}. The radio synchtron emission of TNG-Cluster is explored in more detail, and with more sophisticated modeling, in \citep{lee24}.


\section{The TNG-Cluster Sample} \label{sec_sample}

In Figure \ref{fig_mass_function} we show the combined cluster mass functions of TNG300 (in red) and TNG-Cluster (in blue) at redshift zero. The TNG-Cluster simulation samples tens of objects more massive than even the most massive cluster present in TNG300, with halo mass $M_{\rm 200c} > 10^{15.2}$\msun. For massive clusters with total mass $M_{\rm 200c} > 10^{15.0}$\msun, TNG-Cluster improves upon the statistics of TNG300 by a factor of \textit{thirty}, increasing the sample from just three objects to ninety. Likewise, for intermediate mass halos with $M_{\rm 200c} > 10^{14.5}$\msun, we boost the number of simulated clusters by one order of magnitude, from $\sim$ 30 halos to $\sim$ 300. Ultimately we compensate the drop-off of the TNG300 mass function starting at $10^{14.3}$\msun, sampling to achieve a roughly flat log mass spectrum until $10^{15.0}$\msun, where we let the statistics of a 1 Gpc volume limit the sample. All halos above $10^{15.0}$\msun total mass are re-simulated, providing a complete volume-limited sample of the highest-mass clusters: the exact numbers are given in Table \ref{simTable}.

Figure \ref{fig_sim_sample} gives an overview of the simulated cluster sample available in TNG-Cluster, in comparison to several observational samples and as a function of redshift. We compare to the following large surveys: a subset of 169 clusters from the Planck SZ-selected (PSZ1-cosmo) catalog with X-ray followup \citep{rossetti17}; the ACT DR5 SZ-selected cluster sample \citep{hilton20}; the ROSAT MCXC meta-catalog \citep{piffaretti11}; the SPTpol extended cluster survey \citep[SPT-ECS;][]{bleem20}; the XMM-Newton based XXL 365 cluster catalog \citep{adami18}; and the PSZ2-selected CHEX-MATE cluster sample \citep{chexmate21}. We also show the Coma cluster at $M_{\rm 500c} = 5.7^{+1.5}_{-1.1} \times 10^{14}$\msun \citep{okabe14}, the Perseus cluster at $M_{\rm 500c} = 4.7^{+0.3}_{-0.3} \times 10^{14}$\msun \citep{simionescu11}, the Virgo cluster at $z=0$ and $M_{\rm 500c} = 0.8^{+0.1}_{-0.1} \times 10^{14}$\msun \citep[moved slightly in mass for visibility;][]{simionescu17}, the Bullet Cluster at $z=0.296$ with $M_{\rm 500c} = 1.1^{+0.2}_{-0.2} \times 10^{15}$\msun \citep{clowe06}, the Pheonix cluster at $z=0.6$ and $M_{\rm 500c} = 2.3^{+0.7}_{-0.7} \times 10^{15}$\msun \citep{tozzi15}, the El Gordo cluster with $M_{\rm 500c} = 8.8^{+1.2}_{-1.2} \times 10^{14}$\msun \citep[moved from $z=0.87$ to $z=0.8$ for visibility;][]{botteon16}, and the eROSITA threshold mass above which every halo on the sky will be X-ray detected \citep{pillepich12, merloni12}. Figure \ref{fig_sim_sample} demonstrates that TNG-Cluster well samples the mass-redshift space of recent observed cluster samples. 

\begin{figure}
\centering
\includegraphics[angle=0,width=3.2in]{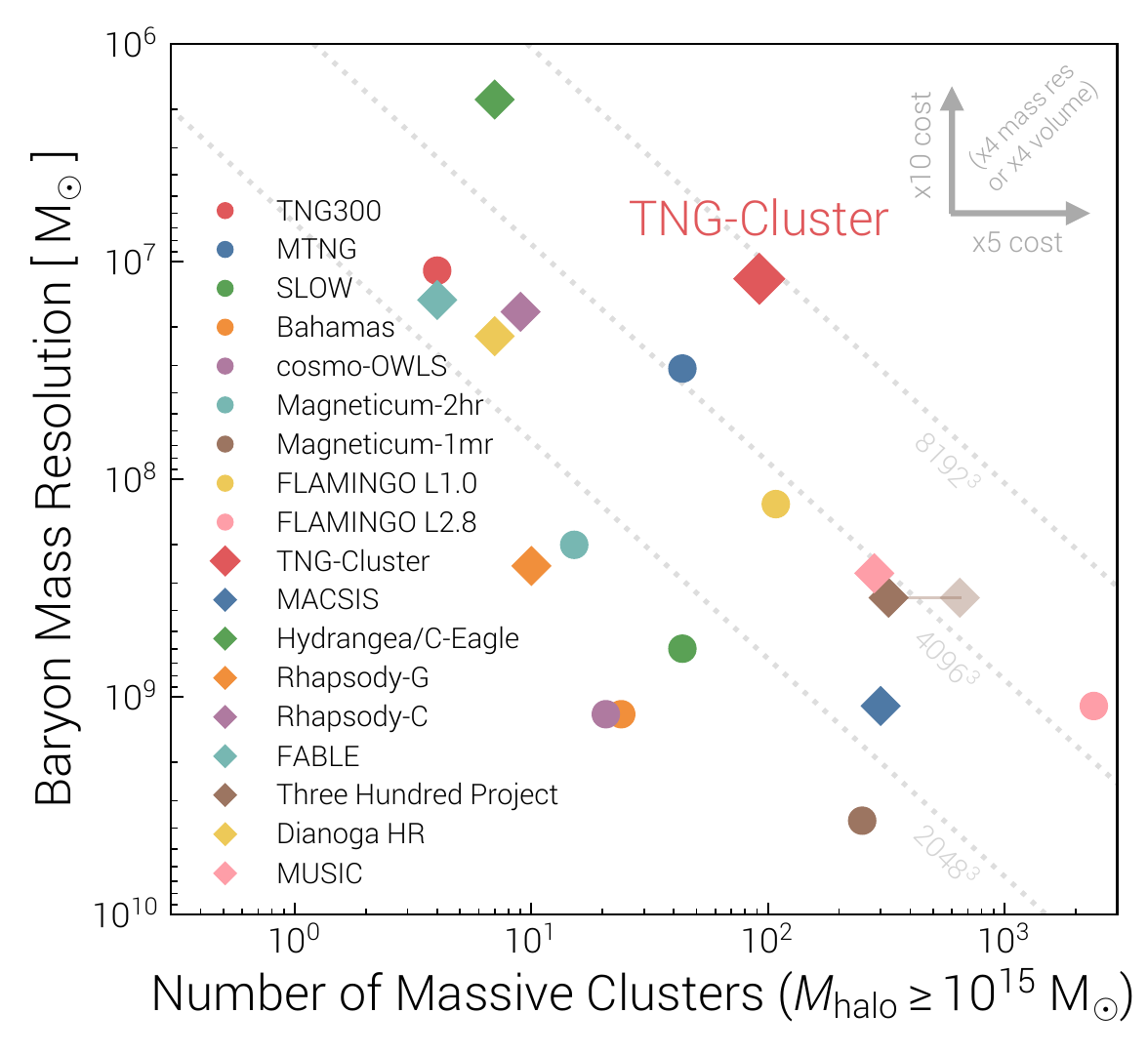}
\caption{ Comparison of the TNG-Cluster sample, introduced here, to other simulated cluster ensembles. For each simulation we contrast the numerical resolution, in terms of gas/baryonic mass resolution (y-axis), versus the number of simulated massive halos with $M_{\rm 200c} \geq 10^{15}$\msun (x-axis). We include only cosmological hydrodynamical simulations of galaxies, both uniform boxes (circles) and zoom-in suites (diamonds). The vast majority of these simulations include models for both stellar and supermassive black hole feedback physics, although they vary in many key aspects (see text for details).
 \label{fig_sim_comparison}}
\end{figure}

In order to compare to other samples of simulated clusters, Figure \ref{fig_sim_comparison} shows TNG-Cluster in terms of the volume versus resolution compromise: on the plane of baryonic mass resolution (better upwards) against the number of simulated massive clusters. We define massive clusters as having $M_{\rm halo,200c} > 10^{15}$\,\msun and only include cosmological (non-idealized) hydrodynamical (with gas) simulations run to $z=0$. Filled circles represent uniform volume (i.e. `full box') cosmological simulations, while diamonds represent suites of zoom-in cluster halos.

TNG-Cluster occupies a unique region of this parameter space, and even more so with the joint TNG300 plus TNG-Cluster sample. In terms of statistics, it is most similar to The Three Hundred cluster simulation project \citep{cui18,cui22}, albeit with only one physical model realised, but with 100 times improved mass resolution. The Hydrangea/C-EAGLE simulations \citep{bahe17b,barnes17b} provide much smaller samples, albeit at higher TNG100-like resolution. The MACSIS sample \citep{barnes17a} as well as large hydrodynamical boxes, particularly cosmo-OWLS \citep{lebrun14}, BAHAMAS \citep{mccarthy17}, and Magneticum \citep{dolag16} also provide comparable, to far larger, samples, but at mass resolutions orders of magnitude worse than TNG-Cluster. This is also the case for the original MUSIC sample \citep{sembolini13} and Omega500 NR \citep{nelsonk14}, although these simulations contain much simpler physics, i.e. without AGN feedback and/or without broad empirical calibration on basic observables such as gas fractions. The recent MillenniumTNG \citep{pakmor23} and FLAMINGO \citep{schaye23} large-volume simulations include similar (to much larger) cluster samples at similar (to much worse) resolution. Much smaller samples with comprehensive/validated physical models include the DIANOGA simulations \citep{bassini20} and the FABLE cluster zooms \citep{henden18}. Finally, the original Rhapsody-G clusters \citep{hahn17} have been updated with more sophisticated physics, including AGN dynamics, feedback, and thermal conduction with Rhapsody-C \citep{pellissier23}.\footnote{TNG50 and TNG100, as well as the original Illustris simulation, EAGLE, and SIMBA all contain no, to $\mathcal{O}(1)$, high-mass clusters, and we exclude them here. We also exclude YZiCS \citep{choi17}, nIFTy \citep{sembolini16b}, and ISC \citep{vazza17}, which contain zero or one such clusters.}

Even with `basic' galaxy formation physics, simulating high-mass galaxy clusters in the full cosmological context at high numerical resolution -- i.e., that achievable at Milky Way halo-mass scales \citep[e.g.][]{guedes11,grand16a,pillepich23,wetzel23}, much less dwarf galaxy scales \citep[e.g.][]{wang15,hopkins18,smith19,agertz20} -- is impossible. Clusters simply contain so much mass that resolving these structures at high resolution is computationally demanding. In this context, TNG-Cluster strikes a balance between detail and statistics. It remains at a baryon mass resolution of $\sim 10^7$\msun (equivalent to TNG300-1), which although low in the context of the IllustrisTNG model and simulations, is high in the context of other cluster simulations. As a result, it is able to include a population of $\sim$\,100 clusters with $M_{\rm 200c} \geq 10^{15}$\msun, enabling unbiased and population-wide comparisons and results. A selection of key physical properties of the main systems simulated within TNG-Cluster are given in Table~\ref{table_clusters}.

Throughout this paper, we exclusively focus on the 352 systems that are the primary zoom targets of the simulation suite, namely the most massive halos in the zoom regions at $z=0$ and their progenitors. In addition to those halos, TNG-Cluster contains many `bonus' halos at high (i.e. uncontaminated) resolution. In particular, there are 241 (9110) additional group (Milky Way) mass halos with $M_{\rm 200c} \geq 10^{13} \,(10^{12})$\msun, i.e. TNG-Cluster also contains large populations of nearby halos in the high density environments centered on clusters.


\begin{figure*}
\centering
\includegraphics[angle=0,width=6.8in]{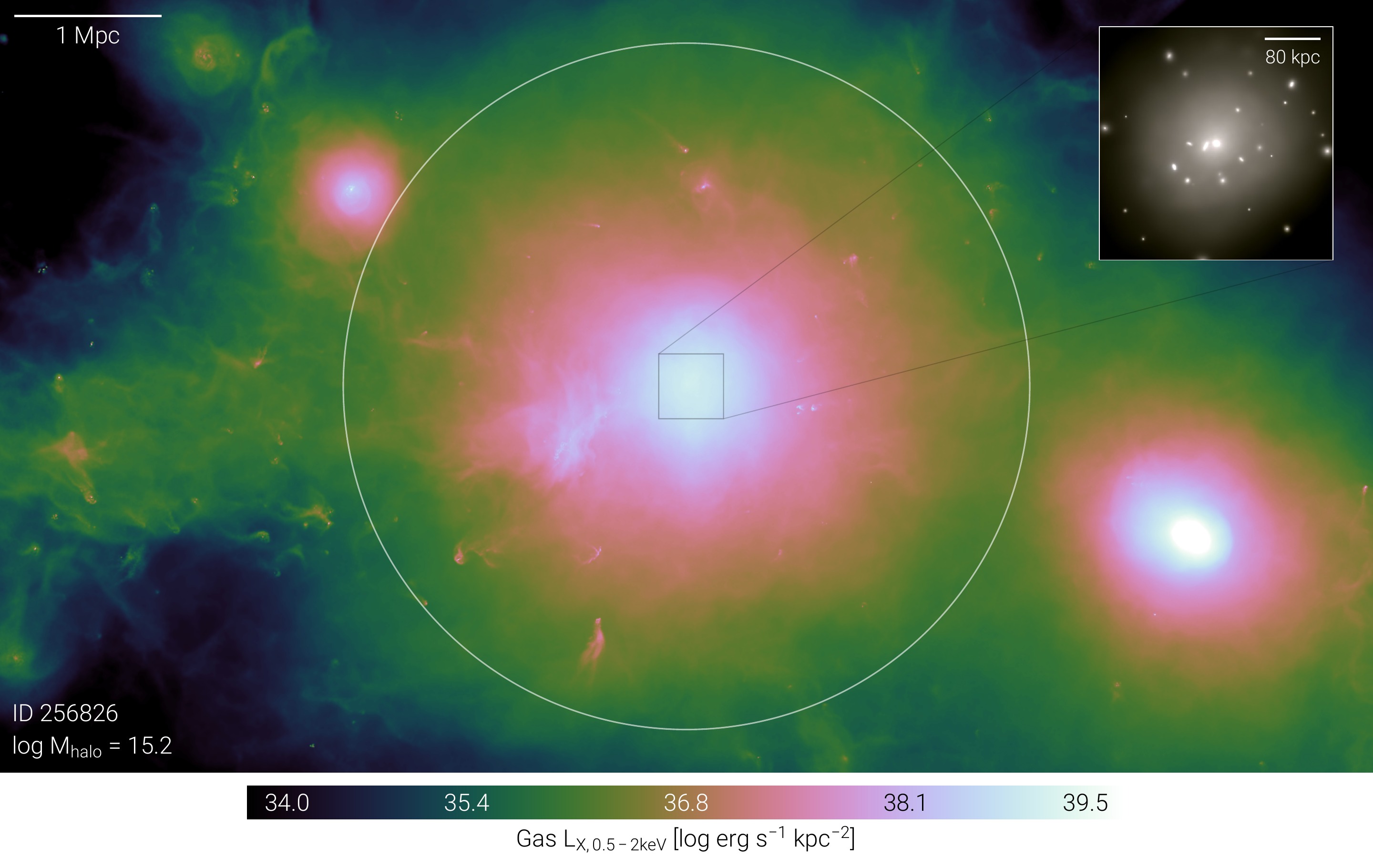}
\includegraphics[angle=0,width=2.26in]{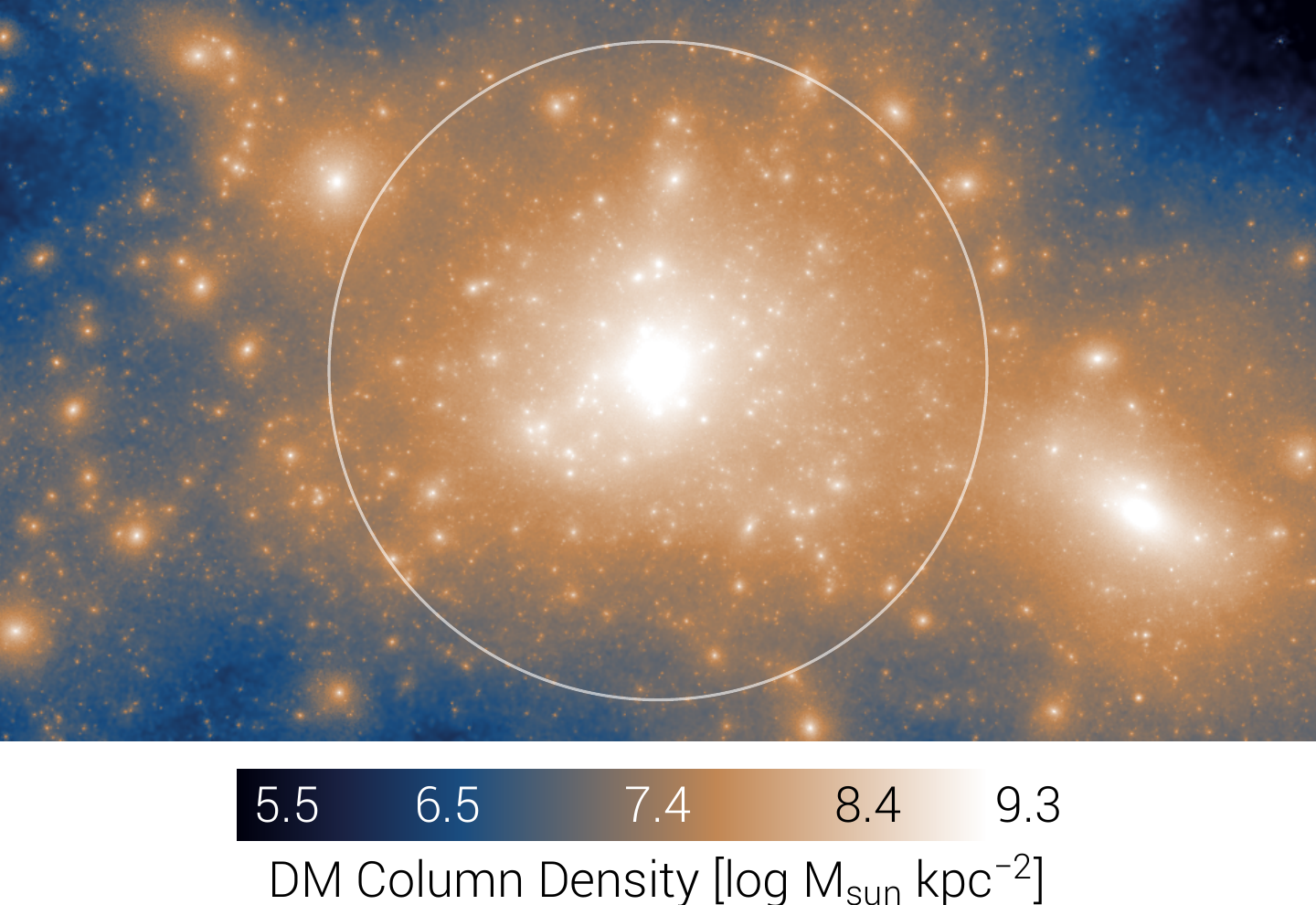}
\includegraphics[angle=0,width=2.26in]{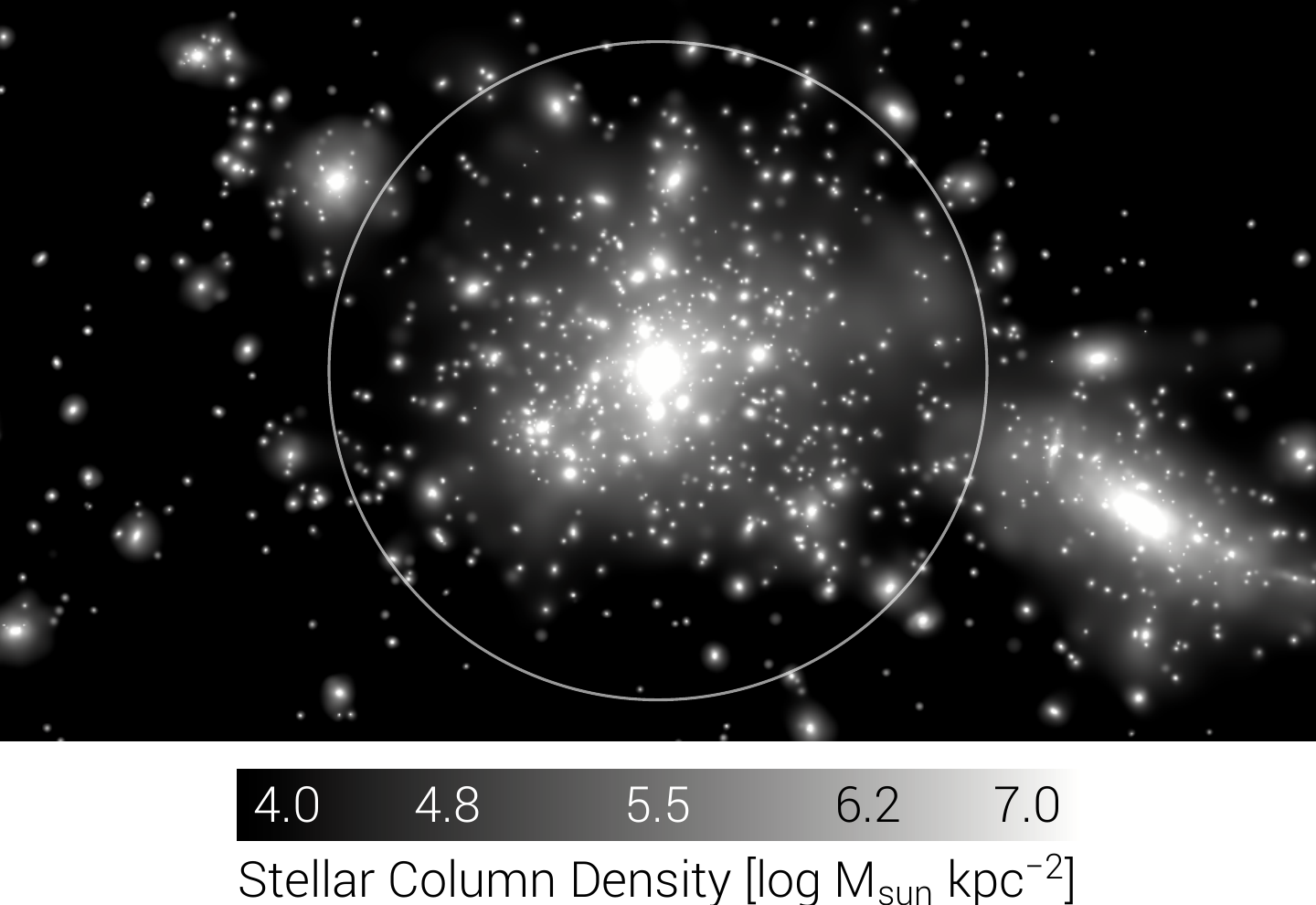}
\includegraphics[angle=0,width=2.26in]{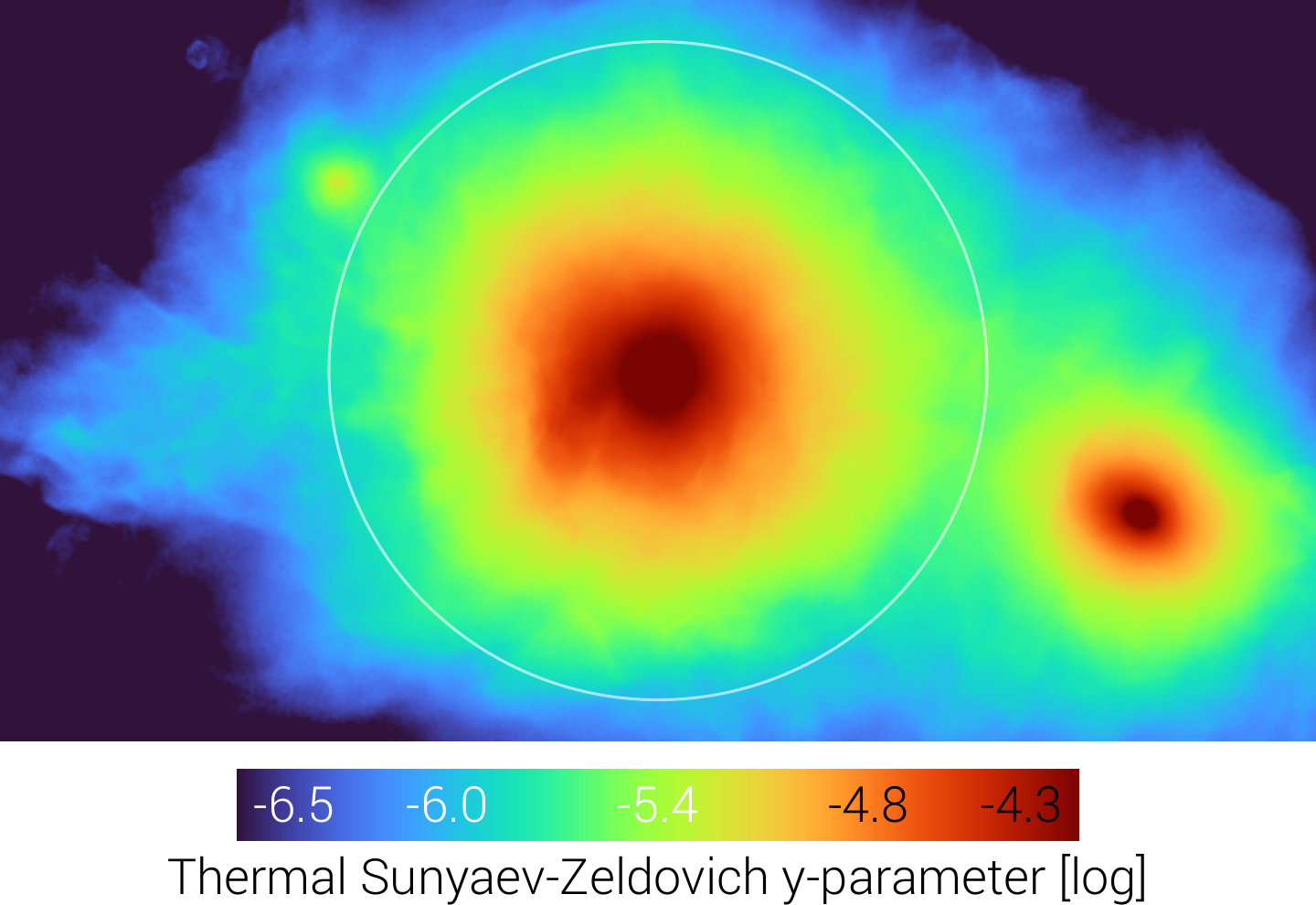}
\includegraphics[angle=0,width=2.26in]{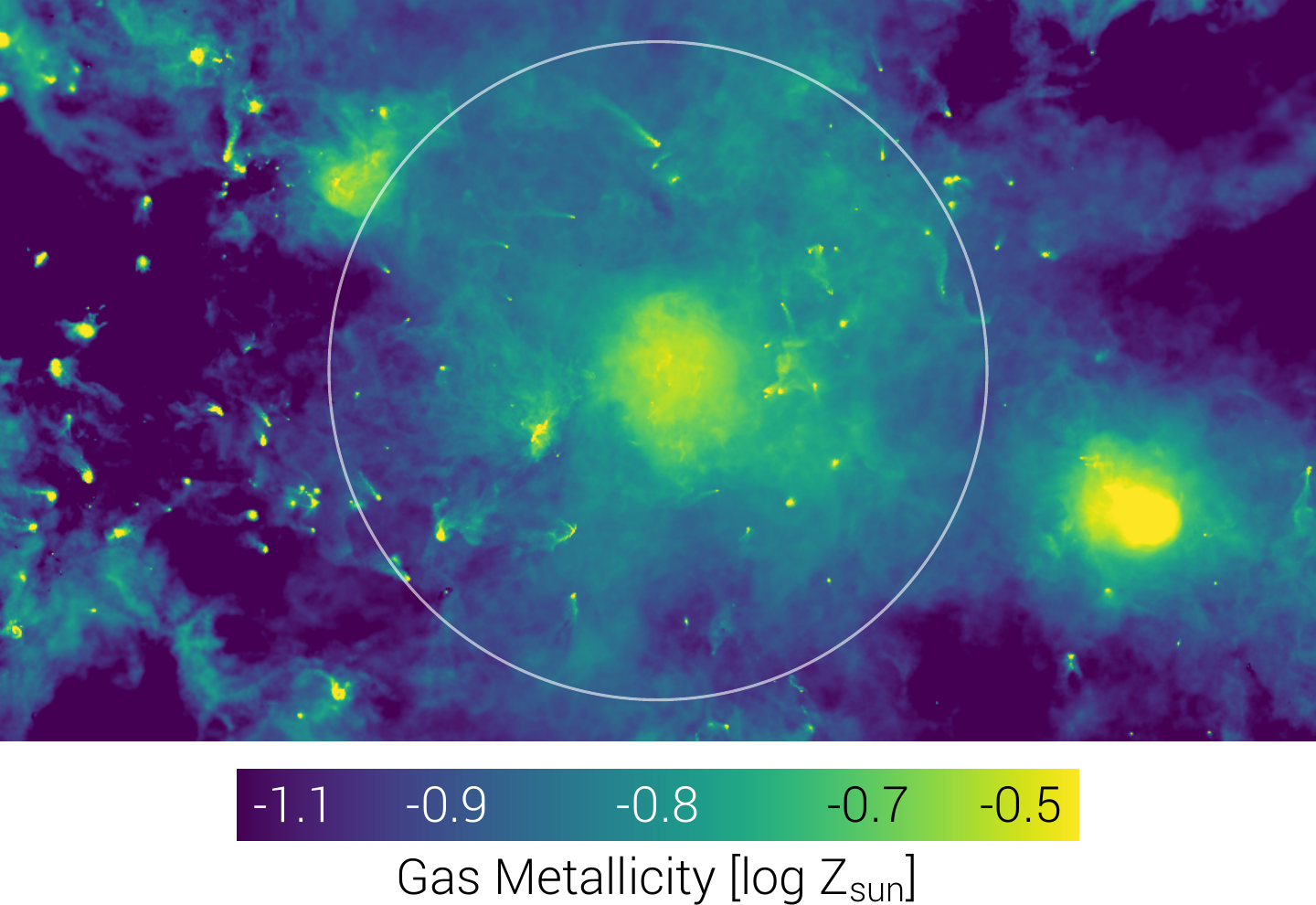}
\includegraphics[angle=0,width=2.26in]{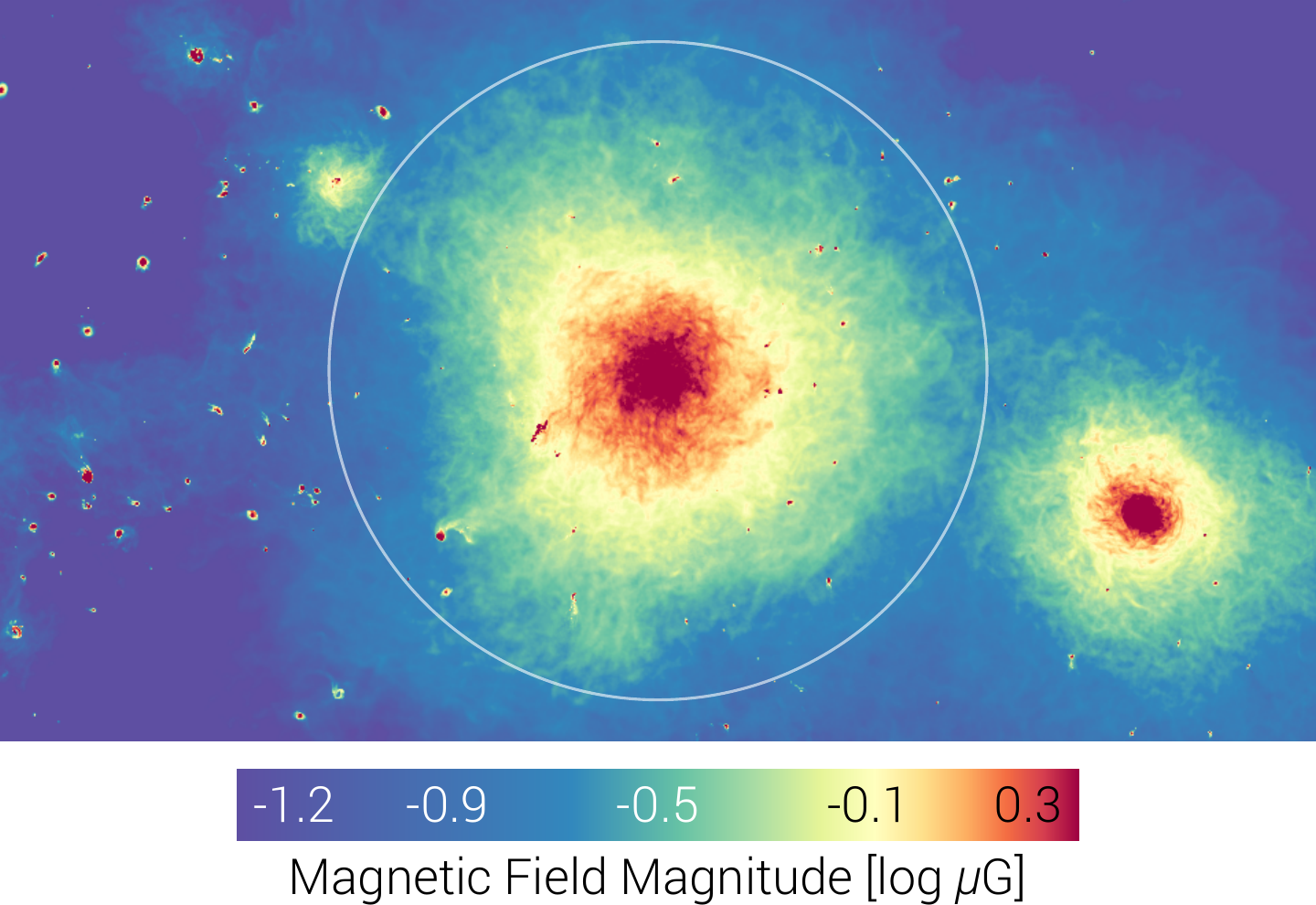}
\includegraphics[angle=0,width=2.26in]{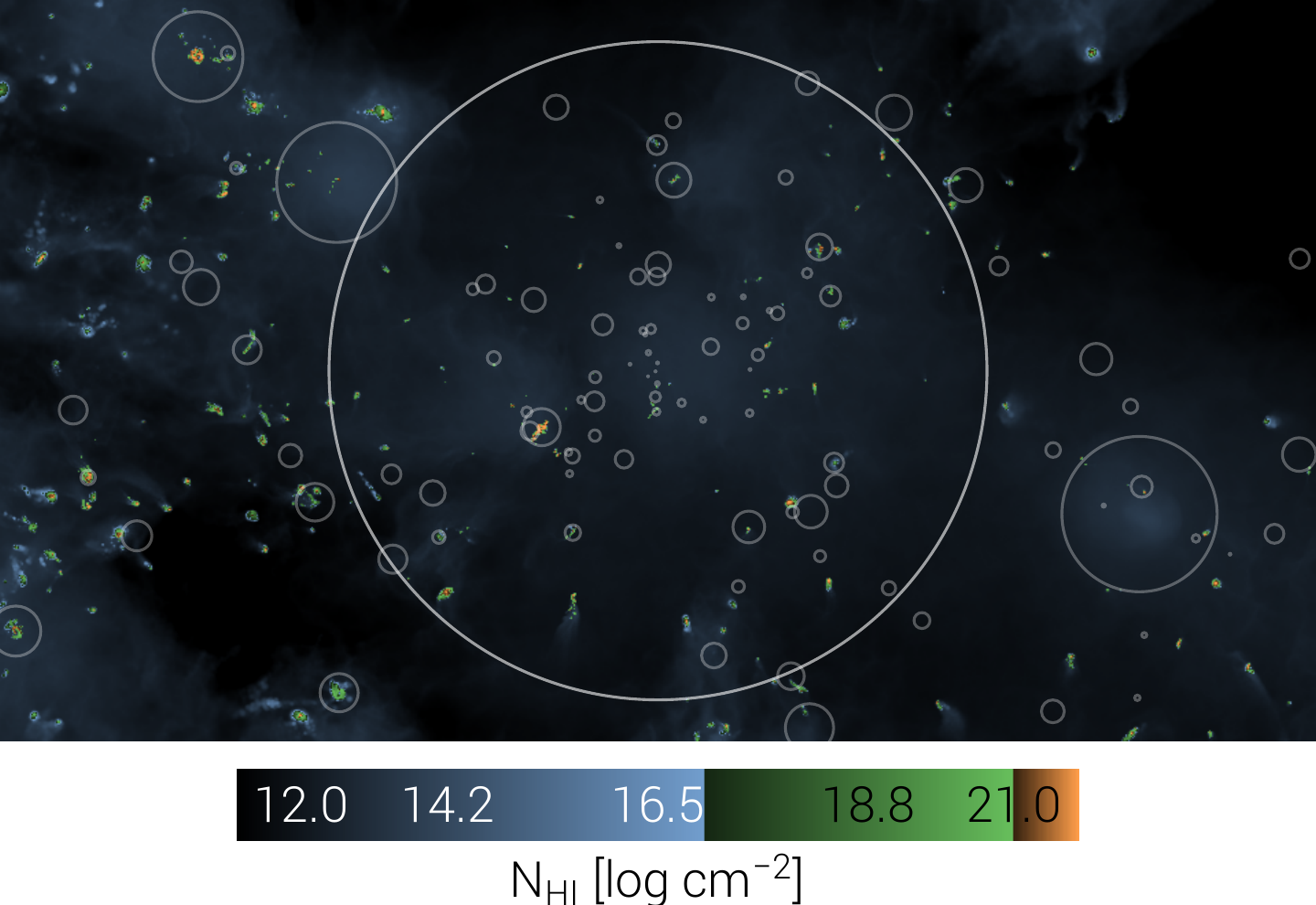}
\caption{ Visualization of the gaseous, stellar, and dark matter distribution and physical properties for the third most massive halo in TNG-Cluster at $z=0$, with a total mass $M_{\rm 200c} \simeq 10^{15.2}$\msun. The main panel shows soft-band ($0.5-2$\,keV) X-ray surface brightness, while the zoomed inset shows a mock stellar light image of the brightest central/cluster galaxy (BCG). The smaller panels show dark matter density, stellar density, the Sunyaev-Zeldovich y-parameter map, gas-phase metallicity, magnetic field strength, and neutral HI column density. In this final panel, the location and size of the 100 most massive subhalos are also indicated with circles. In all cases, the white circles mark $R_{\rm 200c}$. This single halo is resolved in TNG-Cluster with $\sim$ 100,000,000 resolution elements, split between dark matter, stars, gas cells, and SMBHs. It contains $\sim$ 20,000 resolved substructures, including $\sim$ 3,500 luminous (satellite) galaxies. Not pictured: the cluster also hosts $\sim 800$ SMBHs with $M \geq 10^6$\msun, and the central SMBH has a total mass of $\sim 10^{10.2}$\msun. Our \href{https://www.tng-project.org/cluster/gallery/}{online TNG-Cluster `infinite galleries'} show similar projections across the full sample of 352 halos.
 \label{fig_vis_single}}
\end{figure*}

\section{Properties of the Intracluster Medium Gas} \label{sec_gas}

\subsection{Visual Overview}

We begin to explore the richness and detail of the TNG-Cluster results by visualizing a number of spatially resolved physical properties and observable quantities of a single halo. 

Figure \ref{fig_vis_single} shows seven projections for the third most massive cluster at $z=0$, with a total halo mass of $M_{\rm 200c} \simeq 10^{15.2}$\msun. The main panel focuses on a key observable of galaxy clusters: X-ray emission from the ICM. We show the soft-band ($0.5-2$\,keV) X-ray surface brightness on the scale of the virial radius (white circle). While centrally concentrated, we also see: (i) a bright merger front feature to the lower left of the core \citep{zhang20a}, (ii) a filamentary bridge of extended emission extending off to the left \citep{reiprich21}, (iii) two infalling substructures and early sloshing fronts just outside $R_{\rm 200c}$ \citep{zuhone11}, and (iv) small-scale, whispy, filamentary, and tail-like features throughout the ICM, signposts of lower mass accreting satellites, gas stripping, and jellyfish galaxies \citep{yun19}.

The smaller six panels below show, from left to right: dark matter projected density, stellar mass density, the Sunyaev-Zeldovich Compton-y map, gas-phase metallicity, magnetic field strength, and cool gas as traced by neutral hydrogen column density. All are shown on the same scale as the main panel.

The distribution of dark matter (middle left panel) reveals the substructure richness of massive clusters \citep{angulo12}. In TNG-Cluster, this single halo is resolved with $\sim 100,000,000$ particles/cells, and contains a total of 19,700 resolved subhalos, with total masses between $8.2 < M_{\rm sub} / \rm{M}_\odot < 14.6$, of which 3,500 are luminous galaxies with stellar masses spanning $6.8 < M_{\rm sub,\star} / \rm{M}_\odot < 12.6$. The spatial distribution of these stars (middle center panel) reveals the massive central galaxy (i.e. BCG), a large satellite population, and the diffuse intracluster light \citep{lin04,conroy07,pillepich18b}.

These two collisionless components assemble within the gaseous component. In addition to X-ray emission, this hot plasma of the ICM of clusters is also commonly observed via the Sunyaev-Zeldovich effect, whereby cosmic microwave background photons interact with free electrons via inverse Compton scattering \citep{sunyaev70}. The resulting spectral distortions include thermal, kinetic, and polarization signatures -- all are modulated by the impact of galaxy formation i.e. feedback physics \citep{nagai06}. We show the dimensionless Compton y-parameter map (middle right panel) as a measure of the thermal SZ effect (tSZ; see Section \ref{sec_methods2}). The central cluster and the substructure to the lower right are both clearly visible, while the lower mass infaller to the upper left has a factor of 10 smaller integrated $Y_{\rm SZ}$. The spatially resolved SZ maps are significantly smoother than for X-ray emission \citep{battaglia12}, that traces the same hot phase but with a stronger density dependence.

The SZ morphology is also significantly smoother than the distribution of gas-phase metals (lower left panel), which is inhomogeneous and clumpy throughout the cluster volume. The cluster core reaches metallicities of $\sim Z_\odot / 2$. The bulk of the cluster ICM is also highly enriched, with a shallow gradient and still $\gtrsim 0.1 Z_\odot$ out to the virial radius \citep{ettori15,yates17,mernier17}. This reflects an interplay between metal expulsion through feedback-driven outflows and environmental pre-enrichment of subsequent accretion \citep{vog18a}.

\begin{figure*}
\centering
\includegraphics[angle=0,width=7.0in]{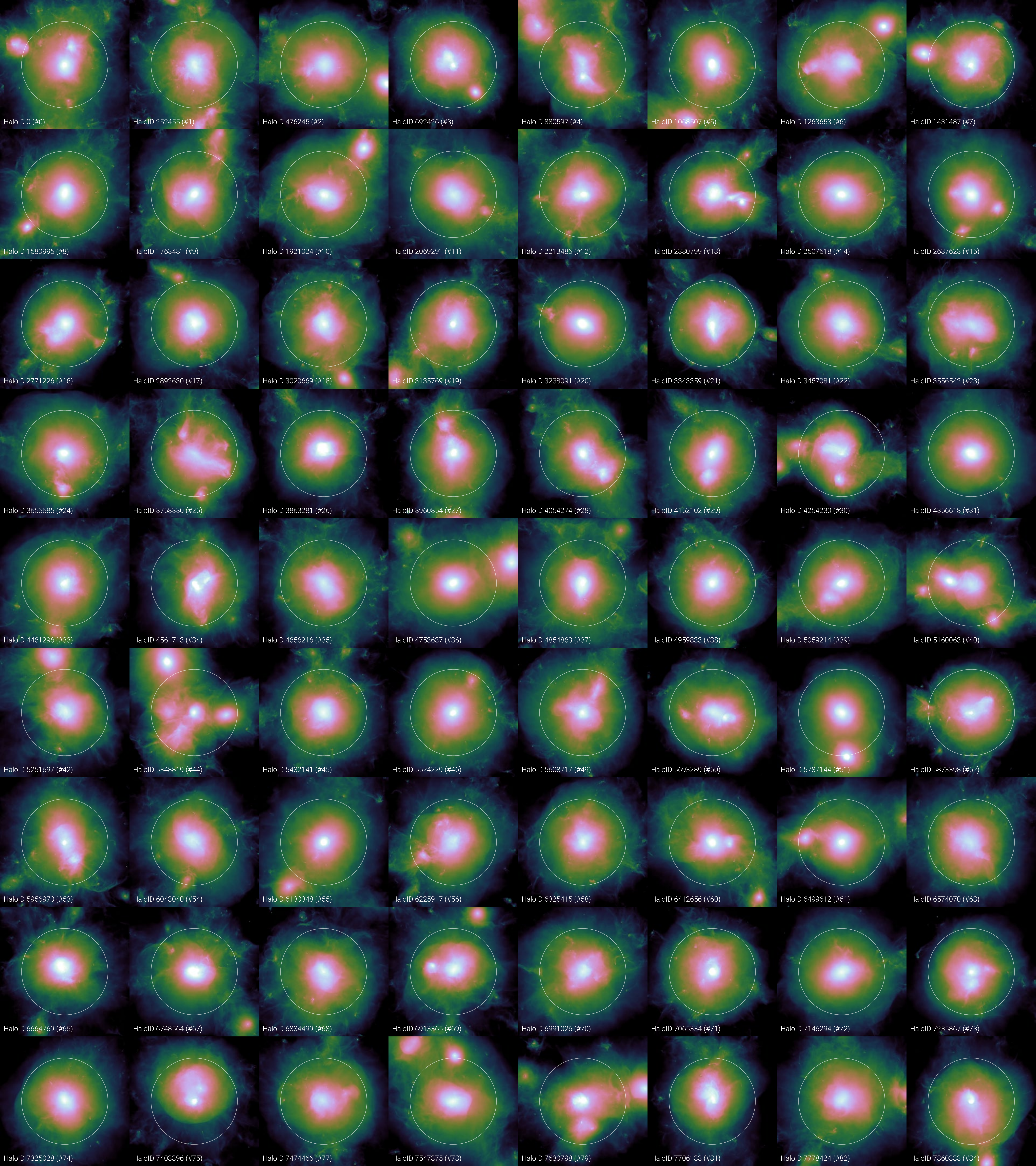}
\caption{ Gallery of the first 72 halos of TNG-Cluster at $z=0$, showing soft-band ($0.5-2$\,keV) X-ray luminosity. The colormap is the same as Figure \ref{fig_vis_single}. Each panel is $2 r_{\rm vir}$ across, while the white circles mark $r_{\rm vir}$, and halo IDs are labeled. A variety of X-ray morphologies are evident, from relaxed, centrally peaked surface brightness distributions, to un-relaxed and binary merging clusters, triple and multi-component merging systems, and large scale sloshing and gas displacement effects.
 \label{fig_vis_gallery}}
\end{figure*}

Magnetic fields in TNG-Cluster are self-consistently amplified from a vanishingly small primordial seed field in the initial conditions to their present day values due to a combination of turbulent and small-scale dynamo processes \citep{pakmor14,pakmor17}.\footnote{Magnetic field amplification with the AREPO ideal MHD solver has been well studied in the Auriga simulations \citep{pakmor23b}. B fields were omitted from MillenniumTNG due to memory constraints, making TNG-Cluster uniquely informative at the high-mass end.} In the dense cores of clusters, magnetic fields reach values of $\gtrsim 10 \mu G$, rapidly decreasing with density and distance towards cluster outskirts \citep{marinacci18}. The resolved magnetic field structure hints to a turbulent and complex small-scale topology \citep{dominguez19}, large-scale features associated with anisotropy of the ICM and feedback-driven outflows \citep{nelson21,ramesh23c}, and dynamical effects from satellite galaxy draping \citep{pfrommer12}.

As satellites infall and begin to orbit within the ICM they are subject to strong environmental processes including ram-pressure stripping \citep{ayromlou21b}. The removal of their cold interstellar medium (ISM) gas is traced by the distribution of neutral hydrogen \citep{rohr23} throughout the cluster (lower right panel). Low column density gas ($12 \lesssim \log N_{\rm HI} [\rm{cm}^{-2}] \lesssim 17$) is present throughout the cluster. The central galaxy is itself devoid of any significant amount of cool gas, although this is not always true, as we discuss below. Lyman-limit systems (LLS; $N_{\rm HI} > 10^{17} \rm{cm}^{-2}$, in green) and damped Lyman-alpha absorbers (DLA; $N_{\rm HI} > 10^{20.3}$, in orange) are strongly localized around satellite galaxies. In some cases, extended tails of stripped gas produce `jellyfish' phase galaxies \citep{poggianti17,zinger23}. In this final panel we also include smaller white circles marking the location and size of the 100 most massive subhalos of this cluster -- while some contain abundant HI, others are entirely devoid of gas due to environmental effects which eventually quench the star formation of cluster member galaxies \citep{bahe15,donnari20b}.

Although not atypical, this single cluster is not necessarily representative of the population, and its properties cannot showcase the full diversity of the TNG-Cluster sample. Figure \ref{fig_vis_gallery} therefore shows a gallery of X-ray emission from the 72 most massive $z=0$ halos in the simulation, with the same spatial scale, and color scale, as above. In every stamp, the white circle shows $R_{\rm 200c}$. Some clusters are clearly isolated, relaxed, non-merging systems. Their X-ray morphologies are to first order circularly symmetric and centrally peaked. However, a large fraction of clusters exhibit strongly perturbed X-ray surface brightness maps indicative of ongoing mergers \citep{poole06,zuhone11}. Many systems also exhibit three or more distinct X-ray components \citep{monteiro22}, suggesting complex merger scenarios and potentially accretion along preferred, large-scale filament directions \citep{colberg05,kartaltepe08}.

\subsection{Properties and Observables of the Hot ICM}

\begin{figure}
\centering
\includegraphics[angle=0,width=3.3in]{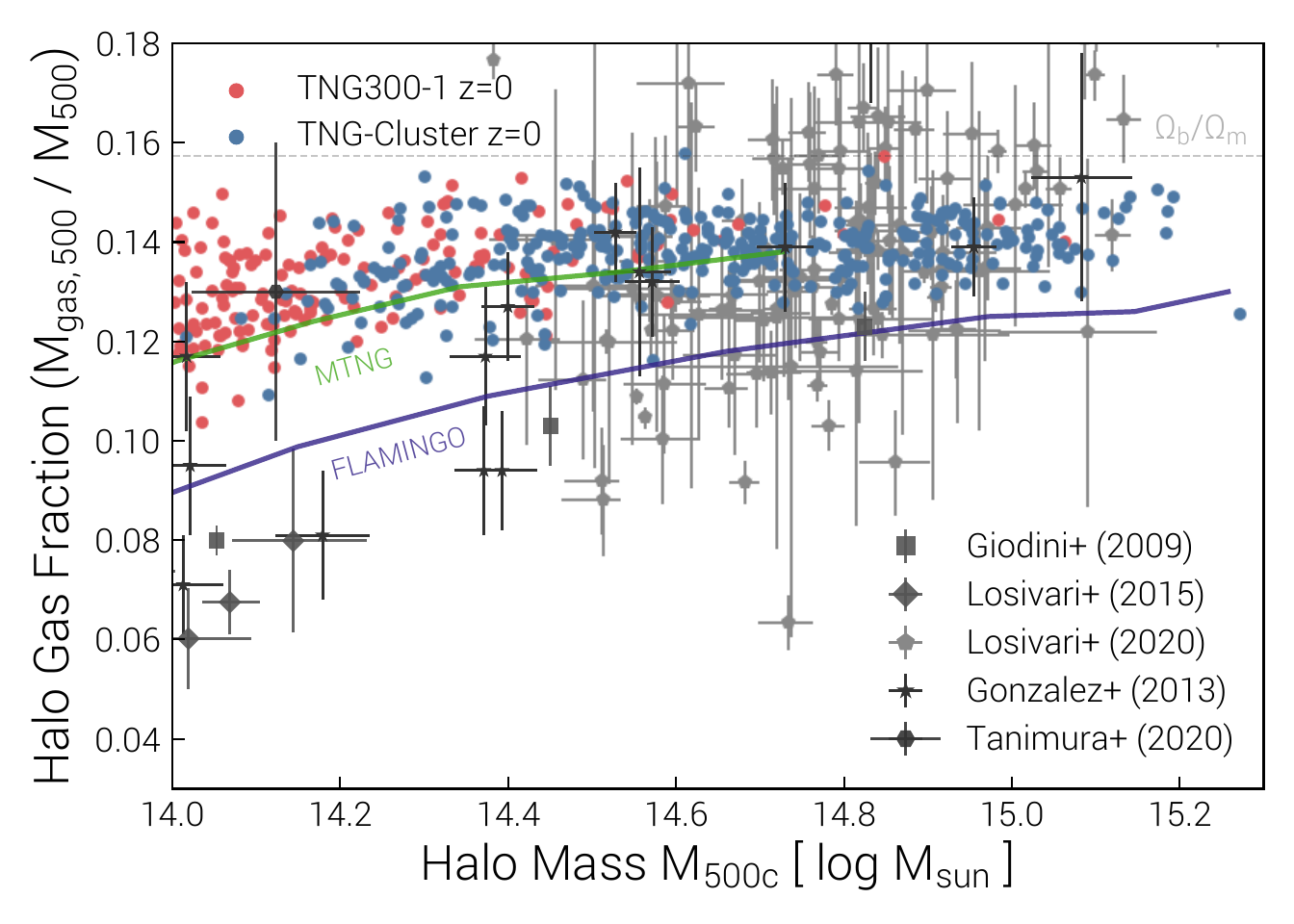}
\caption{ The gas fraction of halos, taken as the ratio of gas mass to total mass, both within a 3D aperture of $R_{500c}$. We compare to observational inferences derived from X-ray emission \protect\citep{giodini09,gonzalez13,lovisari15,lovisari20}, measurements of the kinetic Sunyaev-Zeldovich effect \citep{tanimura21}. In all three cases, in lieu of detailed mock observations enabling apples-to-apples evaluation, this is a qualitative, at face-value comparison only. For comparison, the green and purple solid curves are the running averages from the recent MTNG and FLAMINGO simulations.
 \label{fig_gas_fraction}}
\end{figure}

\subsubsection{Halo gas mass fraction}

We begin our quantitatively exploration of the properties of TNG-Cluster with total halo gas content. Figure \ref{fig_gas_fraction} shows the $z=0$ halo-scale gas fractions ($M_{\rm gas,500} / M_{\rm 500c}$) as a function of halo mass. The cosmic baryon fraction $f_{\rm b,cosmic} = \Omega_{\rm b} / \Omega_{\rm m} \simeq 0.16$ is indicated by the horizontal dotted line. Red and blue symbols show clusters from TNG300 and TNG-Cluster, respectively. This figure extends towards higher masses a similar analysis of TNG100 and TNG300 by \citet{pop22}. We include observational data derived from X-ray measurements \citep{giodini09,gonzalez13,lovisari15,lovisari20}, and kSZ extracted from Planck data \citep{tanimura21}. The TNG-Cluster gas fractions are relatively flat across the cluster mass regime. Data agrees that massive clusters have high $f_{\rm gas,r500c} \gtrsim 90$\% of the cosmic baryon fraction. However, at least some observations suggest stronger trends arising from smaller inferred $f_{\rm gas}$ at $M_{\rm 500c} \lesssim 10^{14.4}$\msun \citep[e.g.][]{chiu18}, suggesting that group gas fractions in TNG may be on the high side.

While the highest mass clusters are baryonically closed, with $f_{\rm b} \simeq f_{\rm b,cosmic}$ \citep{ayromlou23}, this is not the case for low mass clusters and groups, where AGN feedback can significantly reduce the gas content of dark matter halos \citep{mccarthy11,genel14}. Gas fractions (in group-mass halos) correlate well with the impact of baryons on the total matter power spectrum \citep[on certain scales;][]{vandaalen11,arico21,giri21}, implying that realistic halo gas fractions are important for precision cosmology applications. 

Several caveats exist for our comparison with $f_{\rm gas}$ data. X-ray, weak lensing, and SZ-based estimates of gas content and/or total halo mass have different and non-negligible systematic uncertainties involved, and we emphasize that this comparison with data is meant to be qualitative, at face value, only. We have also not added observational uncertainty to the simulated points, such that (i) the simulated scatter is by definition smaller than the apparent observational scatter, and (ii) the simulated points are confined to $f_{\rm gas} < f_{\rm b,cosmic}$, which is not true for the data. Further, X-ray based inferences of halo-scale gas fractions rely on the assumption of hydrostatic equilibrium, which is not generally true \citep{rasia06}. Simulations suggest hydrostatic biases of $b \sim 0.1 - 0.3$ \citep{barnes21,jennings23}. A comparison can approximately correct for this effect \citep{schaye23}, or forward model the X-ray emission to enable true apples-to-apples comparisons \citep{biffi13}. Leaving such detailed synthetic observations for the future, we broadly conclude that the TNG-Cluster gas fractions are qualitatively consistent with data, with quantitative agreement to be demonstrated.

We also compare to the recent FLAMINGO \citep[purple line;][]{schaye23} and MTNG \citep[green line;][]{pakmor23} simulations. As expected, the MTNG gas fraction trend is compatible with TNG-Cluster, albeit without the high-mass end sampling. In contrast, FLAMINGO has noticeably $\sim 15$\% lower $f_{\rm gas,r500c}$ in high-mass clusters, highlighting the impact of different baryonic feedback models even in these massive halos.

\subsubsection{Magnetic field and synchrotron emission}

\begin{figure}
\centering
\includegraphics[angle=0,width=3.4in]{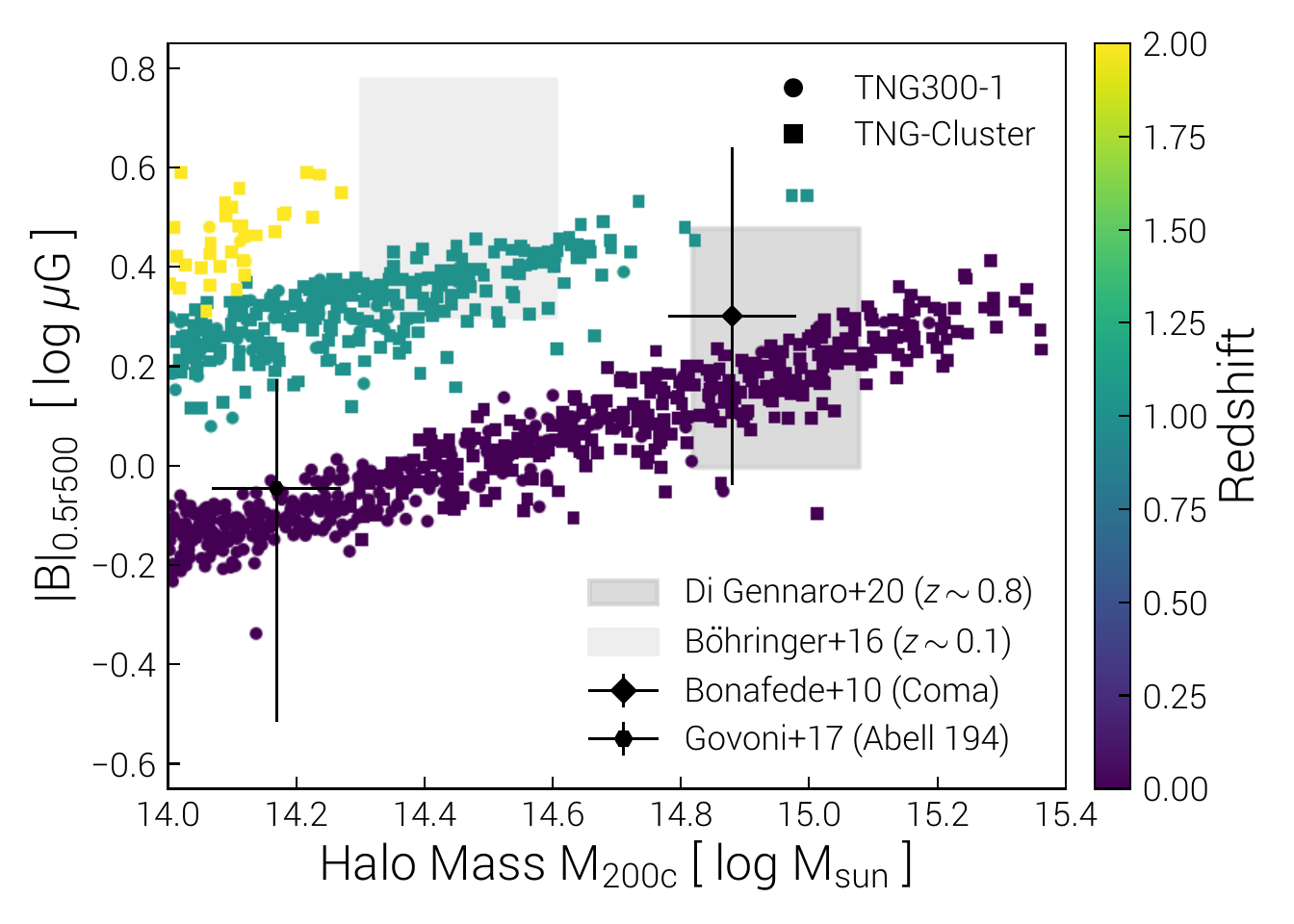}
\caption{ The volume-weighted mean magnetic field strength in $\mu$G, measured for all gas within $0.5\,R_{\rm 500c}$ of halos, as a function of halo mass. Color indicates three distinct redshifts considered, from $z=0$ to $z=2$. Circular (square) markers show halos from TNG300 (TNG-Cluster). We provide several qualitative, face-value comparisons with observational inferences from Faraday rotation measure analyses \citep{bonafede10,bohringer16,govoni17,digennaro20}. The magnetic field strength in the inner ICM has characteristic values of order $\sim \mu$G, increasing for larger halo masses, higher redshifts (at fixed mass), and smaller cluster-centric distances.
 \label{fig_bfields}}
\end{figure}

Figure \ref{fig_bfields} shows a first look at the magnetic field properties of the ICM in TNG-Cluster \citep[see also][for a TNG300 analysis]{marinacci18}. For each halo we derive the mean, volume-weighted magnetic field strength within $0.5\,R_{\rm 500c}$ at $z=0,1,2$ and for all three redshifts show the trend with halo mass. Circular and square symbols show halos from TNG300 and TNG-Cluster, respectively. In the TNG model, the magnetic field strength in gaseous halos declines rapidly with increasing distance \citep{marinacci18, ramesh23c,ramesh23a}, so values of $|B|$ in the core (outskirts) of our simulated clusters are significantly higher (lower). This particular choice roughly approximates an average value over the inner ICM which could be inferred observationally.

We make several comparisons with data, which are meant in the qualitative sense only. For the Coma cluster, Faraday rotation measure (RM) modeling infers a decreasing magnetic field strength with halocentric distance, from $\sim 4 \mu$G to $\sim 1 \mu$G in the outskirts \protect\citep{bonafede10}. For comparison we also include the lower mass Abell 194 cluster, where the uncertainty shows the difference between the central value of $\langle B_0 \rangle = 1.5 \mu$G and the $\sim 1$ Mpc halo volume average $B \sim 0.3 \mu$G \protect\citep{govoni17}. Measurements of diffuse radio emission from clusters with LOFAR now infer that magnetic fields in intermediate redshift clusters $z \sim 0.7-0.8$ are similar to local clusters, with $B \gtrsim 1-3 \mu$G, implying a relatively fast early amplification mechanism \protect\citep[dark gray;][]{digennaro20}. Measurements of RM place a similar constraint of $B \sim 2-6 \,(l/\rm{10 kpc})^{-1/2} \,\mu$G at $z \sim 0.1$, where $l$ is is the magnetic coherence length of the ICM \protect\citep[light gray;][]{bohringer16}. Given the limited observational statistics, different methodologies, and myriad physical assumptions for each method, data broadly infer cluster magnetic fields of $\sim \mu$G with $\sim 1$\,dex uncertainty \citep{carilli01}.

\begin{figure}
\centering
\includegraphics[angle=0,width=3.3in]{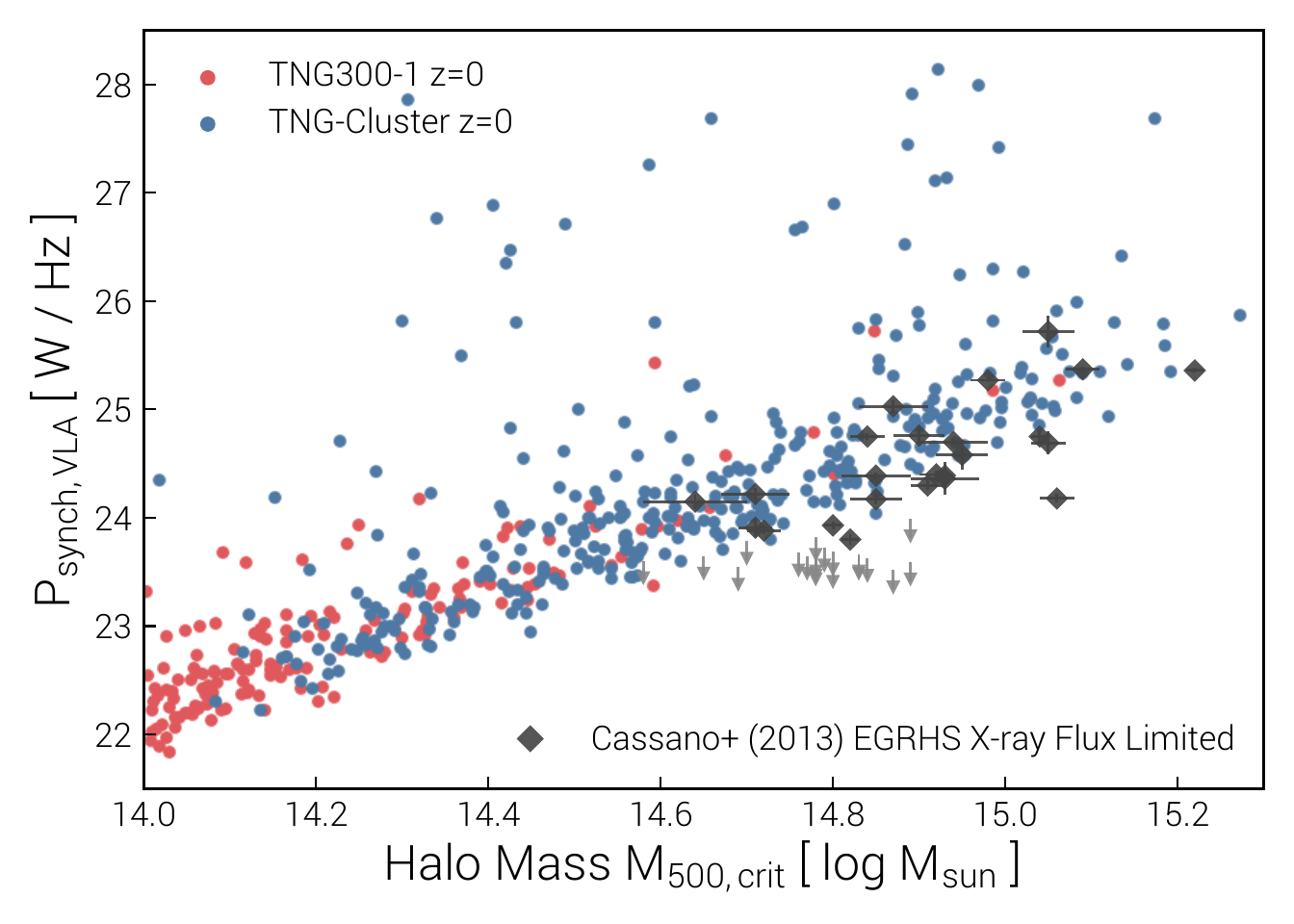}
\caption{ Halo radio synchrotron emission. We show $P_{\rm synch}$ at 1.4\,Ghz for a VLA-like configuration, adopting a simple model for the relativistic electron population that assumes equipartition with the magnetic energy density (see text for details). Radio power rises rapidly with increasing halo mass, following an overall clear relation plus a small fraction of highly emitting outliers. Halo integrated synchrotron emission in TNG-Cluster is in the ballpark of observational results \citep{cassano13}.
 \label{fig_radio}}
\end{figure}

Given the magnetized ICM, Figure \ref{fig_radio} presents a first look at the cluster radio synchrotron emission. An in-depth modeling effort for the diffuse radio emission in cluster outskirts, as accelerated by cluster merger shocks, and the diverse morphologies of radio relics in TNG-Cluster, is presented in a companion paper \citep{lehle24}.

Here we adopt a simple post-processing model to estimate the relativistic electron populations in each gas cell, and thus their observable emission, which is not directly tracked in the simulation (see Section \ref{sec_methods3}). The resulting synchrotron power is a steep function of halo mass, rising from $P \sim 10^{22}$\,W\,Hz$^{-1}$ for $M_{\rm 500c} = 10^{14}$\msun to $P \sim 10^{25}$\,W\,Hz$^{-1}$ by $M_{\rm 500c} = 10^{15}$\msun. Red and blue symbols show halos from TNG300 and TNG-Cluster, respectively. The clear mass trend is accompanied by a small number of outliers which have, surprisingly, significantly enhanced synchrotron emission, by up to several orders of magnitude. We speculate that these may reflect ongoing mergers and strong gas compression events.

In gray symbols with error bars, and gray upper limits, we show results from the Extended GMRT Radio Halo Survey \citep[EGRHS;][]{cassano13}. This is a sample of 67 deep pointed 1.4\,Ghz radio observations, based on X-ray flux limited selections with ROSAT. The detections scatter well within the region occupied by TNG-Cluster halos. At the same time, we have no simulated clusters at much lower values, which would be consistent with the upper limits of that sample. Such clusters lacking large radio halos are typically found to be more relaxed and less dynamically disturbed \citep{brunetti09}, suggesting future modeling improvements \citep[e.g.][]{vazza21}.

\subsubsection{Sunyaev-Zeldovich and X-ray scaling relations}

\begin{figure}
\centering
\includegraphics[angle=0,width=3.4in]{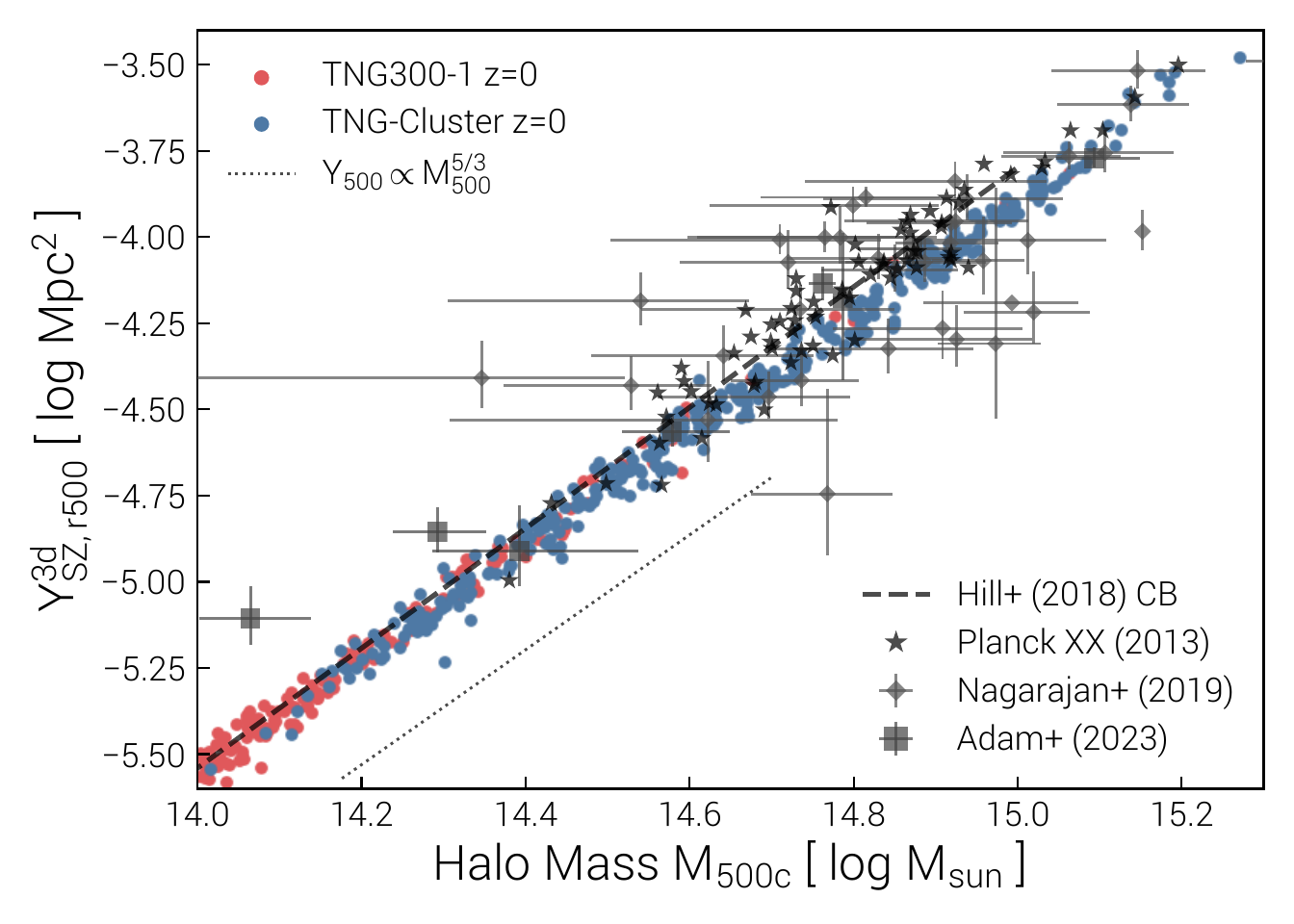}
\includegraphics[angle=0,width=3.4in]{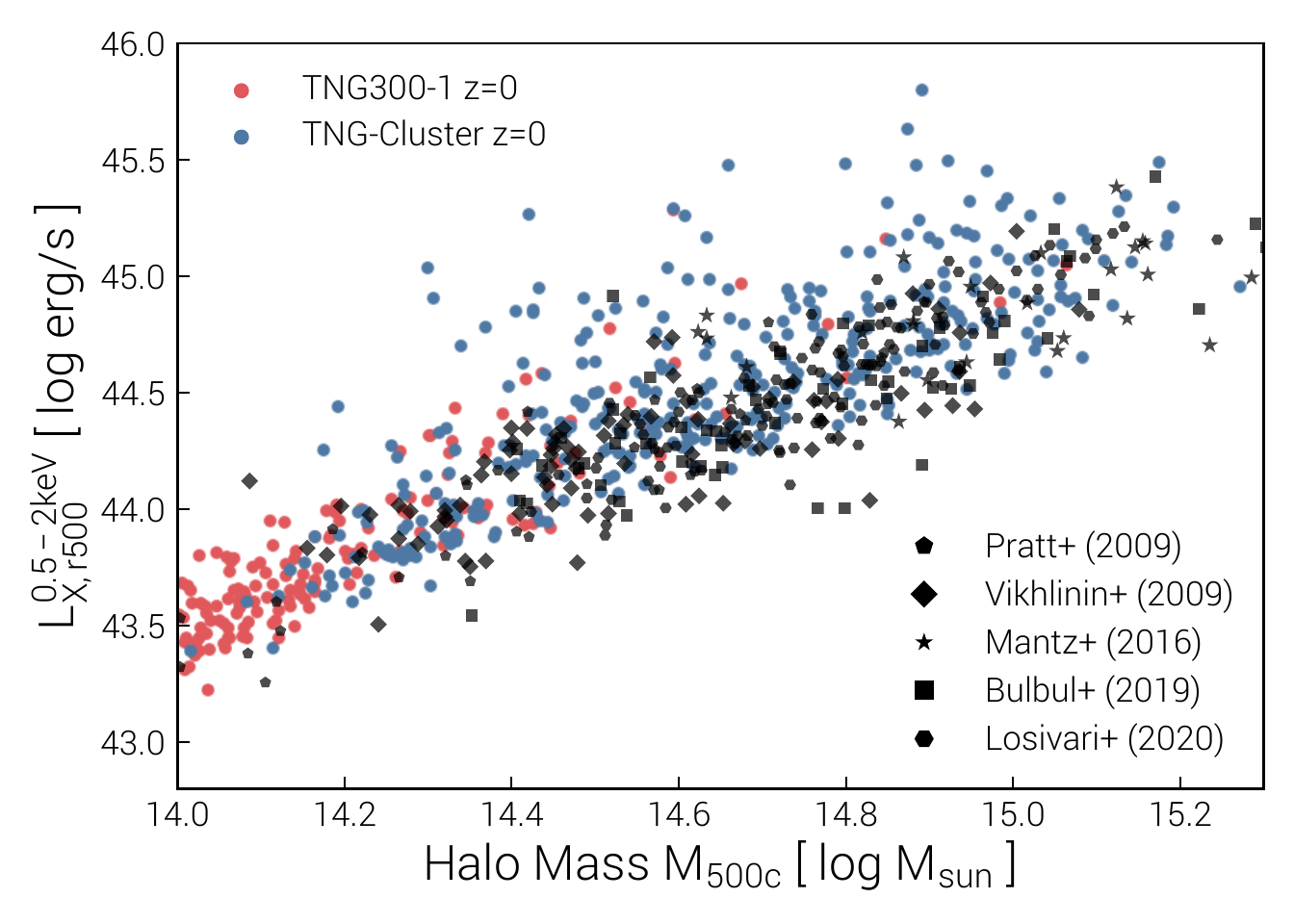}
\caption{ Sunyaev-Zeldovich (top) and X-ray (bottom) scaling relations versus halo mass ($M_{\rm 500c}$) in TNG-Cluster, and comparison to selected data. \textbf{Top:} The $Y_{\rm SZ,500} - M_{\rm 500c}$ relation of the simulated clusters is relatively tight and has a slope consistent with the $5/3$ self-similar expectation. We compare to large survey SZ data from Planck \citep{planck2013_xx,hill18}, weak-lensing based cluster mass estimates \citep{nagarajan19}, and early NIKA2 observations \citep{adam23}. \textbf{Bottom:} The integrated $0.5-2$\,keV cluster luminosity within $R_{\rm 500c}$ rises as a function of mass, and with significant scatter. We compare to available X-ray samples at low redshift \citep{pratt09,vikhlinin09,mantz16,bulbul19,lovisari20}. The SZ and X-ray scaling relations are both in broad qualitative agreement with data.
 \label{fig_xray_sz}}
\end{figure}

We proceed to two key observables of the hot ICM. Figure \ref{fig_xray_sz} shows the classical Sunyaev-Zeldovich and X-ray scaling relations as a function of halo mass \citep[see also][for previous TNG analyses]{lim21, truong21,pop22}.

In the top panel we show the relation between the thermal SZ signal and halo mass. We compute $Y_{\rm SZ}$ at $z=0$ in units of $\rm{pMpc}^2$, and so do not include the $E(z)^{-2/3}$ redshift evolution correction, nor the dependence on angular diameter distance $d_{\rm A}^2$. As before, red and blue symbols show halos from TNG300 and TNG-Cluster, respectively. The simulated clusters follow a tight $Y_{\rm 500}-M_{\rm 500c}$ relation. Its slope is close to the self-similar expectation \citep[dashed line;][]{kaiser86}, being only slightly shallower.

We compare the $Y_{\rm SZ,500}$ scaling relation to observations from \citet[stars;][]{planck2013_xx}, where $M_{\rm 500c}$ values are derived from an empirical mass-proxy relation using X-ray observations of local clusters from XMM-Newton, under the assumption of hydrostatic equilibrium with calibrated bias. An independent re-analysis of the Planck maps, combined with a SDSS DR7-based group catalog results in a largely consistent relation \citep[dashed line;][their `compensated break' model]{hill18}. However, several issues complicate the comparison with simulations. In both cases, the inferred values of $M_{\rm 500c}$ can differ from true intrinsic values. Weak lensing allows independent calibration of cluster mass systematics, and results in consistent scaling relations \citep[gray diamonds with errorbars;][]{nagarajan19}. Further, the $R_{\rm 500c}$ scale for the majority of cluster samples is not well resolved by the large Planck beam, such that the extraction of $Y_{\rm 500}$ is complex. Alternatively, comparisons can be made to the Planck data integrating out to a distance of $5 R_{\rm 500c}$ \citep[e.g.][]{henden18}, but at such scales the total tSZ effect begins to probe larger scales, 2-halo gas clustering, and projected contributions more than the cluster gas content itself \citep{hadzhiyska23}. In all cases, the resulting two-dimensional projected value must then be converted to $Y_{\rm 500}^{\rm 3d}$, which requires additional physical assumptions.

\begin{figure}
\centering
\includegraphics[angle=0,width=3.4in]{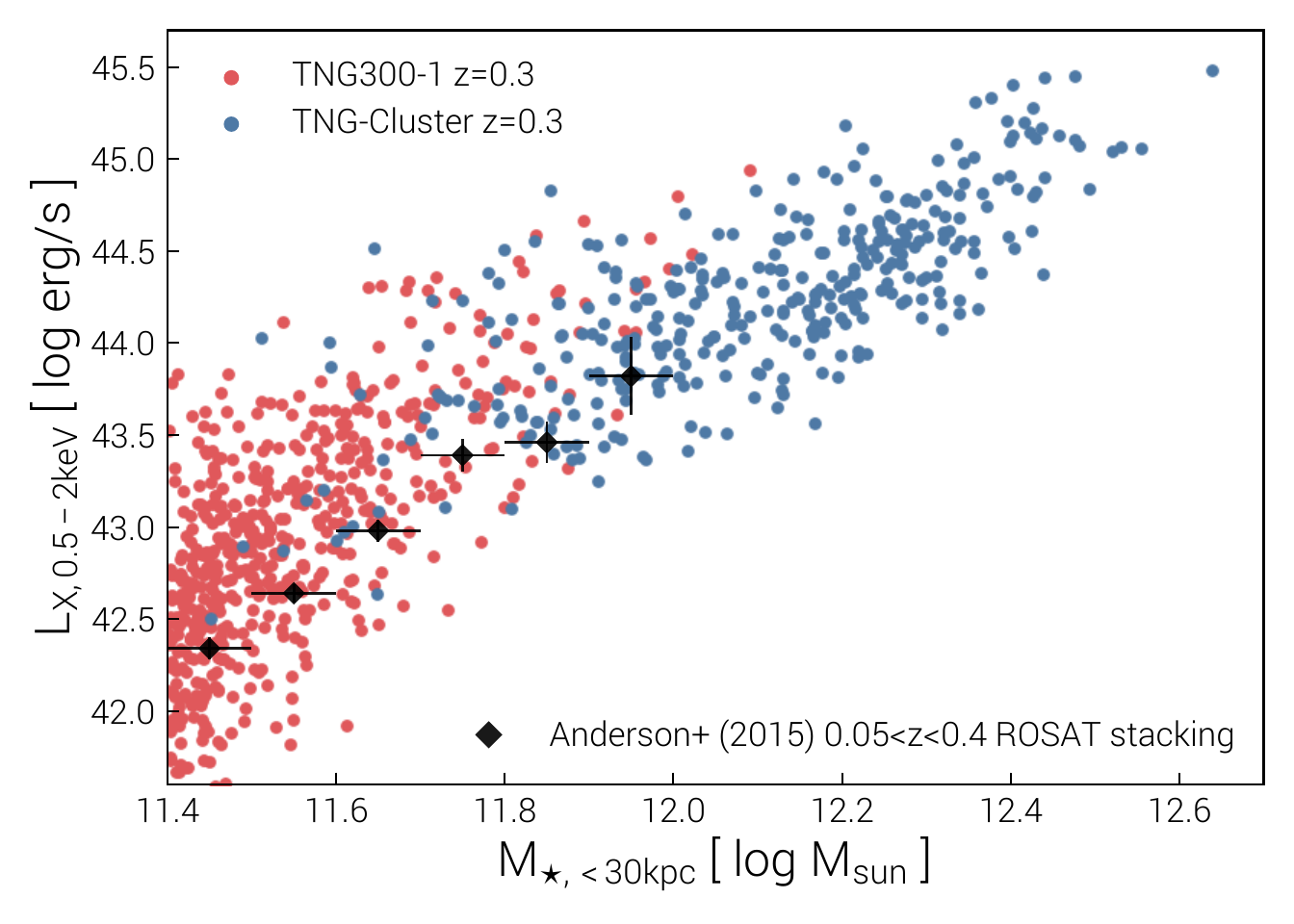}
\caption{ X-ray scaling relation vs galaxy stellar mass. We show TNG300 (red) and TNG-Cluster (blue) both at $z=0.3$, which is the average redshift of the observational sample of \citet{anderson15}, based on ROSAT stacking around locally brightest galaxies from SDSS.
 \label{fig_xray}}
\end{figure}

\begin{figure*}
\centering
\includegraphics[angle=0,width=6.0in]{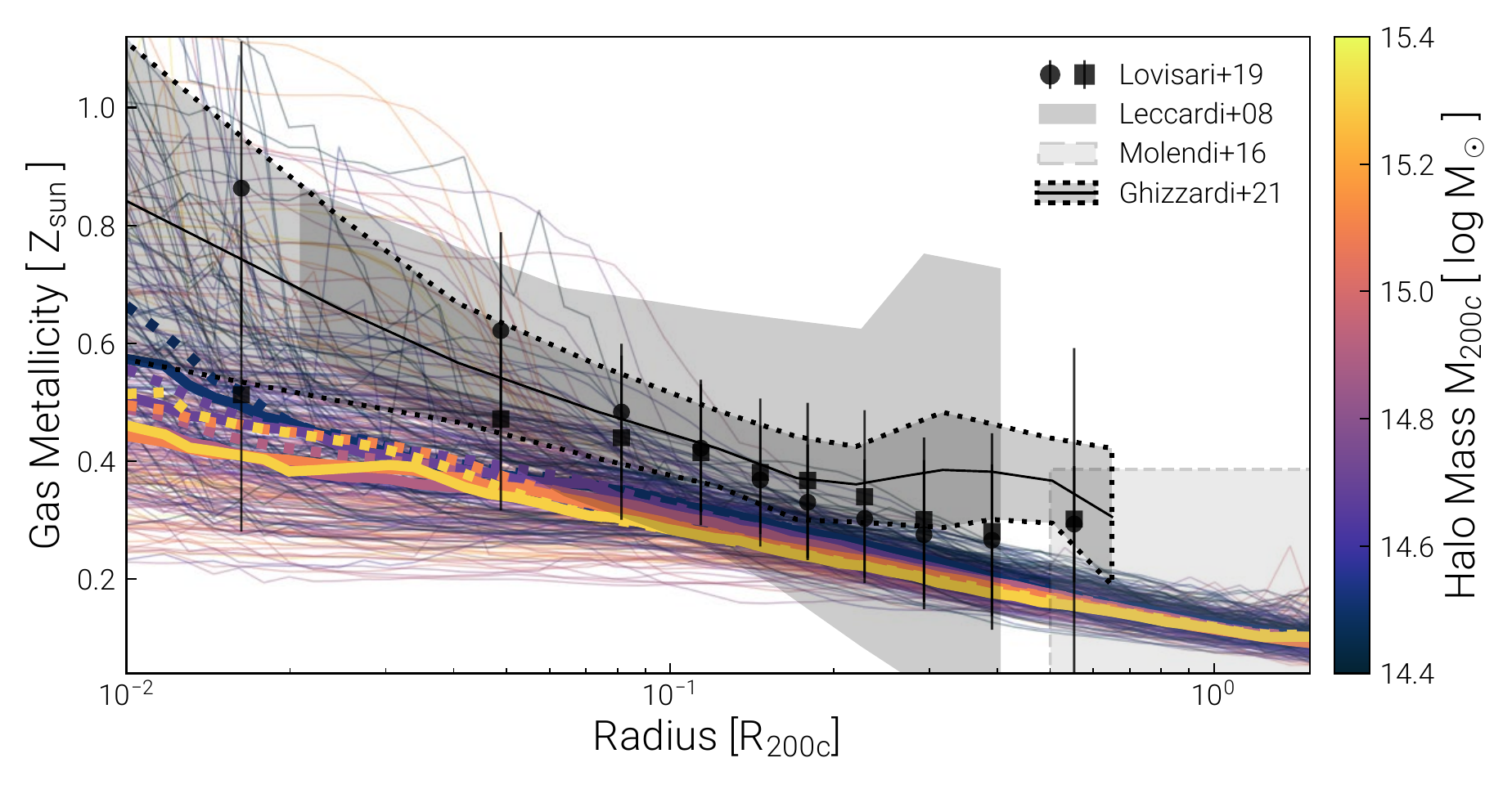}
\includegraphics[angle=0,width=3.4in]{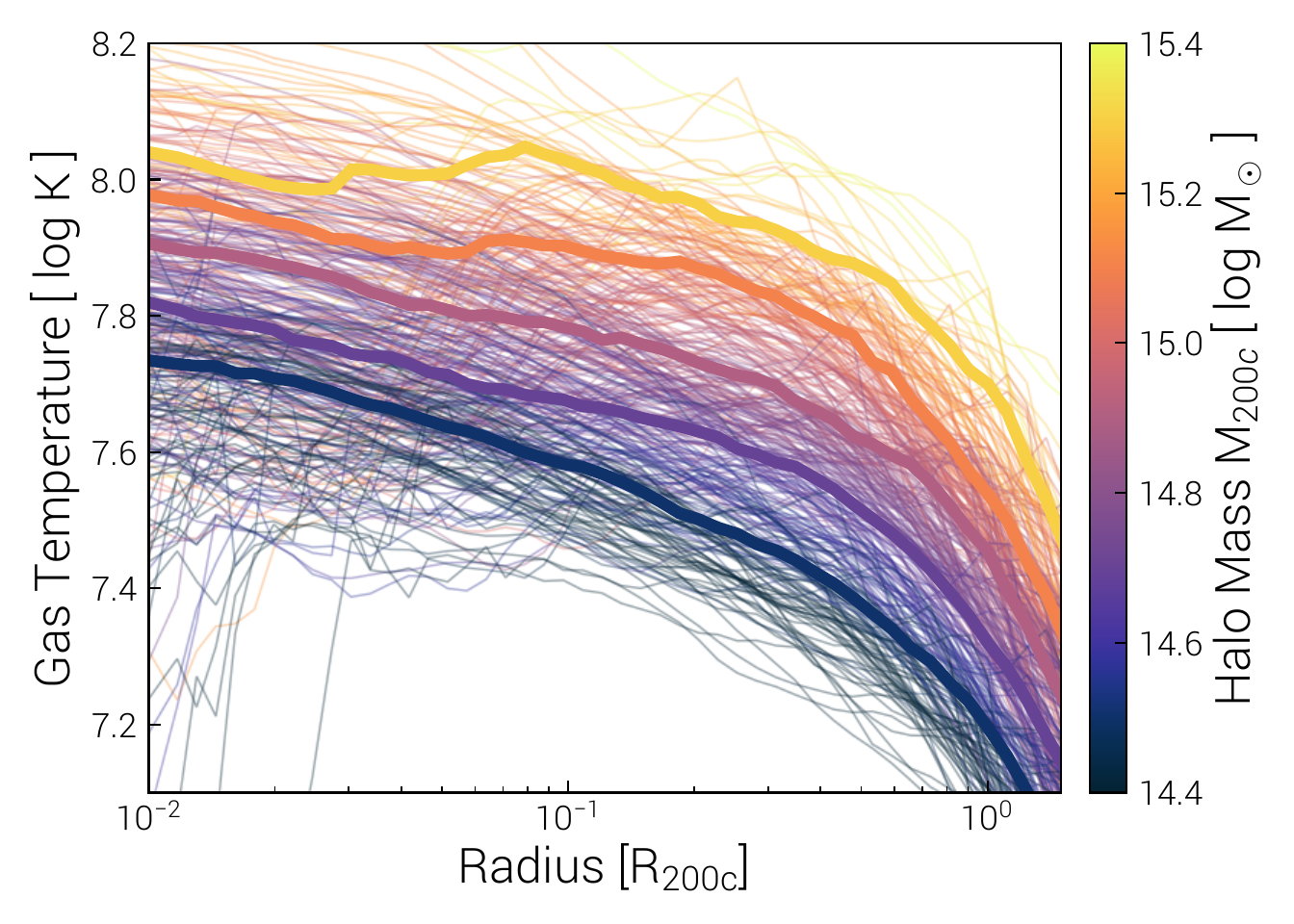}
\includegraphics[angle=0,width=3.4in]{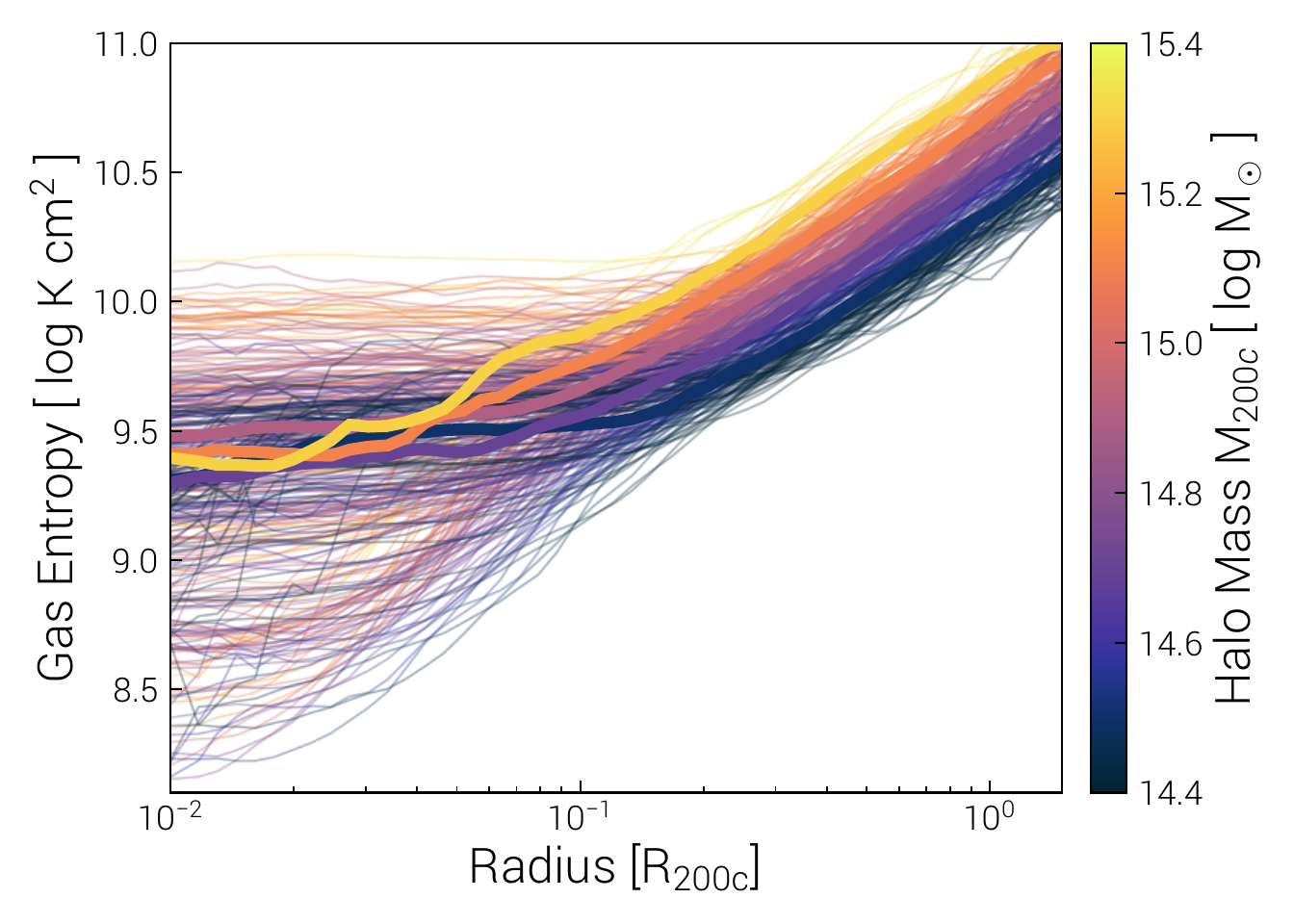}
\caption{ Radial profiles of ICM metallicity (top), temperature (lower left), and entropy (lower right) of the complete TNG-Cluster sample at $z=0$. In all three panels thin lines show the 352 individual clusters, colored by halo mass, while the five thick lines show median provides in $0.2$\,dex halo mass bins. For metallicity, we also include mean stacked profiles as thick dotted lines. Profiles include all gas and are mass-weighted. We compare the metallicity profiles to observational inferences for low-redshift clusters \citep[][see text]{leccardi08,molendi16b,lovisari19,ghizzardi21}.
 \label{fig_profiles}}
\end{figure*}

As a result, we present the comparison of the $Y_{\rm 500}-M_{\rm 500c}$ scaling relation at face value only, and defer consideration of the complex systematics and observational mock details of relevance. In this context, TNG-Cluster is consistent with available observational constraints. Such a coarse-grained agreement in the integral SZ signal is a common success of most recent cluster simulation projects \citep{hahn17,barnes17b,henden18,cui18,pop22}. Due to its weaker dependence on gas density, it is clearly a less sensitive probe of baryonic and feedback physics than X-ray emission.

Figure \ref{fig_xray_sz} (bottom panel) shows the scaling relation between X-ray luminosity and halo mass at $z=0$. We measure the soft-band ($0.5-2$\,keV) integrated luminosity within $R_{\rm 500c}$, as before in 3D (see Section~\ref{sec_methods3}). Total halo X-ray luminosity increases strongly with mass, rising from $L_{\rm X,r500c} \sim 10^{43.5}$\,erg\,s$^{-1}$ for $M_{\rm 500c} \simeq 10^{14}$\msun to $L_{\rm X,r500c} \sim 10^{45.0}$\,erg\,s$^{-1}$ at $M_{\rm 500c} \simeq 10^{15}$\msun. We compare to samples of X-ray observed clusters at low redshift: the 31 X-ray luminosity limited clusters at $z < 0.2$ from the Representative XMM-Newton Cluster Structure Survey \citep[REXCESS;][]{pratt09}, the 36 X-ray flux selected clusters, both the low and high redshift samples, observed with Chandra \citep{vikhlinin09}, the diverse sample of 224 clusters, an X-ray flux limited sample, with Chandra and ROSAT data \citep{mantz16},\footnote{Note that the \citet{mantz16} sample, with luminosities reported in the wider $0.1-2.4$\,keV band, is converted using the mean ratio of $L_{\rm X,r500}^{\rm 0.1-2.4 keV} / L_{\rm X,r500}^{\rm 0.5-2.0 keV} = 1.66$ across the TNG-Cluster sample at $z=0$.} the 59 SPT-SZ selected clusters at $0.2 < z < 1.5$ with XMM-Newton followup \citep{bulbul19}, and the 120 Planck-ESZ selected clusters at $z < 0.55$ also with XMM-Newton data \citep{lovisari20}. We always show and compare against `core-included' X-ray luminosities, without excising the core regions (commonly $< 0.15 R_{\rm 500c}$), which gives rise to larger scatter than core-excised values \citep{pratt09,mantz18}.

Overall, the $L_{X,500c} - M_{\rm 500c}$ scaling relation of TNG-Cluster is broadly consistent with these data sets. Scatter at fixed halo mass is significant at $\sim 1$\,dex, both in observations and in the simulations. The physical origin of this scatter must be understood for cosmological applications of mass-observable relations. While it may be driven primarily by mergers and clusters in different states of relaxedness \citep{randall02}, it may instead be dominated by halo concentration \citep{yang09} or differences in the central thermodynamical properties \citep{ohara06}.

\begin{figure*}
\centering
\includegraphics[angle=0,width=3.4in]{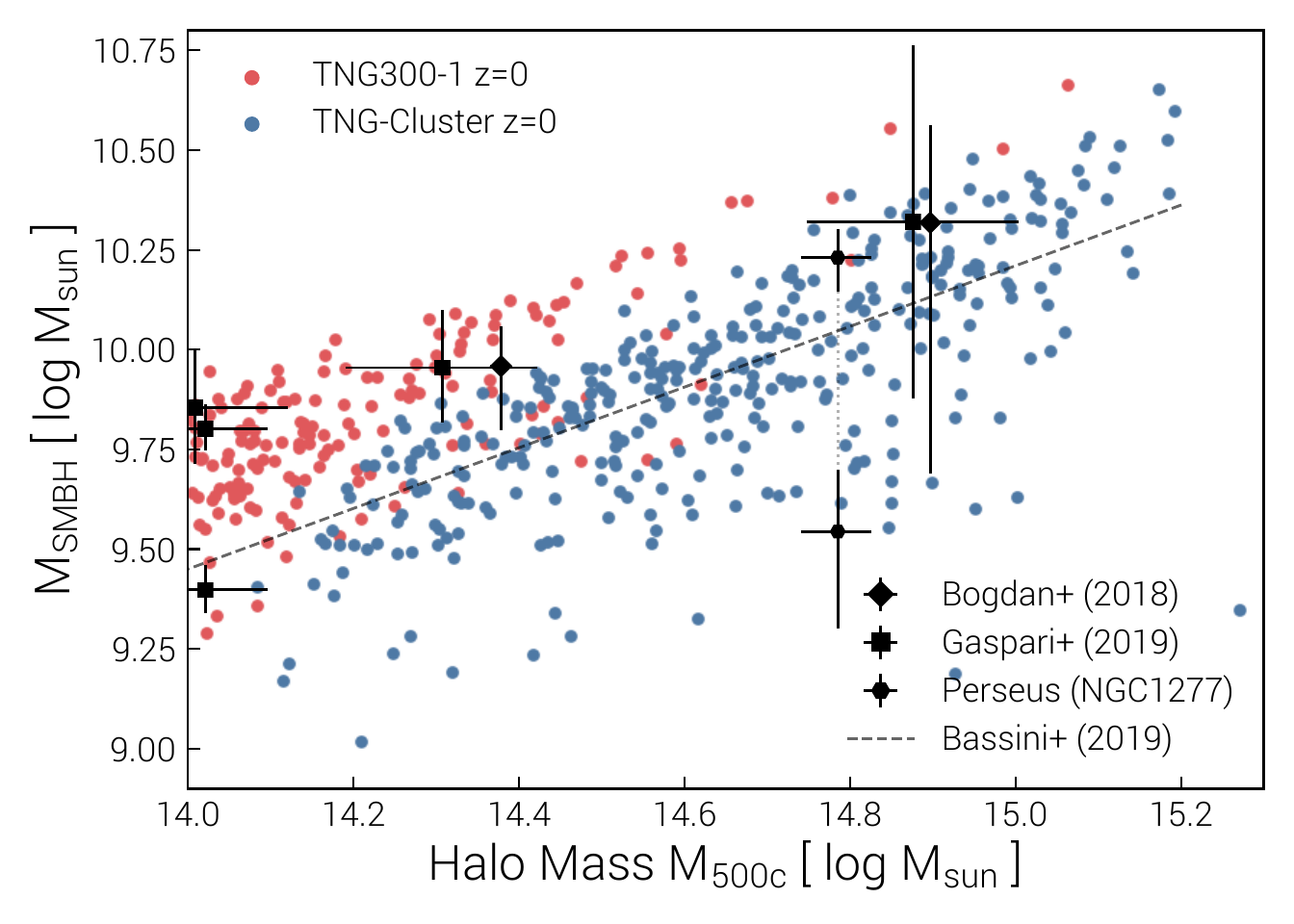}
\includegraphics[angle=0,width=3.4in]{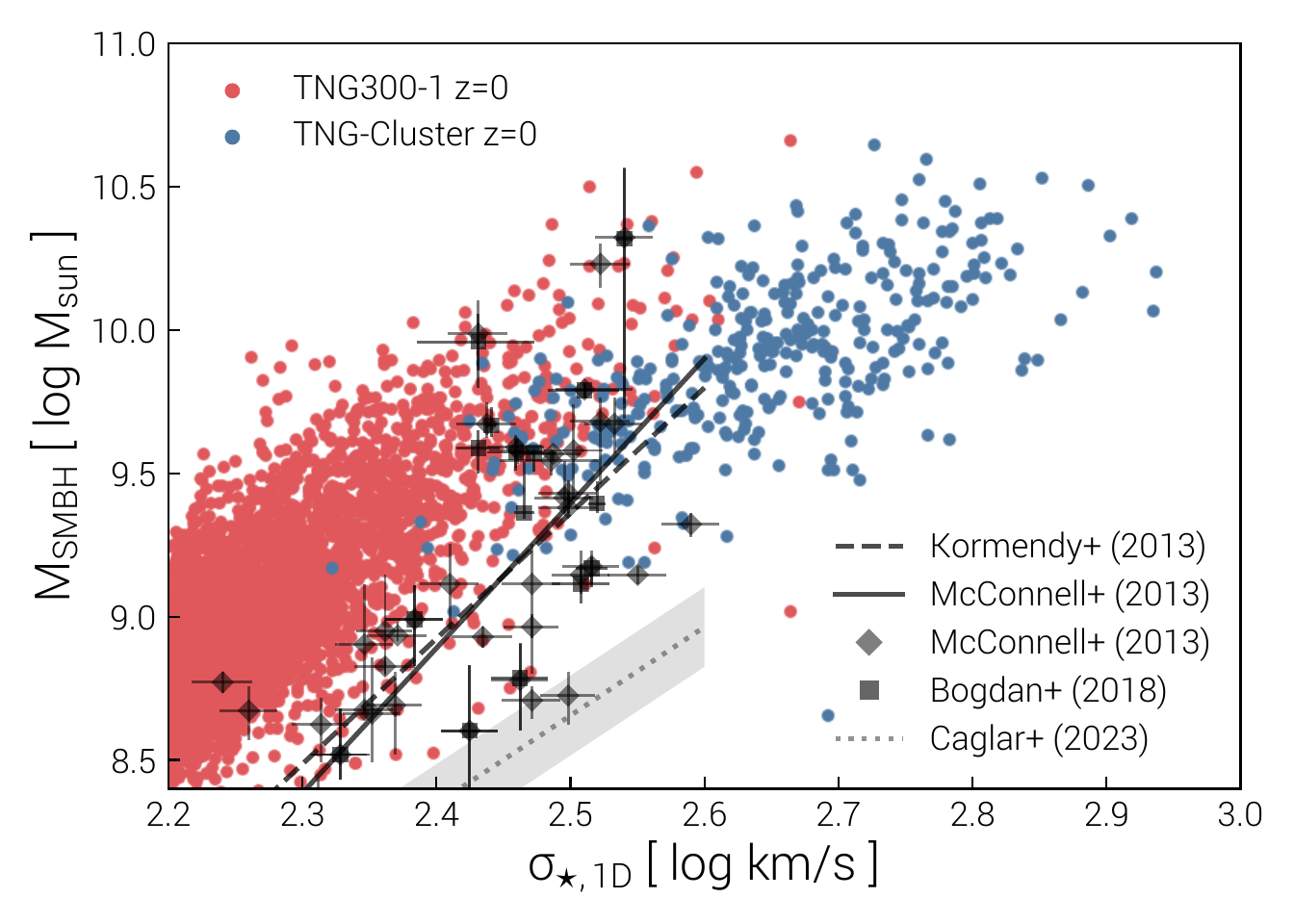}
\caption{ SMBH mass scaling relations, as a function of total halo mass (left) and stellar velocity dispersion (right). We adopt the most massive SMBH of each BCG, if several are present, and measure $\sigma_\star$ based on the 1D line-of-sight stellar kinematics within an aperture of 10 pkpc (see text). We compare to observational scaling relations \protect\citep{kormendy13,mcconnell13}, as well as measurements for group and cluster BCGs \protect\citep{bogdan18,gaspari19}. Note that the scaling relations are largely extrapolations in the TNG-Cluster regime. We also include two inferences for NGC1277 of Perseus (see text), and compare to the simulation results of \protect\citet{bassini19}. The most massive SMBHs in TNG-Cluster reach $4-5 \times 10^{10}$\msun, although SMBH growth slows down in the cluster regime, somewhat flattening the scaling relations.
 \label{fig_smbhs}}
\end{figure*}

Figure \ref{fig_xray} shows one further view of the scaling relation between X-ray luminosity and mass: $L_{\rm X}$ versus galaxy stellar mass. We compare to observational results which stack ROSAT data on the locations of $\sim 250,000$ locally brightest galaxies from SDSS \citep{anderson15}. The average redshift of this sample is $z \sim 0.3$, where we also show the simulation results. Unfortunately the volume probed by the SDSS main galaxy sample is too small to extend to the most massive galaxies, so this data only overlaps with the low-mass end of TNG-Cluster. Nonetheless, the comparison suggests that halo-scale X-ray emission as function of galaxy mass is in reasonable agreement with data, if perhaps somewhat high. Future stacking experiments with eROSITA beyond the currently available eFEDS field \citep{comparat22,chadayammuri22} will significantly expand such results.

\subsubsection{Radial profiles of ICM metallicity} 

Existing X-ray observations of clusters enable spatially resolved inferences of key physical properties of the hot ICM. Figure \ref{fig_profiles} shows radial profiles of three key properties: metallicity (top), temperature (lower left), and entropy (lower right). These are intrinsic, mass-weighted quantities from the simulations, and do not mock the details of any observational measurements. Thin lines show all 352 individual halos from TNG-Cluster at $z=0$, while the five thick lines show median stacked profiles in $0.2$\,dex bins of halo mass. The radial and mass trends are diverse.

Metallicity profiles are relatively flat and show only a small (inverse) dependence on halo mass. Our simulated clusters reach $Z \sim 0.5\, \rm{Z}_\odot$ in their centers, decreasing slowly to $\sim 0.3\, \rm{Z}_\odot$ by $0.1\,R_{\rm 200c}$, and to $\sim 0.1\, \rm{Z}_\odot$ by $R_{\rm 200c}$. However, there is significant halo to halo diversity, and individual clusters can have overall metallicity profiles a factor of two above or below the average \citep[see also][for previous analysis of cluster metallicities in TNG300]{vog18a}.

We compare to a number of observational constraints on the radial gas-phase metallicity profiles of clusters: the sample of 207 groups and clusters at $z < 0.1$ with XMM-Newton coverage \citep{lovisari19},\footnote{We re-scale all normalizations from each assumed solar metallicity value to the TNG value of $Z_\odot = 0.0127$. We also re-scale from $R_{\rm 500c}$ to $R_{\rm 200c}$ values where necessary, using the average ratio of $R_{\rm 500c} / R_{\rm 200c} = 0.65$ for the TNG-Cluster sample at $z=0$.} the mixture of 50 cool-core and non-cool-core clusters at $0.1 < z < 0.3$ also with X-ray data from XMM-Newton \citep{leccardi08}, the outskirts constraint from \citet{molendi16b} which is an extrapolated, rough upper limit, and the X-COP sample \citep{eckert17} of 13 $z < 0.1$ high-mass clusters with $M_{\rm 500} > 3 \times 10^{14}$\msun and large field of view coverage \citep{ghizzardi21}.

Observations broadly infer that cluster metallicity profiles can be high in cluster cores, reaching values \mbox{$\sim 0.5 - 0.8\, \rm{Z}_\odot$} (especially for cool-core systems), but decreasing to $\sim 0.2 - 0.3\, \rm{Z}_\odot$ for $R \gtrsim 0.2 R_{\rm 500c}$, beyond which radial metallicity profiles are relatively flat. This may reflect feedback-driven metal redistribution from the central BCG itself \citep{gitti12}, that most metals are already produced in the cores of cluster progenitors at high redshift \citep{pearce21}, or the collective effect of environmental pre-enrichment in an overdense region at earlier times \citep{vog18a,biffi18}.

Overall, TNG-Cluster is in reasonable agreement with observed ICM metallicity profiles, particularly in the radial trend. The simulated profiles are plausibly too low, by $\sim 0.1 \rm{Z}_\odot$, regardless of distance. However, without a detailed forward modeling effort, we leave more quantitative comparisons for future work. Unfortunately, the systematics involved in observational estimates of ICM metallicity from X-ray spectra are significant, making comparisons beyond the factor of $\sim 2$ level difficult \citep{molendi16b}.\footnote{Although we show TNG-Cluster profiles at $z=0$, the typical redshift of observed clusters is higher. We note that the simulated profiles shift down by $\sim 0.05\, \rm{Z}_\odot$ at $z=0.2$ (not shown). X-ray emission weighting, as opposed to mass weighting, has a negligible impact. Profiles in 2D projection instead of 3D are also lower by $\sim 0.05\, \rm{Z}_\odot$.} Future X-ray instruments with high spectral resolution will improve our observational census of cluster metals significantly \citep{mernier23}.

\subsubsection{Radial profiles of ICM temperature and entropy} 

Figure \ref{fig_profiles} also shows temperature (lower left) and entropy (lower right) profiles of TNG-Cluster at $z=0$. Most temperature profiles rise in overall normalization with the halo virial temperature, and monotonically decrease with increasing radius. However, a subset of clusters show central temperature depressions indicative of strong cool cores \citep[see the companion paper by][]{lehle24}. There is similar diversity in cluster entropy profiles. While many halos have high central entropy ($K_0$), indicative of non-cool-core (NCC) clusters, others have $K_0$ values an order of magnitude lower, highlighting cool-core (CC) clusters. Although $K_0$, as well as central cooling time and density, all increase with halo mass, this CC vs NCC heterogeneity of the cluster population occurs even at fixed halo mass \citep[see][]{lehle24}.

The kinematics of the ICM, and comparisons to observations, are presented in the TNG-Cluster companion papers by \citet{truong24} and \citet{ayromlou24}.


\section{Properties of the Galaxies} \label{sec_stars}

We now turn our attention to the properties of galaxies residing within TNG-Cluster halos. First, in terms of their SMBHs, cool gas content and star formation activity, their stellar masses, and finally to the satellite galaxy populations they host.

\subsection{Supermassive black holes}

Figure \ref{fig_smbhs} shows the scaling relation between central SMBH mass and total halo mass \citep[left;][]{ferrarese02} as well as central stellar velocity dispersion \citep[right;][]{magorrian98}. We include only the central i.e. BCGs of each cluster, take the most massive SMBH if there are several present, and measure the one-dimensional, mass-weighted $\sigma_{\rm \star, 1D}$ along a random line-of-sight direction, within an aperture of 10 kpc, to capture the central dynamics of BCGs.

We compare to a number of observationally inferred relations \citep{kormendy13,mcconnell13,caglar23}, as well as to observational samples of group and cluster BCGs \citep{bogdan18,gaspari19}. In addition, we indicate the SMBH of the Perseus cluster (NGC1277) with $M_{\rm SMBH} = 2-20 \times 10^9$\msun \citep[encompassing $1\sigma$ estimates from][]{vandenbosch12,emsellem13} and $M_{\rm 500c} = 6.1 \pm 0.6 \times 10^{14}$\msun \citep{giacintucci19}. Finally, we include the scaling relation found in the Dianoga cluster zoom simulations for comparison \citep{bassini19}.

Comparison of the trends of SMBH masses with halo and galaxy properties in TNG-Cluster versus data is not straight-forward. The values of $M_{\rm SMBH}$ as a function of halo mass are broadly consistent with the small samples available, although any statistical comparison is impossible. With respect to the $M_{\rm SMBH}-\sigma_{\star}$ relation, we emphasize that observational scaling relations are almost entirely extrapolations in this regime, as direct measurements of SMBH scaling relations for massive BCGs with $M_{\rm 500c} \gtrsim 10^{14}$\msun, or $\sigma_\star \gtrsim 400$\,km\,s$^{-1}$, are exceedingly rare. As a result, small differences in slope lead to large differences in anticipated SMBH mass for such massive BCGs. In addition, the values of $\sigma_\star$ extracted from the simulations depend on, and increase with, aperture \citep{sohn22}. If we instead measure $\sigma_\star$ within the stellar half mass radius (i.e. the effective radius), this would shift values slightly to the right. However, we emphasize that our measurement of stellar velocity dispersion is not forward modeled and differs in numerous respects from methods used to infer $\sigma_\star$ from data, including the use of effective velocity dispersion \citep{sijacki15}, preventing any quantitative comparisons for the time being.

The $M_{\rm SMBH}-\sigma_{\star}$ relation, as well as the $M_{\rm SMBH} - M_{\star}$ relation (not shown) both become shallower and level off in the cluster regime. This is partly due to lower SMBH masses in TNG-Cluster versus TNG300, which we discuss below in the context of stellar masses. This trend may also reflect a shift in the physics of SMBH growth: it coincides with the mass scale where halos retain their full baryon fractions \citep{ayromlou23}, and thus the energetics of SMBH feedback are no longer sufficient to significantly impact the baryon distribution and thermodynamical properties of the ICM \citep{voit23}. Indeed, the slope of the underlying $M_{\rm SMBH}-M_{\rm halo}$ relation \citep{marasco21} evolves from super-linear to sub-linear at high masses, while the scatter decreases, reflecting a natural outcome of the hierarchical assembly process \citep{truong21}.

\begin{figure*}
\centering
\includegraphics[angle=0,width=3.4in]{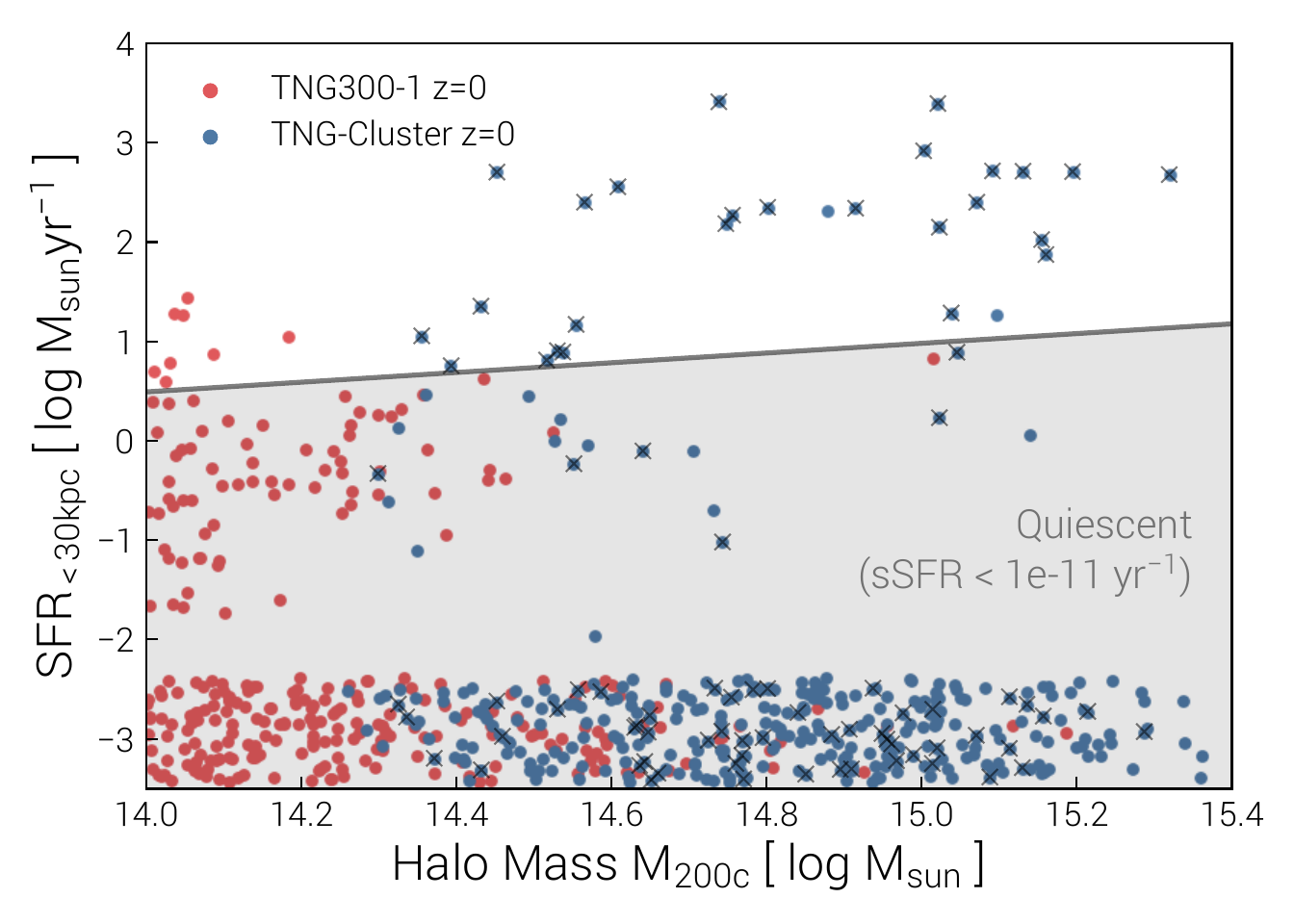}
\includegraphics[angle=0,width=3.4in]{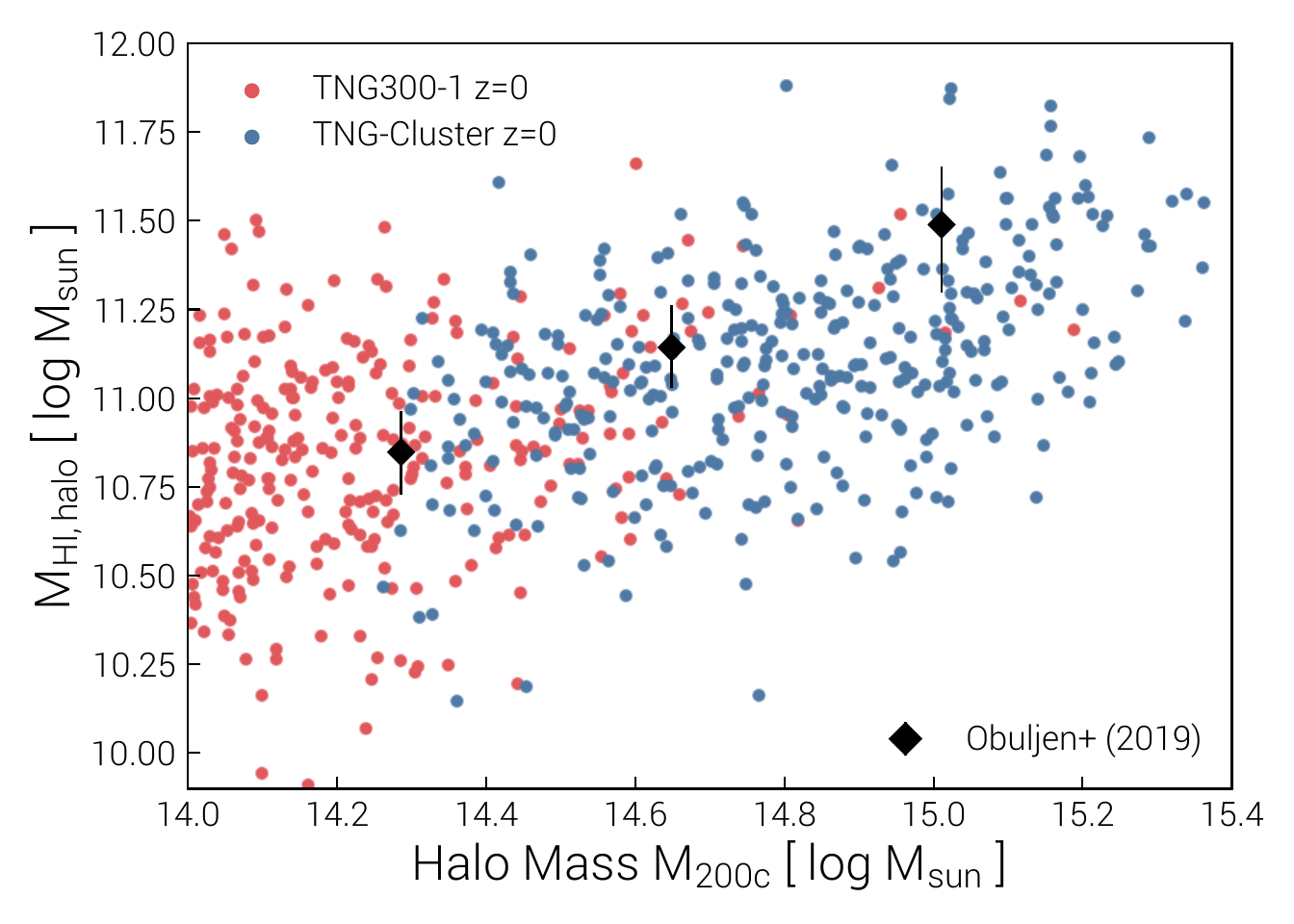}
\caption{ Star formation activity and cold gas contents. \textbf{Left:} Galaxy star formation rate as a function of halo mass, at $z=0$. A quiescent threshold of $\rm{sSFR} < 10^{-11} \rm{yr}^{-1}$ is indicated as the gray triangular region, adopting the stellar mass to halo mass relation of TNG-Cluster. Black crosses mark those clusters identified as strong cool-cores with the fiducial criterion of \protect\citep{lehle24}. The vast majority of central galaxies in clusters are quenched, and only a small number have non-negligible levels of residual star formation. \textbf{Right:} Total cold (HI) mass in the halo as a function of mass, compared to observations stacking ALFALFA data on SDSS groups and clusters \citep{obuljen19}.
 \label{fig_coolgas}}
\end{figure*}
\subsection{Cool Gas and Star Formation Activity}

Figure \ref{fig_coolgas} shows the distribution of galactic star formation rates (SFRs, left panel) as a function of halo mass at $z=0$. We include only the central galaxy (BCG) of each cluster, and show instantaneous SFRs. Values of absolute zero in the simulation result from our finite resolution \citep{donnari19}, and for TNG-Cluster these correspond physically to upper limits of $< 10^{-3} \,\rm{M}_\odot {\rm yr}^{-1}$. These systems are placed at this value, with $\pm 0.5$\,dex scatter added for visual clarity.

The vast majority of central cluster galaxies in TNG-Cluster are quenched. The gray shaded triangular region (left panel) indicates the canonical $z=0$ quiescent threshold of $\rm{sSFR} < 10^{-11} \rm{yr}^{-1}$, and only $\sim 8\%$ of TNG-Cluster BCGs are star-forming according to this definition. Adding in the galaxies with residual (non-zero) SFR, the TNG-Cluster fraction increases to $\sim 13\%$. The vast majority of star-forming BCGs reside in cool-core clusters, as marked by small black crosses in the panel \citep[see the companion paper by][]{lehle24}. There is no strong halo mass trend in star formation activity \citep[as seen in][]{orellanagonzalez22}, although observations infer an increasing star-forming fraction with cluster richness \citep{liu12}. The overall star-forming fraction from TNG-Cluster is similar to fraction of star-forming cluster BCGs in \citet{mcdonald15} based on a sample of 90 SPT selected clusters, where a strong trend with redshift towards mostly star-forming BCGs at $z \gtrsim 1$ is evident as well \citep[see also][]{bonaventura17}.

The origin of the fuel for this `residual' star-formation is unclear. The relatively small fraction of star-forming BCGs suggests a short duty-cycle of high SFR activity, which could be driven either by episodic -- possibly `hidden' -- cooling flows from the hot ICM \citep{molendi16,fabian23} that are subsequently terminated by AGN feedback \citep{voit15,calzadilla22}, or by cosmologically rare gas-rich mergers \citep{webb15}. In either case, this suggests that in-situ star formation, and not solely ex-situ dry mergers, contribute to cluster BCG formation: whereas this is expected particularly at high redshift and in the protocluster regime \citep{montenegrotaborda23}, our findings show that, in a fraction of BCGs, this is the case also at the current epoch.

Figure \ref{fig_coolgas} also shows the total mass of neutral HI in the halo, as a function of halo mass at $z=0$ (right panel). Although the fraction of neutral HI relative to the total halo gas mass is small, $M_{\rm HI} / M_{\rm gas} \sim 10^{-4} - 5 \times 10^{-3}$, in absolute terms there is an enormous amount of such cool $\sim 10^4$\,K gas in clusters. We therefore compare to observational results that stack large-beam ALFALFA survey data at the locations of $z \sim 0$ SDSS groups and clusters \citep{obuljen19}. We note that this comparison is intended at face-value only, as we measure 3D total halo masses for the simulations, versus 2D projected values over large beam sizes from ALFALFA. For any non-zero line-of-sight contribution, the former is a lower limit on the latter. Regardless, the mass of HI gas in TNG-Cluster halos is in the ballpark of low-redshift observational inferences \citep[see also][]{zhangh20}.

Observations suggest that significant amounts of molecular gas are common although not ubiquitous in star-forming BCGs \citep{castignani20,dunne21}. Unsurprisingly, large amounts of cool gas correlate with central cooling flows as well as starbursts \citep{edge01}. Cool gas is also prevalent throughout clusters, as mapped in detail for local examples including Virgo with H$\alpha$ narrow-band imaging \citep[VESTIGE;][]{boselli18a} and Fornax with MeerKAT \citep{serra16}. High-mass halos hosting rich satellite populations in the TNG simulations similarly have large amounts of cool gas with complex morphologies and frequent ram-pressure stripping origins \citep{yun19,nelson20,rohr23,zinger23}. The contributions of satellites to the cluster ICM in TNG-Cluster is the subject of a companion paper \citep{rohr24}.

\begin{figure*}
\centering
\includegraphics[angle=0,width=3.4in]{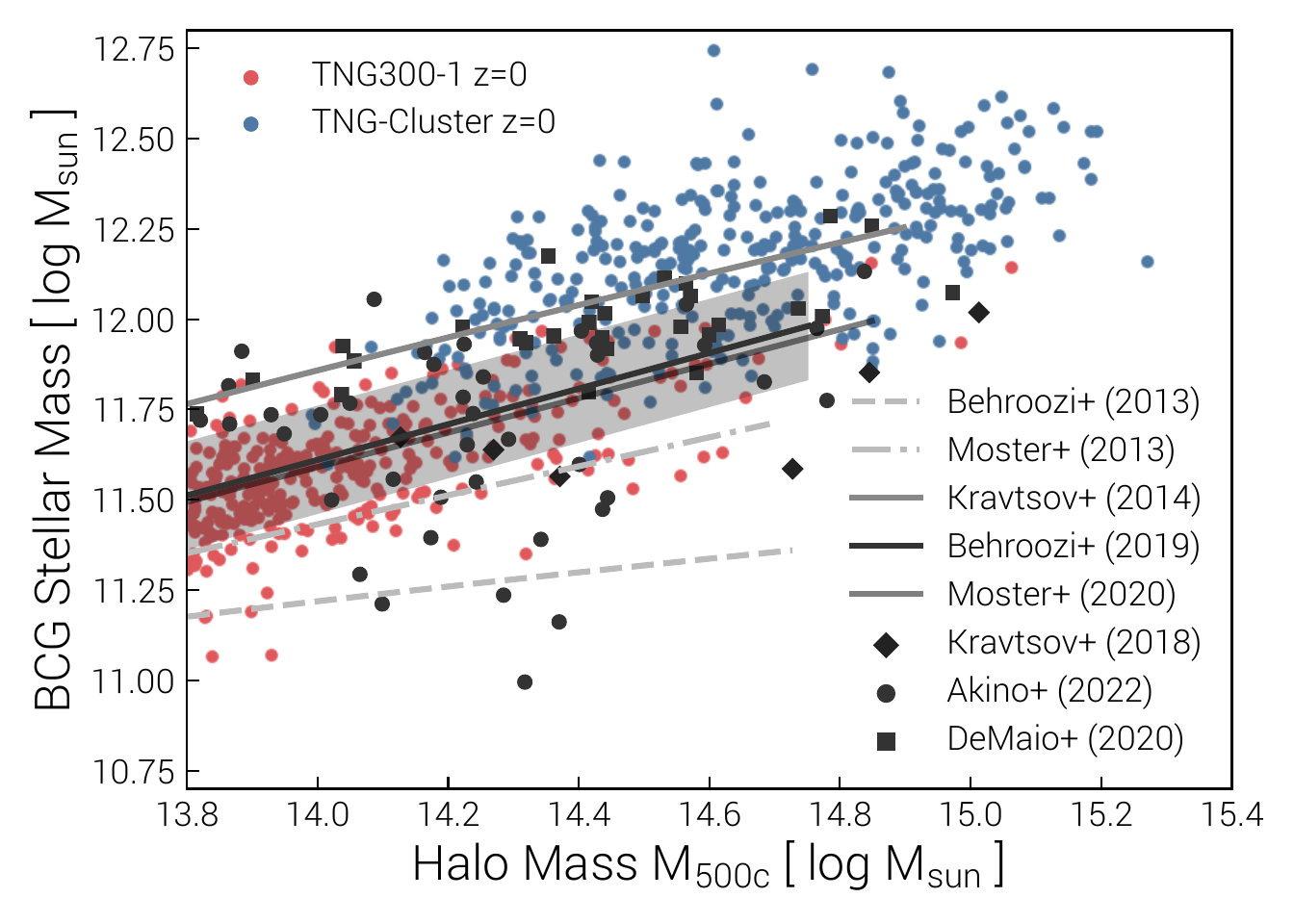}
\includegraphics[angle=0,width=3.4in]{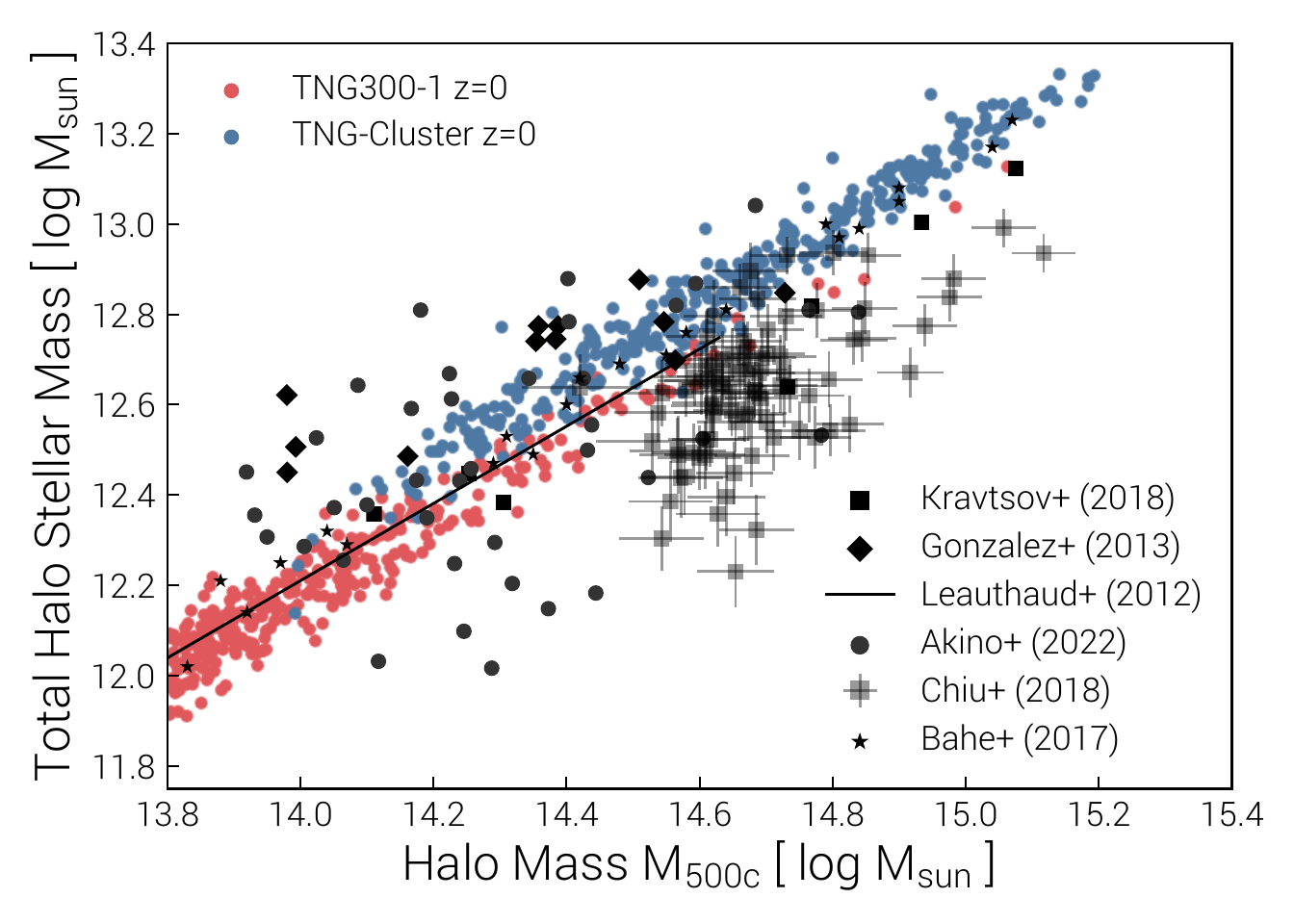}
\caption{ The stellar content of TNG-Cluster halos: stellar mass of the central galaxy i.e. BCG (left), and total stellar mass of the halo i.e. the central galaxy together with the extended ICL as well as satellites (right). For central galaxy stellar mass we sum stars within $r < 30$ pkpc, and compare to abundance matching results \citep{behroozi13,moster13,behroozi19,moster20} and observational inferences using similar definitions \protect\citep{kravtsov18,demaio20,akino22}. For total halo stellar mass we sum stars within $r < r_{\rm 500c}$, and also compare to data using similar definitions \protect\citep{leauthaud12,gonzalez13,kravtsov18,chiu18,akino22} as well as the Hydrangea simulations \protect\citep{bahe17b}.
 \label{fig_stellar_mass}}
\end{figure*}

\subsection{Stellar Content} 

We conclude this overview of TNG-Cluster by quantifying the stellar mass content of our simulated halos and their satellites \citep[see][for previous TNG analyses]{pillepich18b} Figure \ref{fig_stellar_mass} shows the stellar masses of TNG-Cluster central galaxies (left) as well as total halo stellar masses (right). We define the former using a 3D radial aperture of 30\,pkpc, and the latter as all stars within $R_{\rm 500c}$ (excluding satellites).

We compare galaxy stellar masses to a number of observational constraints (left panel), noting however that at such high masses BCG stellar mass measurements are becoming less statistical as well as more sensitive to measurement technique and observational details \citep{dsouza15}. We compare to semi-empirical inferences i.e. older abundance matching type results \citep{behroozi13,moster13} as well as their newer counterparts \citep{behroozi19,moster20}. The updated models infer significantly higher stellar masses at fixed halo mass. However, the semi-empirical models are all based on stellar mass functions derived from photometry which is non-ideal for massive galaxies and underestimates their total stellar masses considerably. The 30\,pkpc aperture is chosen to roughly replicate these measurements \citep{schaye15,pillepich18b}. We also compare to direct optical/infrared measurements of BCGs in 21 clusters at $z \lesssim 0.1$, integrating stellar light out to $\sim 100 - 300$\,kpc, with total halo masses derived from X-ray data \citep{kravtsov18}. In addition, we contrast against the X-ray selected HSC-XXL sample with weak lensing based halo mass estimates \citep{akino22}, as well as the low and intermediate redshift BCG sample of \citet{demaio20}, including some CLASH targets \citep[see also][]{lin12}.

Broadly speaking, observational inferences of BCG stellar masses are broadly consistent with the low-mass end of the cluster distribution. This is the case in particular where our sample is dominated by TNG300. The situation is less clear for the most massive clusters with $M_{\rm 500c} \gtrsim 10^{14.5}$\msun where observations become rare. Nonetheless, the TNG-Cluster sample has BCG stellar masses that are plausibly too large, by $\sim 0.1-0.2$\,dex in the mean. If we compare data to the simulated clusters at $z = 0.3$ instead of $z=0$ (not shown), the typical median redshift of observed cluster samples, the situation is improved but not entirely so. This could suggest that the SMBH feedback model in TNG, which has never been evaluated in such massive halos, does not suppress star formation in these BCGs. More likely, since the stellar mass content of such massive galaxies are $> 90$\% ex-situ in origin \citep{rodriguezgomez15, pillepich18b}, it may indicate too much stellar mass growth during earlier i.e. proto-cluster phases. Comparison of resolved stellar mass profiles will shed light on this \citep{ardila21}, although targeted imaging campaigns are needed to build a sample of the highest mass clusters.

We also compare total halo stellar masses to a number of observational constraints (right panel). In particular, the same samples as previously \citep{gonzalez13,kravtsov18,akino22}, as well as a halo occupation distribution analysis of the COSMOS data at slightly higher redshift $z \sim 0.4$ \citep{leauthaud12}. The observational points again scatter about the combined simulation results, albeit with significantly larger scatter. We also compare to the SZ-selected sample of 91 clusters with $0.2 < z < 1.25$ with photometric estimates on combined BCG and satellite stellar masses \citep{chiu18}. This data falls $\sim 0.2$\,dex below the simulations, but does not detect nor include the ICL component, and has a notably higher mean redshift. Constraints on the total halo stellar mass at $M_{\rm 500c} \gtrsim 10^{15}$\msun are unfortunately exceedingly sparse, and it is hard to draw any statistical comparisons.

Both panels of Figure \ref{fig_stellar_mass} also hint towards a steepening and/or normalization offset of the stellar mass to halo mass relations between TNG300 and TNG-Cluster. Given that they employ the same physical galaxy formation model, at the same resolution, this is unexpected. The recent MTNG simulation found a similar effect in their large hydrodynamical box \citep[MTNG740;][]{pakmor23}, namely that stellar masses were more similar to TNG100 than TNG300, at odds with our analysis of stellar mass convergence with resolution in the TNG model \citep{pillepich18b}. Instead, galaxy stellar masses are consistent with a higher, TNG100-like resolution. \citet{pakmor23} attribute this to the exclusion of magnetic fields in MTNG, and the interaction of magnetic fields with the kinetic SMBH feedback mode of the TNG model. However, since we include magnetic fields in TNG-Cluster, we can rule out this possibility.

In TNG-Cluster we find that the `steeper' star-formation timescale change, originally developed and applied in TNG50 \citep[see Section 2.2 of][]{nelson19a}, is the clearest cause. While we verified that this change made no difference to galaxy properties for TNG50, this was only possible to study for $M_{\rm 200c} < 10^{14}$\msun halos. In massive clusters, it appears that the acceleration of star-formation for the densest gas leads to a more rapid conversion of nuclear gas into stars. As a result, central gas densities are slightly lower, which causes lower accretion rates onto SMBHs, and lower overall kinetic-mode energy output. In turn, SMBH masses are reduced, and stellar masses increased, with respect to a TNG300 resolution simulation. In net effect, this means that many of the physical properties of TNG-Cluster galaxies are in fact more similar to a TNG100 resolution simulation, which is largely advantageous as this is the resolution that the model was calibrated at \citep{pillepich18a}. The magnitude of this effect is $\sim 30-40$\% in mass, i.e. the same as the change in stellar masses from TNG100 to TNG300 resolution, and we do not expect it to significantly affect any conclusions or analyses of TNG-Cluster.

\subsection{Satellites}

\begin{figure}
\centering
\includegraphics[angle=0,width=3.4in]{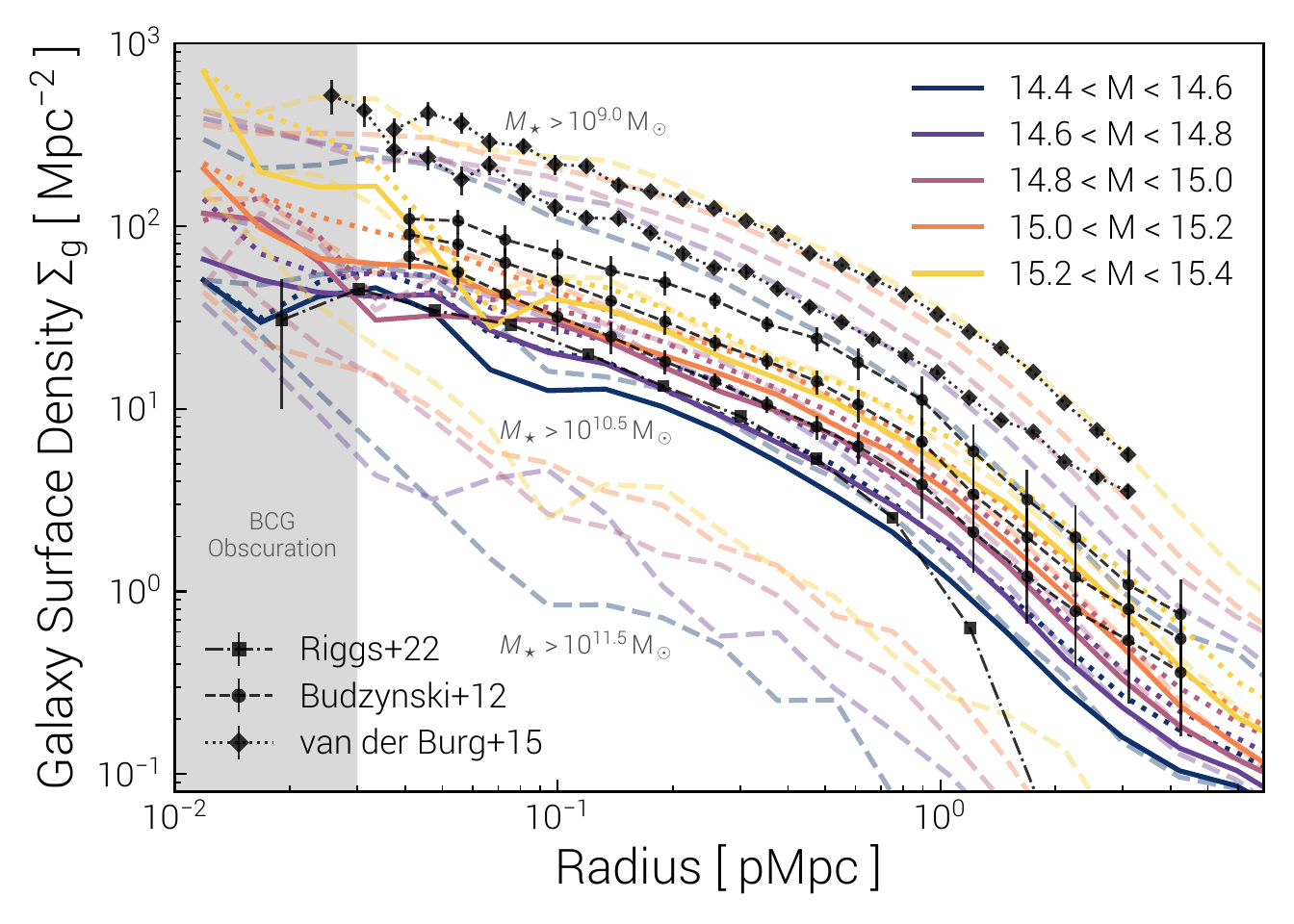}
\caption{ The radial distribution of nearby (satellite) galaxies, from the centers of TNG-Cluster halos at $z=0$. We show the surface number density, i.e. number of galaxies per $\rm{Mpc}^2$, as a function of distance. We stack around clusters in five halo mass bins, as indicated in the legend for $M_{\rm 200c}$ in $\log \rm{M}_\odot$. We include galaxies with $M_{\rm r} < -20.5\,$mag (thick solid lines including dust modeling, dotted lines without; see text) in order to match observational constraints from SDSS \protect\citep[][black circles and dashed lines;]{budzynski12}. We also compare to data from GAMA \protect\citep[black squares and dot-dash line;][]{riggs22}. In addition we include stellar mass based constraints with three additional sets of dashed colored lines: $M_\star > \{10^9, 10^{10.5}, 10^{11.5}\}$\msun. The first i.e. highest set of colored curves are comparable to observational constraints from \citet[][see text]{vanderburg15}.
 \label{fig_sat_profiles}}
\end{figure}

Returning to the results of TNG-Cluster, we briefly consider the satellite galaxy populations. Figure \ref{fig_sat_profiles} shows the galaxy surface density i.e. radial number density profiles of galaxies surrounding TNG-Cluster halos at $z=0$. We stack mean profiles in five halo mass bins, as indicated in the legend \citep[see also][for related analysis with TNG300]{riggs22} Solid lines include galaxies satisfying $M_{\rm r} < -20.5$, in order to match observational selections, using the fiducial resolved dust modeling of \citet{nelson18a} to derive stellar light.\footnote{For comparison, the dotted lines show stellar magnitudes neglecting dust. Different color modeling assumptions have only a minor impact. Neglecting dust increases galaxy counts by $\sim 50$\%. Different projection depths also have marginal impact, slightly increasing counts at $\gtrsim 1$\,Mpc.} We project along a line-of-sight depth of $\pm 5$\,Mpc and average over three independent projection directions for each curve. The three sets of fainter, dashed lines show three stellar mass cuts. The lowest set of dashed curves corresponds to $M_\star > 10^{11.5}$\msun, while the middle set includes galaxies with $M_\star > 10^{10.5}$\msun -- the median stellar mass at this absolute magnitude threshold. Finally, the highest set of dashed curves shows lower mass galaxies with $M_\star > 10^{9}$\msun, a regime inaccessible in large sky surveys \citep{lin04b} but available with targeted cluster surveys \citep{vanderburg15}.

The three have notably different radial behavior, with $\Sigma_{\rm gal}(r)$ declining most slowly (rapidly) for the lower (higher) stellar mass samples. Galaxies with a Milky-Way like stellar mass of $M_\star > 10^{10.5}$\msun or higher are found with a density of $\sim 100\,$Mpc$^{-2}$ in the centers of clusters, where obscuration by the central galaxy at $\lesssim 30\,$kpc makes their identification in data challenging. Their density drops by a factor of one hundred, to $\sim 1$\,Mpc$^{-2}$ at a halocentric distance of 1\,Mpc. In all cases, a host halo mass trend is present: more massive clusters are surrounded by more satellite and neighboring galaxies. $\Sigma_{\rm gal}$ increases by roughly a factor of two, at any distance, for a factor of ten increase in host $M_{\rm 200c}$. This reflects the underlying dependence of richness on cluster mass -- we refer to the companion paper \citep{rohr24} for an analysis of satellite counts i.e. cluster richness and the corresponding comparison with observations.

With respect to data, broad agreement with the radial number density and color profiles of satellite galaxies around massive hosts was a noted success of the original Illustris simulation \citep{vog14a,sales15}, in contrast to semi-analytical models at the time \citep{wang14}. Here, we compare to observational measurements based on stacking of clusters from SDSS at $0.15 < z < 0.4$, with halo mass estimates from optical richness \citep{budzynski12}. From that work we show three sets of black markers, corresponding to three $M_{\rm 500}$ halo mass bins in $\log \rm{M}_\odot$: $14.0-14.4$, $14.4-14.7$, and $14.7-15.0$ (from lowest to highest). In all cases neighboring galaxies are included if they satisfy an absolute magnitude threshold of $M_{\rm r} < -20.5$, and should therefore be compared to the thick colored lines from TNG-Cluster. While the radial shape of $\Sigma_{\rm gal}(r)$ is remarkably similar with the data, the simulations appear offset to lower amplitudes. However, the trends with threshold stellar mass (or $r$-band magnitude) suggest that such differences can reflect biases as small as $\sim 0.1$\,dex in $M_\star$ or $M_{\rm r}$.

We therefore also compare to radial satellite profiles inferred from the GAMA survey, using a substantially different, friends-of-friends halo member identification methodology \citep[$r < 19.8$;][including a comparison with TNG300]{riggs22}. In particular, we show the highest halo mass bin available, for $13.7 < \log M_{\rm 200m} / \rm{M}_\odot < 14.8$ at $z < 0.27$ (black squares and line). We do not try to reproduce the methodology in detail, and offer it as a face value comparison. The overall shape of $\Sigma_{\rm gal}(r)$ is similar to TNG-Cluster, while the normalization suggests a best-match with our $M_{\rm 200c} \sim 10^{14.7}$\msun bin.\footnote{The rapid drop in the GAMA profiles at distances $\gtrsim 1$\,Mpc are understood to stem from the FoF algorithm \protect\citep[see][]{riggs22}.} 

We also compare to the radial galaxy number density measured around a sample of 60 massive clusters at $z < 0.26$, based on deep \textit{ugri}-band photometry \citep[][diamonds and dotted lines]{vanderburg15}. Halo masses are based on the \citet{evrard08} scaling relation from member velocity dispersion $\sigma_{\rm v}$, which is itself inferred from spectroscopically confirmed cluster members. We show two results: for galaxies with $10^9 < M_\star / \rm{M}_\odot < 10^{10}$ (upper curve) and for $M_\star > 10^{10}$\msun (lower curve).\footnote{$R_{\rm 200}$ scaled observational values are converted with the mean $R_{\rm 200} = 1.7$\,Mpc of the sample.} With a mean halo mass of $M_{\rm 200} \simeq 10^{14.8}$\msun, the sum of the two lines is best compared to the middle colored line in the $M_\star > 10^9$\msun group from TNG-Cluster. As above, the radial shape of $\Sigma_{\rm gal}(r)$ is in excellent agreement, while the observations suggest a somewhat higher normalization. In general, galaxy radial profiles are understood to be quite similar to lensing i.e. total matter profiles in cluster outskirts, though not necessarily in their cores, suggesting inside-out i.e. hierarchical assembly \citep{wang18b,shin21}.

These radial profiles of galaxy counts reflect the rich satellite galaxy populations of clusters. Simultaneously, the physical properties of those galaxies capture the process of galaxy evolution in such dense environments. By self-consistently modeling the stellar and gaseous bodies of the galaxy populations embedded within the ICM and dark matter halos of galaxy clusters, TNG-Cluster sheds light on the intertwined physical processes that shape some of the most massive structures in the Universe.


\section{Summary and Conclusions} \label{sec_conclusions}

In this paper we introduce the new TNG-Cluster project, a suite of cosmological magnetohydrodynamical simulations of high-mass galaxy clusters. Adopting the IllustrisTNG galaxy formation model, we simulate 352 clusters spanning a halo mass range of $\log{(M_{\rm 200c} / \rm{M}_\odot)} \gtrsim 14.5 - 15.4$ at $z=0$. We present the simulations, and first results on the sample and key physical properties, spanning halo assembly and dark matter, to cluster galaxies and their stars, the intracluster medium (ICM), and the supermassive black hole (SMBH) population. Our main results are:

\begin{itemize}
    \item The statistics of TNG-Cluster (Figure \ref{fig_mass_function}) are well matched to compare to the largest current observational surveys of galaxy clusters (Figure \ref{fig_sim_sample}), as detected in Sunyaev-Zeldovich (SZ) effect surveys including Planck, ACT-SZ DR5, and SPT-ECS; in X-ray emission with ROSAT, eROSITA, XMM-Newton XXL 365; and in the optical with SDSS, HSC x SpARCS, and more. The $z=0$ cluster progenitors reveal the assembly of these massive structures at higher redshifts, including protoclusters phases at $z \sim 1$ and $z \sim 2$.
    \item TNG-Cluster is currently the largest sample of simulated clusters at comparable numerical resolution, with $\sim 90$ halos at $M_{\rm 200c} > 10^{15}$\msun (Table \ref{simTable}). Compared to projects with similar statistics, its (mass) resolution is 10x to 100x better, with $m_{\rm baryon} = 10^7$\msun (Figure \ref{fig_sim_comparison}). It also includes the comprehensive, well-validated, and empirically constrained TNG galaxy formation model. By increasing the cluster sample previously available in the TNG300 simulation at unchanged model and resolution, we are able to quantitatively test this model in the high-mass regime, and explore a rich variety of physical phenomena in galaxy clusters (Figure \ref{fig_vis_single}).
    \item Halo gas fractions rapidly approach the cosmic baryon fraction, as expected for the most massive halos. There is no mass trend for clusters with $M_{\rm 500c} \gtrsim 10^{14.5}$\msun. This is consistent with X-ray and SZ data (Figure \ref{fig_gas_fraction}).
    \item Halo magnetic field strength increases with cluster mass, and towards higher redshift. Volume-weighted mean $|B|$ field strengths in the inner halo reach $\sim \mu\rm{G}$ values, consistent with the few available observational measurements (Figure \ref{fig_bfields}).
    \item Radio synchrotron emission ($P_{\rm 1.4 Ghz}$), modeled in postprocessing, shows a tight relation with halo mass, reaching $10^{25}$ W\,Hz$^{-1}$ for our most massive clusters, consistent with data (Figure \ref{fig_radio}). Radio relics and the structure of radio emission in TNG-Cluster is studied in detail in a companion paper \citep{lee24}.
    \item The scaling relations of Sunyaev-Zeldovich y-parameter, and soft-band X-ray luminosity from $0.5-2$\,keV, both measured within $R_{\rm 500c}$, are in reasonable agreement with available data (Figure \ref{fig_xray_sz}), as is halo X-ray luminosity as a function of galaxy stellar mass (Figure \ref{fig_xray}). These relations have non-negligible scatter that encodes the physics of halo assembly and baryonic physics.
    \item Radial profiles of ICM metallicity are remarkably flat, and broadly consistent with observations. Profiles of temperature and entropy are diverse, reflecting the diversity of the cluster population and the co-existence of cool-core and non-cool-core ICM structure, even at fixed halo mass (Figure \ref{fig_profiles}).
    \item Cluster BCGs host SMBHs with masses as high as $4-5 \times 10^{10}$\msun, and the relation between SMBH mass and $M_{\rm 500c}$ is consistent with the small observational samples at such high masses (Figure \ref{fig_smbhs}, left panel). The relation between $M_{\rm SMBH}$ and stellar velocity dispersion $\sigma_\star$ flattens in the cluster regime (Figure \ref{fig_smbhs}, right panel), potentially reflecting a shift in the physics of SMBH growth and how AGN feedback interacts with gaseous halos. The impact of SMBH feedback on the resolved structure of the ICM is the subject of a companion paper (\textcolor{blue}{Pillepich et al. in prep}).
    \item The vast majority of central galaxies in TNG-Cluster are quiescent, with specific star formation rates $\rm{sSFR} < 10^{-14}$\,yr$^{-1}$. However, a small but non-negligible fraction exhibits activity up to $\rm{SFR} \gtrsim 100 - 1000 \rm{M}_\odot\,\rm{yr}^{-1}$ (Figure \ref{fig_coolgas}). The resulting star-forming BCG fractions (or quenched fractions) as a function of galaxy mass are in broad agreement with observations.
    \item Star-forming BCGs are preferentially in cool-core (CC), as opposed to non-cool-core (NCC) clusters. TNG-Cluster produces a diversity of strong CC, weak CC, and NCC halos, studied in a companion paper \citep{lehle24}.
    \item The ICM is multi-phase. In particular, cluster halos contain non-vanishing amounts of cool gas, including neutral HI. The $M_{\rm HI}$ increases with halo mass, and is the ball park of observational data (Figure \ref{fig_coolgas}, right panel).
    \item Kinematics and turbulence in cluster cores are studied in detail in a companion paper \citep{truong24}, while velocity structure across the halo and into the outskirts is the focus of \citet{ayromlou24}.
    \item Stellar masses of cluster BCGs occupy a similar parameter space as a function of halo mass as observations, although they are above some semi-empirical models and mean values for available samples by $\sim 0.2$ dex (Figure \ref{fig_stellar_mass}). This may indicate insufficient star-formation suppression by the SMBH feedback model of TNG, either at late times or in proto-cluster phases at $z \sim 1-2$.
    \item Satellite populations are rich, and their overall abundance and radial distribution increases with halo mass in agreement with data (Figure \ref{fig_sat_profiles}). The satellite galaxy population, and their gaseous contents in particular, is the focus of a companion paper \citep{rohr24}.
\end{itemize}

By surveying a broad range of physics, physical processes, physical properties, and observable outcomes, we emphasize the breadth of TNG-Cluster and its many potential applications. In fact, the majority of the comparisons to observations presented in this paper are intended at face value only, as we have not replicated in detail observational effects. As a result, we expect that each of the scientific topics explored here will be revisited in more depth in future work. 

Practically, TNG-Cluster achieves high resolution due to its nature as a suite of individual zoom simulations. This is in contrast to large uniform volume simulations such as MillenniumTNG and FLAMINGO. By trading resolution for volume they can probe questions that TNG-Cluster cannot, such as large-scale galaxy clustering, and cosmological projection i.e. lightcone effects on observables. In contrast, TNG-Cluster focuses on the details and structure of cluster halos, the ICM, the galaxy populations in and near high-mass halos, including their SMBHs, and the assembly of all these components across cosmic time.

TNG-Cluster is unique in several model aspects, including in its modeling of magnetic fields. An assessment of the level of (dis)agreement with observational data in the cluster mass regime will benefit from more detailed, quantitative comparisons in the future. In addition, other potentially important (plasma) physics processes of interest for the ICM are not yet treated. These include (anisotropic) thermal conduction, radiative transfer, and non-continuum i.e. kinetic effects, as well as relativistic particle populations such as cosmic rays, and relativistic outflows/jets launched by SMBHs. While these complex physical processes have been explored in idealized contexts, their inclusion in full cosmological simulations remains a future goal.


\section*{Data Availability}

The IllustrisTNG simulations themselves are publicly available and accessible at \url{www.tng-project.org/data}, as described in \cite{nelson19a}, where the TNG-Cluster simulation will also be made public in 2024. Data directly related to this publication is available on request from the corresponding author.

\section*{Acknowledgements}

DN and MA acknowledge funding from the Deutsche Forschungsgemeinschaft (DFG) through an Emmy Noether Research Group (grant number NE 2441/1-1). This work is co-funded by the European Union (ERC, COSMIC-KEY, 101087822, PI: Pillepich). KL acknowledges funding from the Hector Fellow Academy through a Research Career Development Award. KL and ER are fellows of the International Max Planck Research School for Astronomy and Cosmic Physics at the University of Heidelberg (IMPRS-HD). NT acknowledges that the material is based upon work supported by NASA under award number 80GSFC21M0002. This work is supported by the Deutsche Forschungsgemeinschaft (DFG, German Research Foundation) under Germany's Excellence Strategy EXC 2181/1 - 390900948 (the Heidelberg STRUCTURES Excellence Cluster).

The TNG-Cluster simulation suite has been executed on several machines: with compute time awarded under the TNG-Cluster project on the HoreKa supercomputer, funded by the Ministry of Science, Research and the Arts Baden-Württemberg and by the Federal Ministry of Education and Research; the bwForCluster Helix supercomputer, supported by the state of Baden-Württemberg through bwHPC and the German Research Foundation (DFG) through grant INST 35/1597-1 FUGG; the Vera cluster of the Max Planck Institute for Astronomy (MPIA), as well as the Cobra and Raven clusters, all three operated by the Max Planck Computational Data Facility (MPCDF); and the BinAC cluster, supported by the High Performance and Cloud Computing Group at the Zentrum für Datenverarbeitung of the University of Tübingen, the state of Baden-Württemberg through bwHPC and the German Research Foundation (DFG) through grant no INST 37/935-1 FUGG. The three original TNG simulations were run with compute time awarded by the Gauss Centre for Supercomputing (GCS) under GCS Large-Scale Projects GCS-ILLU and GCS-DWAR on the Hazel Hen supercomputer at the High Performance Computing Center Stuttgart (HLRS). 

\bibliographystyle{aa}
\bibliography{refs}

\clearpage
\begin{appendix}

\section{Technical Aspects}
\label{sec_app1}

\begin{figure}
\centering
\includegraphics[angle=0,width=3.5in]{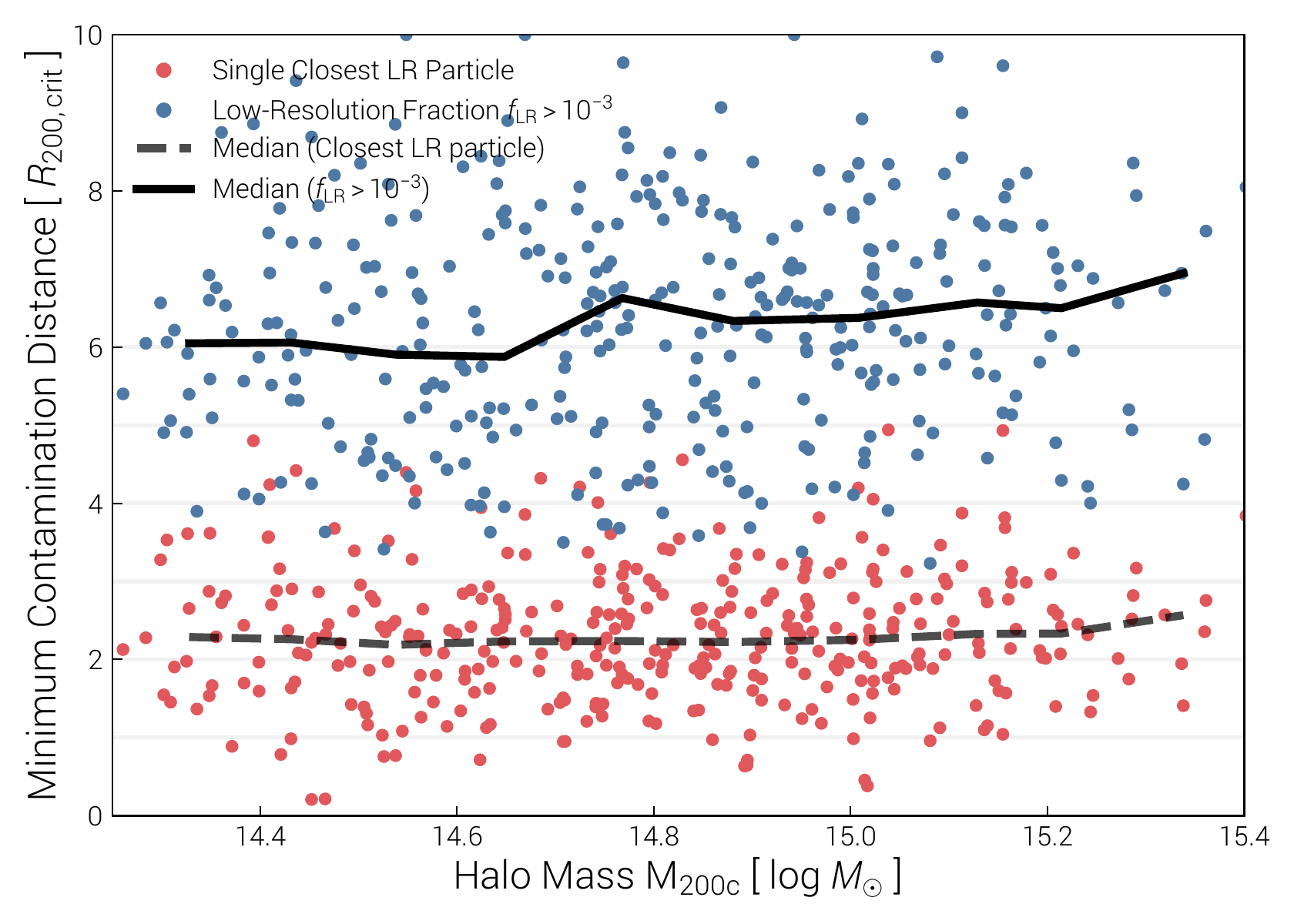}
\includegraphics[angle=0,width=3.45in]{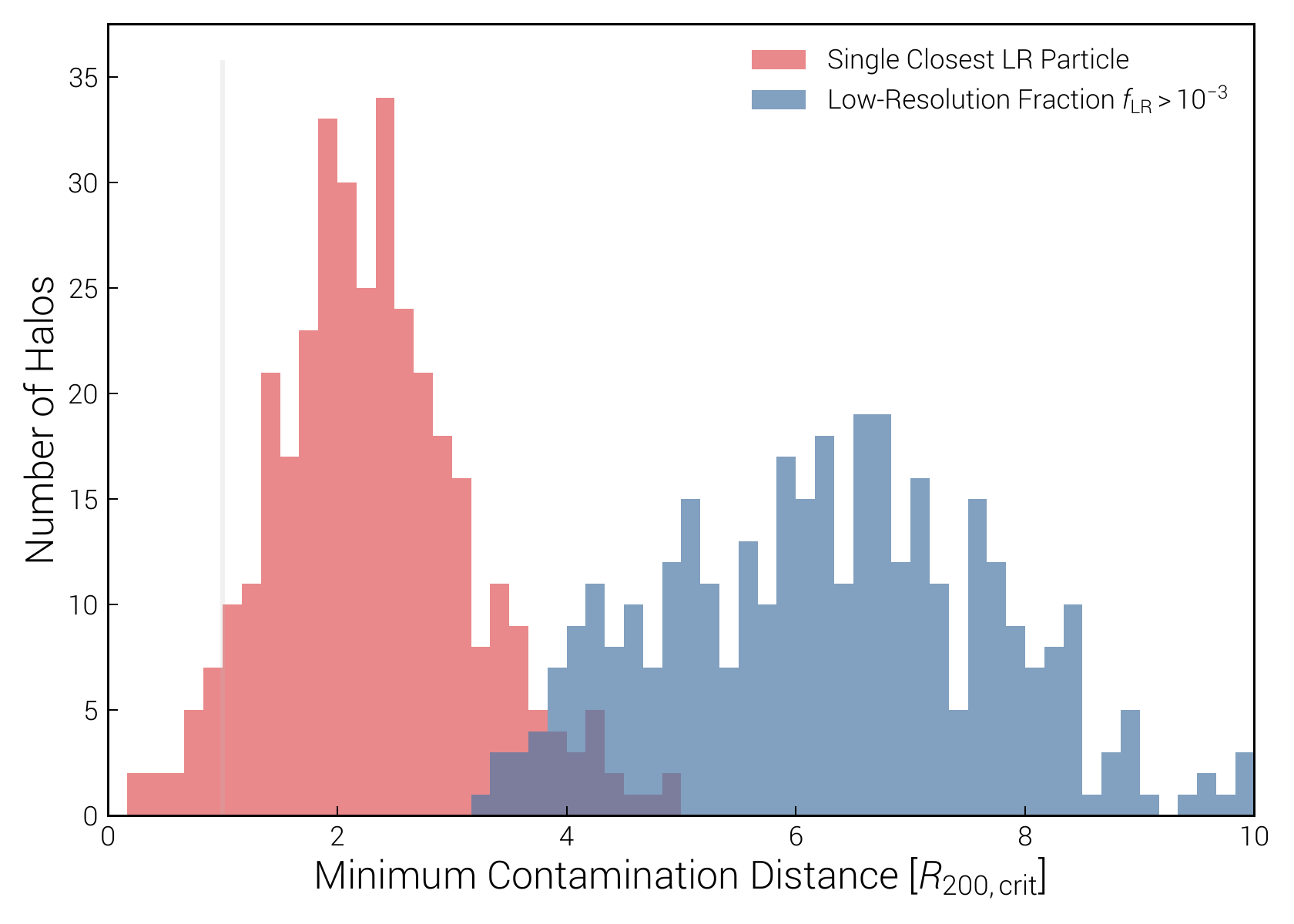}
\caption{ Numerical diagnostic of contamination of the high-resolution zoom regions in TNG-Cluster by lower resolution mass elements, at $z=0$. \textbf{Top:} The minimum distance to the closest low-resolution dark matter particle from the center of each halo (red markers, dashed line), and the distance at which the low-resolution particle fraction exceeds one per thousand (blue markers, solid line). \textbf{Bottom:} Histograms of the number of halos with given minimum contamination radii, as measured by the single closest low-resolution dark matter particle (red) or as when the low-resolution particle fraction exceeds $10^{-3}$ (blue). A handful of halos have one or a few interloping low-resolution particles within one or two virial radii. However, the median distance out to which halos are more or less fully resolved by high-resolution particles is $\sim 4-8$ virial radii ($\gtrsim 5-10$ pMpc).
 \label{fig_contam}}
\end{figure}

Figure \ref{fig_contam} explores the level of low-resolution `contamination' in and near our target clusters. In any zoom simulation, the presence of low-resolution particles in the regions of interest (i.e. within the targeted halos) is undesirable and signifies a degradation of the numerical fidelity of the simulation \citep{onorbe14}. Although not a strict boundary, distances out to which no, or little, contamination is present represent the volumes of interest when using TNG-Cluster.

For the fourteen halos with nonzero contamination within $1.0 r_{\rm vir}$, the median number of low resolution particles in that radius is just one, i.e. insignificant. The median distance out to which halos are more or less fully resolved by high-resolution particles is $\sim 3 R_{\rm 200c}$ ($\gtrsim 4-5$ pMpc). However, if we relax our criterion to one low-resolution particle thousand, i.e. still insignificant, this distance roughly doubles. We also note that contaminating particles are less and less of a concern at high redshift. For example, at $z=7$, the `minimum contamination distances' shift to $> 10 R_{\rm 200c}$ for all halos. Contamination of the TNG-Cluster halos is not problematic.

By construction, large e.g. $\sim 10$\,Mpc scale environmental effects, and large-scale projection effects (e.g. cosmological foregrounds or backgrounds along the lightcone) cannot be directly studied with TNG-Cluster. As designed, however, TNG-Cluster is well-suited not only for studies of the clusters themselves, but also their outskirts, nearby galaxy populations, and large-scale environments out to at least several virial radii.

\section{The TNG-Cluster Catalog: physical properties and observables}
\label{sec_app2}

Table \ref{table_clusters} presents the `TNG-Cluster Catalog' where we collect a wide range of physical properties for each simulated galaxy cluster. These range from theoretical quantities such as formation redshift and dark matter halo concentration to observables such as X-ray luminosity or richness. Basic properties, such as halo mass and $R_{\rm 500c}$ are accompanied by predictions for specific observational quantities with detailed forward-modeling pipelines.

This is a dynamic table which is available, and will be continually updated, online (\url{www.tng-project.org/cluster}). In particular, analyses from the other companion papers presenting first results from TNG-Cluster, as well as results from future publications, will be collected and unified. It is downloadable online in various machine readable formats.


\onecolumn
{\setlength{\tabcolsep}{3.0pt}
\begin{longtable}{llccccccccccccrr}
    \hline\hline
    $\vphantom{X^{X^X}}$ ID$^{\rm \dagger}$ & 
    Halo ID & 
    $M_{\rm 200c}$ & 
    $M_{\rm 500c}$ & 
    $R_{\rm 200c}$ & 
    $R_{\rm 500c}$ & 
    $M_{\rm \star,30 kpc}$ &
    $M_{\rm HI}$ &
    $f_{\rm gas,500}$ &
    SFR & 
    $M_{\rm SMBH}$ &
    $L_{\rm X,500}$ &
    $Y_{500}$ & 
    $z_{\rm form}$ &
    $\lambda$ &
    $\rm{CC?}$ \\
    &&log&log&&&log&log&&log&log&log&log&&&\\
    $\vphantom{X^{X}_{X_{X}}}$ $-$ & 
    $-$ &  
    $[\rm{M}_\odot]$ & 
    $[\rm{M}_\odot]$ & 
    $[\rm{Mpc}]$ & 
    $[\rm{Mpc}]$ & 
    $[\rm{M}_\odot]$ & 
    $[\rm{M}_\odot]$ & 
    $-$ & 
    $[\rm{M}_\odot/\rm{yr}]$ & 
    $[\rm{M}_\odot]$ & 
    $[\rm{erg/s}]$ & 
    $[\rm{Mpc^2}]$ & 
    $-$ & 
    $-$ & 
    $-$ \\ 
    \hline
    \endhead 
$\vphantom{X^{X^{X}}}$0 &        0 & 15.29 & 15.13 & 2.629 & 1.711 & 12.58 & 11.74 & 0.145 &    -- & 10.51 & 45.28 & -3.86 & 0.49 & 248 &  CC \\
       1 &   252455 & 15.40 & 15.27 & 2.871 & 1.912 & 12.16 & 11.37 & 0.126 &    -- &  9.35 & 44.96 & -3.65 & 0.38 & 202 & NCC \\
       2 &   476245 & 15.16 & 14.99 & 2.378 & 1.541 & 12.16 & 11.82 & 0.132 &    -- & 10.17 & 44.59 & -4.10 & 0.39 & 210 & NCC \\
       3 &   692426 & 15.36 & 15.14 & 2.782 & 1.731 & 12.53 & 11.55 & 0.149 &    -- & 10.19 & 45.06 & -3.78 & 0.08 & 167 & WCC \\
       4 &   880597 & 14.94 & 14.71 & 2.018 & 1.248 & 12.14 & 11.66 & 0.136 &    -- &  9.64 & 44.51 & -4.58 & 0.17 & 180 & WCC \\
       5 &  1068507 & 15.16 & 15.02 & 2.389 & 1.575 & 12.20 & 11.56 & 0.142 &    -- & 10.43 & 45.08 & -4.01 & 0.57 & 178 & WCC \\
       6 &  1263653 & 15.34 & 15.08 & 2.730 & 1.654 & 12.42 & 11.22 & 0.130 &    -- & 10.41 & 44.66 & -3.97 & 0.37 & 147 & NCC \\
       7 &  1431487 & 15.29 & 15.05 & 2.635 & 1.611 & 12.62 & 11.43 & 0.142 &    -- & 10.20 & 45.07 & -4.02 & 0.12 & 151 & WCC \\
       8 &  1580995 & 15.36 & 15.19 & 2.779 & 1.799 & 12.52 & 11.37 & 0.149 &    -- & 10.60 & 45.30 & -3.69 & 0.56 & 132 & WCC \\
       9 &  1763481 & 15.21 & 15.03 & 2.469 & 1.588 & 12.20 & 11.57 & 0.148 &    -- & 10.32 & 45.03 & -3.98 & 0.35 & 149 & WCC \\
      10 &  1921024 & 15.23 & 15.03 & 2.509 & 1.589 & 12.40 & 11.49 & 0.144 &    -- & 10.38 & 44.90 & -4.00 & 0.36 & 144 & WCC \\
      11 &  2069291 & 15.15 & 15.02 & 2.375 & 1.583 & 12.42 & 11.30 & 0.135 &    -- & 10.39 & 44.69 & -4.06 & 0.60 & 120 & NCC \\
      12 &  2213486 & 15.29 & 15.11 & 2.626 & 1.689 & 12.34 & 11.43 & 0.146 &    -- & 10.38 & 45.07 & -3.85 & 0.43 & 132 & WCC \\
      13 &  2380799 & 15.34 & 15.18 & 2.733 & 1.789 & 12.52 & 11.58 & 0.146 &    -- & 10.39 & 45.17 & -3.76 & 0.42 & 110 & WCC \\
      14 &  2507618 & 15.32 & 15.17 & 2.694 & 1.774 & 12.43 & 11.56 & 0.151 &  2.68 & 10.65 & 45.49 & -3.70 & 0.71 & 132 &  CC \\
      15 &  2637623 & 15.20 & 15.07 & 2.465 & 1.645 & 12.56 & 11.60 & 0.139 &    -- & 10.45 & 44.91 & -3.97 & 0.56 & 115 & WCC \\
      16 &  2771226 & 15.19 & 15.06 & 2.444 & 1.622 & 12.22 & 11.57 & 0.143 &    -- & 10.29 & 44.92 & -3.99 & 0.35 & 126 & WCC \\
      17 &  2892630 & 15.28 & 15.18 & 2.620 & 1.788 & 12.39 & 11.46 & 0.142 &    -- & 10.53 & 45.14 & -3.72 & 0.58 & 134 & WCC \\
      18 &  3020669 & 15.24 & 15.07 & 2.543 & 1.634 & 12.47 & 11.10 & 0.137 &    -- & 10.34 & 44.90 & -3.97 & 0.33 &  94 & WCC \\
      19 &  3135769 & 15.19 & 14.97 & 2.448 & 1.515 & 12.47 & 11.68 & 0.152 &  2.71 & 10.37 & 45.45 & -4.09 & 0.39 & 106 &  CC \\
      20 &  3238091 & 15.09 & 14.94 & 2.256 & 1.489 & 12.40 & 11.64 & 0.146 &    -- & 10.40 & 45.19 & -4.11 & 0.91 & 121 &  CC \\
      21 &  3343359 & 15.27 & 15.08 & 2.598 & 1.655 & 12.43 & 11.30 & 0.147 &    -- & 10.51 & 45.20 & -3.95 & 0.18 &  90 & WCC \\
      22 &  3457081 & 15.23 & 15.12 & 2.518 & 1.702 & 12.34 & 11.51 & 0.136 &    -- & 10.46 & 44.88 & -3.90 & 0.60 & 112 & NCC \\
      23 &  3556542 & 15.25 & 15.04 & 2.548 & 1.601 & 12.35 & 11.10 & 0.131 &    -- & 10.11 & 44.59 & -4.09 & 0.33 &  88 & NCC \\
      24 &  3656685 & 15.13 & 14.92 & 2.331 & 1.462 & 12.54 & 11.49 & 0.146 &  2.71 & 10.35 & 45.50 & -4.14 & 0.39 &  97 &  CC \\
      25 &  3758330 & 15.21 & 14.94 & 2.474 & 1.489 & 12.30 & 10.99 & 0.132 &    -- & 10.06 & 44.52 & -4.21 & 0.29 & 114 & NCC \\
      26 &  3863281 & 15.24 & 15.13 & 2.537 & 1.723 & 12.23 & 11.18 & 0.147 &    -- & 10.25 & 45.35 & -3.75 & 0.32 & 112 & NCC \\
      27 &  3960854 & 15.10 & 14.92 & 2.269 & 1.458 & 12.24 & 11.49 & 0.146 &    -- & 10.25 & 44.96 & -4.20 & 0.26 & 108 & WCC \\
      28 &  4054274 & 15.21 & 14.99 & 2.485 & 1.547 & 12.28 & 11.16 & 0.142 &    -- & 10.13 & 45.07 & -4.08 & 0.34 & 101 &  CC \\
      29 &  4152102 & 15.20 & 15.06 & 2.455 & 1.622 & 12.54 & 11.25 & 0.138 &    -- & 10.31 & 44.97 & -4.00 & 0.42 &  80 & WCC \\
      30 &  4254230 & 15.21 & 15.04 & 2.484 & 1.605 & 12.40 & 11.52 & 0.147 &    -- & 10.00 & 44.86 & -4.03 & 0.19 & 115 & WCC \\
      31 &  4356618 & 15.21 & 15.09 & 2.478 & 1.664 & 12.52 & 11.07 & 0.138 &    -- & 10.53 & 45.16 & -3.90 & 0.69 &  91 & WCC \\
      33 &  4461296 & 15.07 & 14.92 & 2.225 & 1.457 & 12.35 & 11.39 & 0.146 &    -- & 10.22 & 44.81 & -4.19 & 0.45 &  90 & WCC \\
      34 &  4561713 & 15.11 & 14.98 & 2.300 & 1.531 & 12.23 & 11.36 & 0.140 &    -- &  9.83 & 45.11 & -4.09 & 0.96 &  89 &  CC \\
      35 &  4656216 & 15.18 & 15.00 & 2.418 & 1.548 & 12.53 & 11.02 & 0.142 &    -- & 10.30 & 44.67 & -4.07 & 0.51 &  92 & WCC \\
      36 &  4753637 & 14.81 & 14.68 & 1.821 & 1.213 & 12.38 & 11.21 & 0.140 &    -- & 10.10 & 44.80 & -4.57 & 0.77 & 115 & WCC \\
      37 &  4854863 & 15.13 & 14.98 & 2.330 & 1.536 & 12.52 & 11.35 & 0.148 &    -- & 10.38 & 45.30 & -4.05 & 0.43 &  93 &  CC \\
      38 &  4959833 & 15.16 & 15.02 & 2.378 & 1.577 & 12.59 & 11.77 & 0.141 &    -- & 10.33 & 45.26 & -4.02 & 0.69 & 107 &  CC \\
      39 &  5059214 & 15.09 & 14.92 & 2.262 & 1.458 & 12.50 & 11.23 & 0.143 &  2.72 & 10.23 & 45.02 & -4.17 & 0.54 &  83 &  CC \\
      40 &  5160063 & 15.15 & 14.95 & 2.367 & 1.497 & 11.94 & 11.69 & 0.142 &    -- &  9.60 & 45.18 & -4.21 & 0.20 &  92 & NCC \\
      42 &  5251697 & 14.86 & 14.74 & 1.888 & 1.269 & 12.21 & 11.27 & 0.135 &    -- &  9.81 & 44.53 & -4.51 & 0.48 &  65 & NCC \\
      44 &  5348819 & 15.01 & 14.57 & 2.128 & 1.119 & 12.01 & 11.10 & 0.116 &    -- &  9.65 & 44.33 & -4.84 & 0.12 &  71 & WCC \\
      45 &  5432141 & 15.10 & 14.97 & 2.273 & 1.517 & 12.30 & 11.57 & 0.142 &    -- & 10.28 & 44.82 & -4.10 & 0.56 &  93 & WCC \\
    ... & ... & ... & ... & ... & ... & ... & ... & ... & ... & ... & ... & ... & ... & ... & ... \\
    4274 & 19372389 & 14.31 & 14.16 & 1.244 & 0.816 & 11.81 & 11.23 & 0.128 & -0.61 &  9.53 & 43.67 & -5.47 & 0.63 &  24 & WCC \\
    4369 & 19389199 & 14.30 & 13.99 & 1.234 & 0.716 & 11.73 & 11.02 & 0.102 &    -- &  9.33 & 43.18 & -5.78 & 0.22 &  18 & WCC \\
    4394 & 19402069 & 14.31 & 14.13 & 1.237 & 0.800 & 11.80 &  9.35 & 0.130 &    -- &  9.65 & 43.75 & -5.50 & 0.77 &  15 & WCC \\
    4414 & 19416927 & 14.35 & 14.21 & 1.279 & 0.847 & 11.62 & 11.05 & 0.119 &    -- &  9.02 & 43.60 & -5.43 & 0.18 &  15 & WCC \\
    5122 & 19429412 & 14.26 & 14.00 & 1.195 & 0.719 & 11.71 & 10.47 & 0.118 &    -- &  9.24 & 43.37 & -5.74 & 0.23 &  12 & WCC \\
    5711 & 19441351 & 14.31 & 14.19 & 1.241 & 0.837 & 11.92 & 10.39 & 0.132 &    -- &  9.63 & 43.89 & -5.42 & 1.00 &   8 & WCC \\
    \hline
    \hline
\caption{$\vphantom{X^{X^{X}}}$ Selection of physical properties of the primary TNG-Cluster sample. Each of the 352 clusters is given as a single row. All values are quoted at $z=0$ unless specified otherwise, and galaxy properties are provided for the central/BCG of each halo. The columns are: ID$^\dagger$ the original Halo ID (from the parent simulation), Halo ID (in TNG-Cluster), halo mass $M_{\rm 200c}$, halo mass $M_{\rm 500c}$, virial radius $R_{\rm 200c}$, $R_{\rm 500c}$, stellar mass measured within a 30 pkpc aperture, neutral HI gas mass in the halo, gas fraction within $R_{\rm 500c}$, star formation rate (`$-$' denotes an upper limit of $< 10^{-3} \,\rm{M}_\odot {\rm yr}^{-1}$), SMBH mass, X-ray luminosity within $R_{\rm 500c}$ in the $0.5-2$\,keV soft-band, integrated Sunyaev-Zeldovich y-parameter within $R_{\rm 500c}$ (both 3D spherical apertures), formation redshift, richness defined as the number of satellite galaxies with $M_\star > 10^{10.5}$\msun, and cool-core state (CC for cool-core, WCC for weak cool-core, and NCC for non-cool core halos, from \textcolor{blue}{Lehle et al. 2023}, as an example). This is a \textbf{dynamic, online table which contains many additional properties}. We will continually update this `TNG-Cluster Catalog' with specialized and contributed analyses. The full, unabridged table is available for \href{https://www.tng-project.org/cluster}{download online} in various formats.}
\label{table_clusters}
\end{longtable}
} 
\twocolumn

\end{appendix}

\end{document}